\documentclass[10pt,preprint]{aastex}
\tighten

\newcommand{\Barnes}{{C. Barnes}}
\newcommand{\Bennett}{{C. L. Bennett}}
\newcommand{\Greason}{{M. R. Greason}}
\newcommand{\Halpern}{{M. Halpern}}
\newcommand{\Hill}{{R. S. Hill}}
\newcommand{\Hinshaw}{{G. Hinshaw}}
\newcommand{\Jarosik}{{N. Jarosik}}
\newcommand{\Kogut}{{A. Kogut}}

\newcommand{\Limon}{{M. Limon}}
\newcommand{\Meyer}{{S. S. Meyer}}

\newcommand{\Odegard}{{N. Odegard}}
\newcommand{\Page}{{L. Page}}

\newcommand{\Spergel}{{D. N. Spergel}}
\newcommand{\Tucker}{{G. S. Tucker}}

\newcommand{\Weiland}{{J. L. Weiland}}
\newcommand{\Wollack}{{E. Wollack}}
\newcommand{\Wright}{{E. L. Wright}}

\newcommand{\Brown}{{Dept. of Physics, Brown University, %
            Providence, RI 02912}}
\newcommand{\Goddard}{{Code 685, Goddard Space Flight Center, %
            Greenbelt, MD 20771}}
\newcommand{\NRCFellow}{{National Research Council (NRC) Fellow}}
\newcommand{\PrincetonPhysics}{{Dept. of Physics, Jadwin Hall, %
            Princeton, NJ 08544}}
\newcommand{\PrincetonAstro}{{Dept of Astrophysical Sciences, %
            Princeton University, Princeton, NJ 08544}}
\newcommand{\SSAI}{{Science Systems and Applications, Inc. (SSAI), %
            10210 Greenbelt Road, Suite 600 Lanham, Maryland 20706}}
\newcommand{\UBC}{{Dept. of Physics and Astronomy, University of %
            British Columbia, Vancouver, BC  Canada V6T 1Z1}}
\newcommand{\UChicago}{{Depts. of Astrophysics and Physics, EFI and CfCP, %
            University of Chicago, Chicago, IL 60637}}
\newcommand{\UCLA}{{UCLA Astronomy, PO Box 951562, Los Angeles, CA 90095-1562}}

\shorttitle{{\sl WMAP} Data Processing}
\shortauthors{Hinshaw et al.}

\newcommand{\map}          {{\sl WMAP}}
\newcommand{\cobe}         {{\sl COBE}}
\newcommand{\dg}           {\mbox{$^{\circ}$}}
\newcommand{\uth}          {\mbox{$^{\rm th}$}}
\newcommand{\lsim}         {\mbox{$_<\atop^{\sim}$}}

\newcommand{\lt}           {\mbox{$<$}}

\newcommand{\asec}         {\mbox{$^{\prime\prime}\ $}}
\newcommand{\ddeg}         {\mbox{${\rlap.}^\circ$}}
\newcommand{\damin}        {\mbox{${\rlap.}^\prime$}}

\newcommand{\per}          {$^{-1}$}

\slugcomment{To be submitted to the Astrophysical Journal}

\begin{document}

\title{First Year {\sl Wilkinson Microwave Anisotropy Probe} ({\sl 
WMAP}\altaffilmark{1}) Observations:
Data Processing Methods and Systematic Errors Limits}

\author{
\Hinshaw \altaffilmark{2},
\Barnes \altaffilmark{3},
\Bennett \altaffilmark{2},
\Greason \altaffilmark{4},
\Halpern \altaffilmark{5},
\Hill \altaffilmark{4},
\Jarosik \altaffilmark{3},
\Kogut \altaffilmark{2}, 
\Limon \altaffilmark{2,6}, 
\Meyer \altaffilmark{7},
\Odegard \altaffilmark{4},
\Page \altaffilmark{3},
\Spergel \altaffilmark{8},
\Tucker \altaffilmark{2,6,9},
\Weiland \altaffilmark{4},
\Wollack \altaffilmark{2},
\Wright \altaffilmark{10}}

\altaffiltext{1}{\map\ is the result of a partnership between Princeton 
                 University and NASA's Goddard Space Flight Center. Scientific 
		 guidance is provided by the \map\ Science Team.}
\altaffiltext{2}{\Goddard}
\altaffiltext{3}{\PrincetonPhysics}
\altaffiltext{4}{\SSAI}
\altaffiltext{5}{\UBC}
\altaffiltext{6}{\NRCFellow}
\altaffiltext{7}{\UChicago}
\altaffiltext{8}{\PrincetonAstro}
\altaffiltext{9}{\Brown}
\altaffiltext{10}{\UCLA}

\email{Gary.F.Hinshaw@nasa.gov}

\begin{abstract}
We describe the calibration and data processing methods used to generate
full-sky  maps of the cosmic microwave background (CMB) from the first year of 
{\sl Wilkinson Microwave Anisotropy Probe (WMAP)} observations.  Detailed
limits on residual systematic errors are assigned based largely on analyses of
the flight data supplemented, where necessary, with results from ground tests. 
The data are calibrated in flight using the dipole modulation of the CMB due to
the observatory's motion around the Sun. This constitutes a full-beam
calibration source. An iterative algorithm simultaneously fits the time-ordered
data to obtain calibration parameters and pixelized sky map temperatures. The
noise properties are determined by analyzing the time-ordered data with this
sky signal estimate subtracted. Based on this, we apply a pre-whitening filter
to the time-ordered data to remove a low level of $1/f$ noise. We infer and
correct for a small ($\sim$1\%) transmission imbalance between the two sky
inputs to each differential radiometer, and we subtract a small sidelobe 
correction from the 23 GHz (K band) map prior to further analysis. {\em No
other systematic error corrections are applied to the data.} Calibration and
baseline artifacts, including the response to environmental perturbations, are
negligible. Systematic uncertainties are comparable to statistical
uncertainties in the characterization of the beam response. Both are accounted
for in the covariance matrix of the window function and are propagated to
uncertainties in the final power spectrum. We characterize the combined upper
limits to residual systematic uncertainties through the pixel covariance
matrix.
\end{abstract}


\section{INTRODUCTION}
\label{sec:intro}

The {\sl Wilkinson Microwave Anisotropy Probe (WMAP)} has produced full-sky
maps of the cosmic microwave background (CMB) of unprecedented precision and
accuracy. On angular scales larger than $\sim0\ddeg5$, the dominant uncertainty
is not the instrument noise but rather the ``cosmic variance'' inherent when
analyzing a single realization (the observable universe) of a random process.
In the cosmic variance limit, no further improvement can be made by reducing
instrument noise, placing even greater importance on the  minimization of
non-random instrumental effects in the data.

The \map\ design emphasizes control of systematic errors
\citep{bennett/etal:2003}.  The observatory was designed with a detailed 
systematic error budget in place, and a mature data analysis  pipeline was
written early to help inform many of the design decisions.  Differential
radiometers compare the temperature from widely-separated regions of the sky
through back-to-back matched optics. Common-mode signals thus cancel before
affecting the sky maps. The radiometer feed horns only illuminate a fraction of
the primary mirrors, reducing the sidelobe response in the beam patterns. The
instrument was designed to have minimal response to electrical or thermal
perturbations and operates in an exceptionally stable environment at the second
Earth-Sun Lagrange point. The observatory's compound spin and precession allow
rapid inter-comparison of different positions on the sky, greatly reducing the
coupling of systematic error signals into the sky maps and effectively
symmetrizing the beam response. \map\ data are calibrated in flight using the
dipole modulation of the CMB from the observatory's orbital motion around the
Sun as a full-beam calibration source. We measure the beam pattern in flight
using observations of the planet Jupiter.

We characterize or limit systematic errors in the \map\ first-year data using
flight data supplemented where necessary with results from ground tests.
Systematic errors may be classified into several broad categories including
the following:

{\em Calibration Errors}.--- The time-ordered data is simultaneously fit for
the calibration and sky map. An iterative algorithm updates the calibration
solution based on the previous iteration of the sky map solution,  and
vice-versa. The most important source of error is confusion between the dipole
signal and higher-order sky signal, especially bright Galactic foreground 
emission in the low frequency \map\ bands. See \S\ref{sec:dipole_cal}.

{\em Map-making errors}.---  These are due to poor convergence in the sky map
solution, or to errors in the determination of the spacecraft  pointing. See
\S\ref{sec:iterate_map} and \S\ref{sec:pointing}.

{\em Beam Errors}.--- Instrument noise, background subtraction, and pointing
errors can limit the in-flight measurement of the beam response from Jupiter. 
Although the beams are not symmetric, the observatory's compound spin and 
precession effectively symmetrize the beam response. Uncertainties in both the
beam solid angle and the window functions must be characterized. See 
\citet{page/etal:2003b} for a complete discussion of beam mapping. We
summarize and incorporate their results in \S\ref{sec:beam_window}.

{\em Sidelobe Response}.---  Sidelobe pickup of bright sources (e.g. the
Galactic plane) introduces an additive signal dependent on the orientation of
the beams on the sky. \citet{barnes/etal:2003} discuss the sidelobe response of
each radiometer and estimate the effect on the first-year sky maps.

{\em Baseline Errors}.--- Thermal or electrical perturbations can produce
signals dominated by an additive term in the time domain.   Slow drifts are
removed as part of the calibration procedure, but signals near the spin period
can couple to the sky maps with some efficiency. See \S\ref{sec:environmental}.

{\em  Striping}.--- Correlations in the time-ordered data from sources not
fixed on the sky (e.g. $1/f$ noise or post-detection filtering) introduce
correlated noise in the sky maps. Application of a pre-whitening filter to the
time-ordered data reduces this effect. See \S\ref{sec:filter}.

We have constructed a detailed model of the instrument that successfully
reproduces all major aspects of the instrument performance. Software
simulations using this model validate the map-making algorithm and allow us to
assess the effect in the sky maps  of various signals in the time-ordered data.
Based on this, we apply a pre-whitening filter to the time-ordered data to
remove a low level of $1/f$ noise. We infer and correct for a small ($\sim$1\%)
transmission imbalance between the two sky inputs to each differential
radiometer, and we subtract a small sidelobe  correction from the 23 GHz (K
band) map prior to further analysis. {\em No other systematic error corrections
are applied to the data.} Calibration and baseline artifacts, including the
response to environmental perturbations, are negligible. Systematic
uncertainties are comparable to statistical uncertainties in the
characterization of the beam response. Both are accounted for in the covariance
matrix of the window function and are propagated to uncertainties in the final
power spectrum. We characterize the combined upper limits to residual
systematic uncertainties through the pixel-pixel covariance

This paper is organized as follows. In \S\ref{sec:overview} we define the terms
and notation used throughout the paper.  In \S\ref{sec:pipeline} we discuss the
iterative algorithm for making maps from time-ordered data, then generalize to
the case of simultaneous calibration and sky map estimation. (Appendix
\ref{app:map_pol} further generalizes to map-making with polarization data.) 
We also discuss the noise properties of the time-ordered data.  In
\S\ref{sec:syserr} we discuss combined systematic error limits from calibration
and map-making.  We also present the noise properties of the sky maps in terms
of their pixel-pixel covariance. Finally, we derive upper limits to
environmental perturbations and summarize the combined systematic error
budget.  In \S\ref{sec:conclude} we present our conclusions.

\subsection{Notation and Overview}
\label{sec:overview}

Throughout this paper, we denote vectors and scalars with bold and plain
lowercase letters, respectively.  Matrices and operators are denoted with 
uppercase bold letters.  Following \citet{stompor/etal:2002} we denote vector
and matrix component indices in parentheses, saving subscripts and 
superscripts to further identify quantities.  A summary of the most frequently
used symbols is given in Table~\ref{tab:notation}.  Unless otherwise stated,
all temperatures are specified in thermodynamic units.

\map\ measures the brightness temperature of the sky as a function of position,
${\bf t}(\theta,\phi) \rightarrow {\bf t}(p)$, where $p$ denotes the sky map
pixel number, indexed from 0, in HEALPix nested format
\citep{gorski/hivon/wandelt:1998}.  To make this measurement, \map\ scans the
sky and measures the temperature  difference between two points at time $t$. The
resulting time-ordered  differential data (TOD) is denoted ${\bf d}$.  The main
goal of the map-making is to obtain the minimum variance estimate of the sky
map, ${\bf \tilde t}$, by inverting the raw differential data. Note that  ${\bf
\tilde t}$ is the true sky temperature convolved with the nominal  instrument
beam, plus instrument noise.  In the  process of solving for the  map, we
calibrate the data by estimating the gain and baseline from the  flight data
itself; characterize the full instrument beam response function  from
observations of the planets; characterize the noise spectrum of the instrument,
and place limits on residual systematic errors.

In order to produce stable data with a nearly-white noise spectrum \map\ 
employs 20 high-electron-mobility-transistor (HEMT) based differential 
radiometers.  Each radiometer measures the brightness difference between two
inputs, one fed by an A-side beam, the other by a B-side beam approximately
141\dg apart. A detailed description of their design and fabrication may be
found in \citet{jarosik/etal:2003}; a summary of their in-flight  performance
is presented in \citet{jarosik/etal:2003b}.  The 20 radiometers form 10
polarization-sensitive ``differencing assemblies'' (DA) which are designated
based on their frequency or waveguide band: K1, Ka1, Q1, Q2, V1, V2, W1, W2,
W3, W4.  The two radiometers in a DA are sensitive to orthogonal linear
polarization modes; the radiometers are designated  1 or 2 (e.g., K11 or K12)
depending on which polarization mode is being sensed.  Each of the 20
radiometers is intrinsically a 2-channel device, with channels  designated 3
and 4 in the flight telemetry, e.g., K113 or K114.  [Channels 3 and 4 were
designated left and right, respectively, in \citet{jarosik/etal:2003}.]  There
are 40 such data channels in the flight telemetry.  As discussed below,  each
of the 40 channels is individually calibrated, then the 4 channels from a 
single differencing assembly are combined to form differential intensity and 
polarization signals as follows.  Let ${\bf d}_{ij}$ be the calibrated 
differential signal from a single channel, $j$, of radiometer $i$.  The
differential intensity data (Stokes parameter $I$) is the average of all  4
channels
\begin{equation}
{\bf d} = \frac{1}{2}({\bf d}_{13} + {\bf d}_{14}) 
        + \frac{1}{2}({\bf d}_{23} + {\bf d}_{24}). 
\label{eq:dt_i}
\end{equation}
The differential polarization data is obtained by taking the difference between 
the two radiometer signals 
\begin{equation}
{\bf p} = \frac{1}{2}({\bf d}_{13} + {\bf d}_{14}) 
        - \frac{1}{2}({\bf d}_{23} + {\bf d}_{24}).
\label{eq:dt_p}
\end{equation}
In Appendix \ref{app:map_pol} on polarization map making, we relate the 
differential polarization signal to the Stokes parameters $Q$ and $U$.  
\citet{kogut/etal:2003} discuss additional aspects of polarization mapping  and
analyze the first-year temperature-polarization correlation data 
based on these maps. Note that we can also form null channels from the data by
taking channel differences, $({\bf d}_{i3} - {\bf d}_{i4})$.  As discussed in 
\S\ref{sec:diff_maps}, these channel combinations provide valuable
consistency  tests for the final sky maps.

A single channel of {\em uncalibrated} differential data may be modeled as
\begin{equation}
{\bf c} =  {\bf g}\left[ {\bf M\cdot(t + t_s) + n}\right] + {\bf b},
\end{equation}
where each quantity is a function of time:
\begin{list}{}{}
\item ${\bf c}(t)$: raw differential data, in counts or digital units (``du'').
\item ${\bf g}(t)$: instrument responsivity (here called gain), in du mK\per.
\item ${\bf b}(t)$: instrument baseline, in du.
\item ${\bf n}(t)$: instrument noise, in mK.
\item ${\bf M\cdot(t+t_s)} \equiv {\bf\Delta t}(t)$: differential sky signal 
from all sources, convolved with the instrument beam, in mK. This includes 
fixed sources, ${\bf t}$ (e.g., CMB and Galactic emission) and moving 
sources, ${\bf t_s}$ (e.g., planets).
\end{list}
In practice, the differential signal is integrated for a fixed time $\tau$ and 
sampled at discrete times $t_i$, thus we may regard time series data as a
vector with $N_t$ observations.  The integration time per observation is 128.0
ms, 128.0 ms, 102.4 ms, 76.8 ms, and 51.2 ms for bands K through W,
respectively.

The differential temperature at time $t$ is the convolution of a
time-dependent  mapping function, ${\bf M}$, with the sky signal at time $t$,
${\bf t+t_s}(t)$ 
\begin{equation}
{\bf\Delta t}(t) = \int d\Omega_{{\bf n}'} \, {\bf M}({\bf n}(t),{\bf n}') 
\, ({\bf t}({\bf n}') + {\bf t_s}({\bf n}',t)).
\label{eq:map_funct_def}
\end{equation} 
Here ${\bf t} = {\bf t_c} + {\bf t_g}$ is the fixed sky signal consisting of 
CMB anisotropy, ${\bf t_c}$, and Galactic foreground signals, ${\bf t_g}$, 
while ${\bf t_s}$ represents all time dependent sources, especially the Sun, 
Earth, and Moon, which are potentially visible in the far sidelobes of the 
instrument.  The operator ${\bf M}$ can be represented as an $N_t \times  N_p$
matrix where each row is the  normalized, full-sky beam response in  sky-fixed
coordinates as given by the scan pattern. Several features of the mapping
function that pertain to the data processing are discussed in Appendix
\ref{app:map_funct}. The main beam response is mapped  in flight using
observations of Jupiter as a far-field point source  \citep{page/etal:2003b}. 
An important aspect of the \map\  optical design \citep{page/etal:2003} was to
limit sidelobe pickup to negligible levels and to have the effective beam 
response in the final sky maps be approximately circularly symmetric. We
discuss each of  these topics in more detail in separate papers,
\citep{barnes/etal:2003,  page/etal:2003b}, while this paper summarizes the
main results in terms of systematic error limits in the sky maps.

The instrument gain, baseline, and noise are determined from the flight data 
itself.  This is an iterative process that we discuss in detail in 
\S\ref{sec:dipole_cal}.  Here we provide a brief overview of our terminology 
in order to frame the following discussion of systematic errors.  Let the gain 
and baseline measured in flight be ${\bf \tilde g}$ and ${\bf \tilde b}$, 
respectively.  The calibrated differential signal is then
\begin{equation}
{\bf\tilde d} = \frac{(\bf c - \tilde b)}{\bf \tilde g}
                     = \frac{\bf g}{\bf \tilde g}{\bf \Delta t}
                     + \frac{\bf g}{\bf \tilde g}{\bf n}
                     + \frac{\bf b - \tilde b}{\bf \tilde g}.
\label{eq:dt_rec}
\end{equation}
With calibrated data available, the sky map is obtained by evaluating the
linear equation
\begin{equation}
{\bf \tilde t} = {\bf W \tilde d},
\label{eq:map_basic}
\end{equation}
where ${\bf W}$ is a linear operator defined in \S\ref{sec:map_making_pre_cal}.
The properties of ${\bf W}$ are determined by the scan strategy of the 
observatory and the noise properties of the time-ordered data, ${\bf d}$.

The \map\ scan pattern is an integral part of the mission design 
\citep{bennett/etal:2003}. It consists of a compound spin and precession 
centered about the Sun--\map\ line, with parameters as given in 
Table~\ref{tab:acs_req}.  There are several aspects of this scan strategy that
are  important for high quality data: scans of a given pixel cross at many
angles so that the effective beam response is symmetric; a given pixel is
observed on many different time scales from minutes to months; the angular
velocity of a given line of sight is nearly constant on the sky; the instrument
observes  a large fraction ($>$30\%) of the sky each day; and the time-average
of the differential data is approximately  zero over an hourly calibration
period, allowing for robust initial baseline estimation.

\section{THE MAP-MAKING PIPELINE}
\label{sec:pipeline}

A graphical overview of the \map\ data processing and analysis pipeline is
shown in Figure~1 of \citet{bennett/etal:2003b}.  The heart of the pipeline 
is a set of programs that bring science and housekeeping data from the 
Science and Mission Operations Center (SMOC) through to a set of calibrated
full sky maps for each of the 10 \map\ differencing assemblies.  Numerous
additional programs are used to generate ancillary data products such as 
beam response maps, calibration files and analysis products.  In this section
we provide a high-level overview of each program in the map-making pipeline
and provide references to later sections of this paper, or to companion 
papers, for further details.

Raw telemetry data from the satellite is transferred approximately once per day
through NASA's Deep Space Network (DSN), to the SMOC, located at the  Goddard
Space Flight Center. The SMOC monitors the basic health and safety of  the
Observatory, sends all commands, and requests re-transmissions of data that
were missing from a previous transmission.  The data are then ``level-0'' 
processed into a set of time-ordered, daily files which contain science data, 
instrument housekeeping data, spacecraft data (including attitude and ephemeris
data), and event message files.  These files are then transferred to the
Science Team's processing facility, also at the Goddard Space Flight Center.

Every time a new full day of data arrives, a series of automated 
procedures perform the following tasks: 1) Generate a standard set of 
daily plots that are archived and visually inspected; 2) Generate a reduced 
``trending archive'', which consists of subsets of the data sampled once every 
10 minutes.  In the case of the science data, we record the mean and rms of
each channel in a 10 minute interval, whereas the housekeeping data are
sub-sampled, once per 10 minutes;  3) Perform a series of data quality checks 
that search the data for violation of pre-set range limits or excessive 
time-gradients in the telemetry signals.  Limit violations are logged and 
notification is sent to a member of the science team via e-mail. (Initial 
limit tests are performed at the SMOC as well.)

At selected time intervals, the level-0 telemetry files are collated by a 
pre-preprocessor into a master archive of raw (uncalibrated) data.  The major
functions of the pre-processor are to: 1) collate the science and housekeeping
data into single daily files; 2) flag data that is suspected or known to be
unusable; 3) interpolate the attitude and ephemeris data to times that are
commensurate with the science data time stamps; and 4) apply a  coarse flag to
data that is within 7$^{\circ}$ of one of the outer planets (Mars through
Neptune) so that it may be rejected from the initial sky maps, but identified
for beam mapping \citep{page/etal:2003b}.  (Only Jupiter data is used for
making the final beam maps.)

Initial sky maps and calibration data are generated from the raw archive using
the iterative map-making algorithm first described in 
\citet{wright/hinshaw/bennett:1996a}, and further described in 
\S\ref{sec:map_making_pre_cal}.  As discussed in  \S\ref{sec:map_making_cal},
the initial calibration is determined by fitting the raw time-ordered data to
the known signal produced by the CMB dipole.  Because the sky signal contains
significant higher-order power ($l > 1$), the  calibration solution must be
iteratively improved in concert with the  initial sky map iterations.  The
convergence of this simultaneous fit has been demonstrated with end-to-end
simulations, which are also described in \S\ref{sec:map_making_cal}.

The calibration solution converges more rapidly than the sky map does, so 
we freeze the initial calibration solution after $\sim$10 iterations before
proceeding to convergence with the sky maps.  With the initial dipole-based 
calibration data in place, a re-processor generates a refined gain 
and baseline solution and applies this to the data.  The program also updates
the data quality flags, as necessary, then writes the calibrated data to a new
final time-ordered archive.  The refinements to the dipole-based gain solution
are discussed in \citet{jarosik/etal:2003b} and in \S\ref{sec:dipole_cal}.  
The initial baseline solution is refined with a pre-whitening filter 
\citep{wright:1996} which is presented in detail in \S\ref{sec:filter}.

The final sky maps are computed using the final calibrated time-ordered data as
input.  The first-year sky maps required 20  post-calibration iterations to be
sufficiently converged. The map-making algorithm is fundamentally  the same as
is used in the initial estimates, but we add some refinements for this final
stage of processing.  These include: 1) correcting for a small  ($\lsim$1\%)
loss imbalance between the A and B-side sky beams.  \citet{jarosik/etal:2003b}
demonstrate that this effect is nearly orthogonal to the gain solution, so its
inclusion after the calibration processing does not invalidate the gain
solution.  2) We weight individual time-ordered observations by their proper
statistical weight to account for the small change in instrument noise 
($\lt$1\%) over the course of a year due to the 0.9 K seasonal temperature 
variation experienced by the instrument cold stage. 3) We compute the
planet--boresight angle for each observation to minimize  unnecessarily
conservative data loss.  The  criterion used for the first-year maps is a cut
of radius $1\ddeg5$.

We also generate polarization maps using a generalization of the temperature
algorithm, the details of which are presented in Appendix \ref{app:map_pol}. 
For the first-year data processing, we have generated maps of the Stokes
parameters  $Q$ and $U$,  but we have not yet fully characterized all of the
potential systematic errors in these maps.  However, the
temperature-polarization  correlation data are much less prone to systematic
errors than the polarization auto-correlation data. \citet{kogut/etal:2003}
have analyzed the temperature-polarization data and, supported by systematic
error limits from \citet{barnes/etal:2003}, they find  a significant
correlation, including  the signature of a relatively early epoch of cosmic
reionization.

In the remainder of this paper, we present a more detailed description of  each
of the map-making and calibration procedures, including an assessment of their
performance with the first-year \map\ data.  We then derive detailed  systematic
error limits applicable to the time-ordered data, to the sky maps, and to the
angular  power spectrum.

\subsection{Map-Making with Pre-Calibrated Data}
\label{sec:map_making_pre_cal}

While the process of generating the final sky maps from calibrated data comes
last in the map-making pipeline, we discuss the algorithm first because the
algebra of map-making is central to the entire data processing scheme, and it
helps to guide the systematic error analysis.

We consider the problem of estimating a sky map, ${\bf t}$, from calibrated,
differential time-ordered data, ${\bf d}$, which is a linear function of the 
sky map
\begin{equation}
{\bf d} =  {\bf M t} + {\bf n},
\label{eq:map_funct_mat}
\end{equation}
where ${\bf M}$ is the mapping function of the experiment, which has $N_p$
columns and $N_t$ rows, $N_p$ is the number of sky map pixels and 
$N_t$ is the total number of time-ordered observations. In its simplest
form, each row (observation) of the scan matrix contains a $+1$ in the column 
(pixel) seen by the A-side beam, and a $-1$ in the column (pixel) seen by the 
B-side beam.  This matrix can be generalized to include the effects of beam 
convolution, but for \map\ these refinements are small and are being deferred 
to future processing. The effects of a differential loss imbalance between
the A and B-side beams is readily accounted for by using values different from
$\pm$1 in ${\bf M}$.  An analysis of this effect in the \map\ radiometers is 
presented in \citet{jarosik/etal:2003b}.  The details of how we account 
for it in the pipeline are given in Appendices \ref{app:map_funct} and 
\ref{app:map_pol}.

The noise ${\bf n}$ is assumed to have zero mean and covariance ${\bf N}$,
\begin{eqnarray}
\left< {\bf n} \right> & = & 0 \\
\left< {\bf nn}^T \right> & = & {\bf N}.
\end{eqnarray}
We defer a detailed discussion of the \map\ noise properties to 
\S\ref{sec:filter}, but for most radiometers it is reasonable to approximate 
the noise covariance as diagonal, ${\bf N} \simeq \sigma_0^2 \,{\bf I}$, 
where $\sigma_0$ is the rms noise per observation and ${\bf I}$ is the 
identity matrix, though this assumption is not required for the algorithm 
described below to converge.

The least-squares, or maximum-likelihood estimate of the sky map, 
${\bf \tilde t}$, results from solving the normal equations
\begin{equation}
{\bf \tilde t} = ({\bf M}^T {\bf N}^{-1} {\bf M})^{-1}
\cdot ({\bf M}^T {\bf N}^{-1} {\bf d}).
\end{equation}
More generally, we obtain an unbiased estimate of the sky map by choosing any
symmetric matrix ${\bf S}$ in place of ${\bf N}^{-1}$ \citep{tegmark:1997}.  
To see this, substitute ${\bf M t} + {\bf n}$ in place of ${\bf d}$ in 
the above equation to get
\begin{equation}
{\bf \tilde t} = ({\bf M}^T {\bf S} {\bf M})^{-1}
\cdot ({\bf M}^T {\bf S} \, [{\bf M t} + {\bf n}])
               = {\bf t} + ({\bf M}^T {\bf S} {\bf M})^{-1}
\cdot ({\bf M}^T {\bf S} {\bf n}).
\end{equation}
Thus ${\bf \tilde t}$ reduces to ${\bf t}$ plus a noise term that is independent 
of ${\bf t}$ and has zero mean over an ensemble average. For the first-year 
processing we take ${\bf S} = {\bf I}$.  To simplify notation, we define a 
matrix ${\bf W} \equiv ({\bf M}^T {\bf M})^{-1} \cdot {\bf M}^T$ in which case 
the map solution is ${\bf \tilde t} = {\bf W d}$.

The pixel-pixel noise covariance in the sky map solution is
\begin{eqnarray}
{\bf\Sigma} & = & \left< ({\bf \tilde t - t})({\bf \tilde t - t})^T \right> 
              =   \left< ({\bf W n})({\bf W n})^T \right> 
              =   {\bf W N W}^T \nonumber \\
            & = & ({\bf M}^T {\bf M})^{-1} \cdot
                  ({\bf M}^T {\bf N}{\bf M}) \cdot
                  ({\bf M}^T {\bf M})^{-1}.
\label{eq:pixel_noise}
\end{eqnarray}
In the limit that ${\bf N}$ is diagonal and the rms noise per observation, 
$\sigma_0$, is constant, ${\bf \Sigma}$ reduces to $\sigma_0^2  \, 
({\bf M}^T {\bf M})^{-1}$. For the \map\ scan pattern, the matrix  
${\bf M}^T {\bf M}$ is diagonally dominant with diagonal elements 
${\bf n}_{\rm obs}(p)$, the number of observations of pixel $p$ by either the A 
or B-side beam.  Thus, to a very good approximation, the pixel-pixel 
covariance matrix is diagonal
\begin{equation}
{\bf\Sigma}(p_i,p_j) \simeq \frac{\sigma_0^2}{{\bf n}_{\rm obs}(p_i)} \delta_{ij}.
\end{equation}
Values for $\sigma_0$ are given by \citet{bennett/etal:2003b}.

The leading order off-diagonal terms occur at the beam separation angle 
($\theta_{\rm beam} \sim 141\dg$), and are of order 0.3\% of the diagonal
elements.  If the time-ordered noise ${\bf N}$ is not diagonal, then maps
produced with  the above algorithm will have correlated noise (stripes) along
the scan paths  defined by ${\bf M}$.  This is a small, but not negligible,
effect for some of the \map\ radiometers, and is entirely negligible for 
others.  The noise properties of the time-ordered data and sky maps are 
further discussed in \S\ref{sec:filter} and \S\ref{sec:diff_maps}, 
respectively.

\subsubsection{Iterative Map Making}
\label{sec:iterate_map}

The evaluation of the sky map solution ${\bf W d}$ requires the 
inversion of the $N_p \times N_p$ matrix ${\bf D} \equiv {\bf M}^T 
{\bf M}$.  We use the iterative approach introduced by 
\citet{wright/hinshaw/bennett:1996a} to evaluate this expression. Briefly, 
suppose we have an initial guess for the sky map, ${\bf t}_0$, which differs
from the true sky map, ${\bf t}$, by $\delta{\bf t}_0 = {\bf t}_0 - {\bf t}$.
Then ${\bf D}\,{\bf t}_0 = {\bf D}\,({\bf t} + \delta {\bf t}_0)$ can be
recast as
\begin{equation}
{\bf D}\,{\bf \delta t}_0 = {\bf D}\,{\bf t}_0  - {\bf M}^T {\bf d},
\end{equation}
where we have used the fact that ${\bf D}\,{\bf t} = {\bf M}^T {\bf d}$.
As noted above, ${\bf D}$ is diagonally dominant, so a good approximate 
inverse for ${\bf D}$ is
\begin{equation}
\tilde{\bf D}^{-1}(p_i,p_j) \simeq \frac{1}{{\bf n}_{\rm obs}(p_i)} \delta_{ij}.
\end{equation}
This leads to the approximate solution for the residual, ${\bf \delta t}_0 
\simeq \tilde{\bf D}^{-1}[{\bf D}\,{\bf t}_0 - {\bf M}^T {\bf d}]$, 
and suggests the following iterative solution
\begin{eqnarray}
{\bf t}_{n+1} & = & {\bf t}_n - {\bf \delta t}_n \\
          & = & {\bf t}_n - \tilde{\bf D}^{-1}
                [{\bf D}\,{\bf t}_n - {\bf M}^T {\bf d}] \\
          & = & (\tilde{\bf D}^{-1}{\bf M}^T) {\bf d}
                + ({\bf I} - \tilde{\bf D}^{-1}{\bf D})\,{\bf t}_n.
\label{eq:map_wright}
\end{eqnarray}
The interpretation of equation~(\ref{eq:map_wright}) is that for each pixel the
new sky map temperature is the average of all differential observations of that
pixel (accounting for the sign of the observing beam) corrected by an estimate
of the signal in the paired beam, based on the previous sky map iteration.  The
expression in equation~(\ref{eq:map_wright}) can be efficiently  evaluated
because the sums can be accumulated by reading through the  time-ordered data
from disk, each iteration, and accumulating data into  arrays of length $N_p$
\begin{eqnarray}
{\bf n}_{\rm obs}\cdot{\bf t}_{n+1}(p_A) \rightarrow {\bf n}_{\rm obs}\cdot{\bf t}_{n+1}(p_A) 
+ w_i\,[ {\bf d}(t_i) + {\bf t}_n(p_B) ] 
& \mbox{     } & {\bf n}_{\rm obs}(p_A) \rightarrow {\bf n}_{\rm obs}(p_A) + w_i 
\nonumber \\
{\bf n}_{\rm obs}\cdot{\bf t}_{n+1}(p_B) \rightarrow {\bf n}_{\rm obs}\cdot{\bf t}_{n+1}(p_B) 
- w_i\,[ {\bf d}(t_i) - {\bf t}_n(p_A) ]
& \mbox{     } & {\bf n}_{\rm obs}(p_B) \rightarrow {\bf n}_{\rm obs}(p_B) + w_i,
\label{eq:accum_map}
\end{eqnarray}
where $w_i = 1$ in the initial sky map processing and is proportional to a
noise weight (equation~\ref{eq:pol_wt}) in the final sky map processing. Note
that it is never necessary to store or invert an $N_p \times N_p$ matrix.

\begin{figure}[hbt]
\begin{center}
\includegraphics[width=1.0\textwidth]{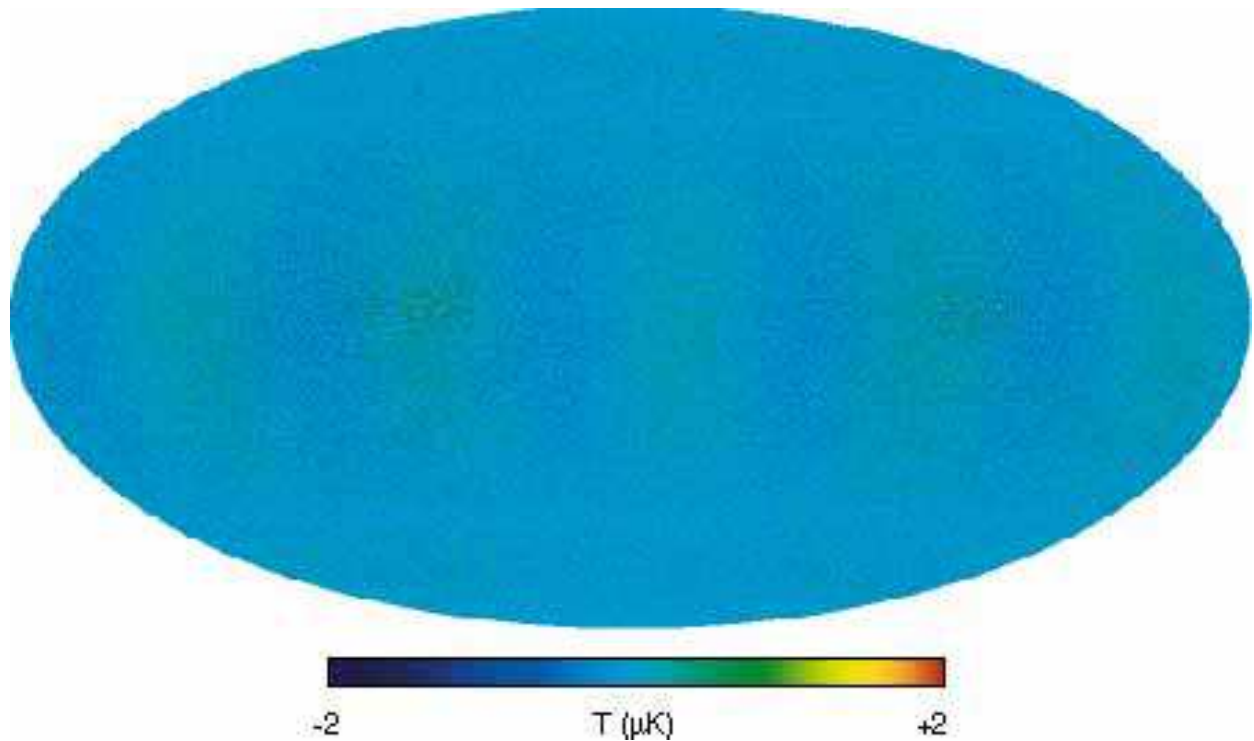}
\end{center}
\caption{A residual sky map, ${\bf t}_{\rm out}-{\bf t}_{\rm in}$, from an
``ideal'' one-year  simulation of Q2 data, designed to test the iterative
map-making algorithm presented in \S\ref{sec:map_making_pre_cal}.  The input
sky map included  realistic CMB signal with a peak-to-peak amplitude of
$\sim\pm$420 $\mu$K, and a  Galactic signal with a peak brightness of $\sim$50
mK.  The rms structure in this map is $<$0.2 $\mu$K, after accounting
for the 0.15 $\mu$K noise that was
introduced to the simulation to dither the  digitized signal. The map is
projected in ecliptic coordinates and shows the anisotropy mode that is least
well measured by \map, due to a  combination of the scan pattern and the beam
separation angle.  This residual level is the result of 50 iterations of the
algorithm -- more iterations would reduce it  even further.}
\label{fig:map_conv_1}
\end{figure}

We have tested this algorithm extensively with flight-like simulations. In this
section we present results for an ``ideal'' noiseless instrument with  circular
beams and perfect calibration to isolate the performance of the  map-making
algorithm from other effects.  More realistic data models are introduced to the
simulation in subsequent sections.  Figure~\ref{fig:map_conv_1}  shows a sample
residual map, ${\bf t}_{out}-{\bf t}_{in}$,  generated from a one-year
simulation of Q2 data. The input sky map included  realistic CMB signal with a
peak-to-peak amplitude of $\sim\pm$420 $\mu$K, and a  Galactic signal with a
peak brightness of $\sim$50 mK.  The output sky map is  recovered with an rms
error of $< 0.2$ $\mu$K, after 50 iterations.  The  dominant structure in the
residual map is a mode aligned with the ecliptic  plane.  The power in this
mode is concentrated in spherical harmonic mode  $l = 4$, due to a combination
of the \map\ scan strategy and the beam separation  angle. This is the mode on
the sky that is least well measured by \map\  (except for the monopole!) and is
thus the slowest to converge, though  additional iterations would reduce its
amplitude even further.  The final  first-year flight sky maps were effectively
run for 80 iterations.  Since the  rms error associated  with this term is very
small, and since we build up  a more realistic data model in subsequent
simulations,  we do  not further  quantify this contribution to the final
systematic error  budget.  Rather,  we subsume it into an overall map-making
and calibration error budget that  includes this and other effects together.

This iterative approach to map-making is readily generalized to polarization
maps as well -- the formalism is presented in Appendix \ref{app:map_pol}.  We
have tested that algorithm with the same simulations used above to test the
temperature algorithm and find that the polarization maps converge even faster
than the temperature maps.  After 10 iterations, the map-making artifacts in a
residual polarization map are $<$0.05 $\mu$K peak-peak.

\subsection{Combined Calibration and Map Making}
\label{sec:map_making_cal}

The processing algorithm described above assumes that the data have already
been calibrated.  In practice, we use the above algorithm in the second stage 
of map-making, after an initial stage in which we  simultaneously solve for the
radiometer calibration and the sky map.  In the initial stage of map-making,
we employ the same iterative algorithm to solve  for the map, but rather than
processing straight through the time-ordered data on each iteration, we process
the data one hour at a time, pausing to solve  for the calibration in each
radiometer channel, before accumulating the calibrated data.   The calibration
solution  then iteratively improves as the sky model improves.  The following
sub-sections lay out the procedure in detail, and  present results for the
flight data with an assessment of its precision and  accuracy based on
flight-like simulations.

\subsubsection{Instrument Calibration from the Dipole Modulation}
\label{sec:dipole_cal}

For a sufficiently short period of time the instrument gain and baseline can be
approximated as constant, ${\bf c} \simeq g_k ({\bf \Delta t + n}) + b_k$,
where $g_k$ and $b_k$ are the gain and baseline during the $k\uth$ 1-hr 
calibration period.  Since the sky signal ${\bf \Delta t}$ is dominated by the
CMB  dipole measured by \cobe, ${\bf \Delta t_d}$, a single channel of raw data
can be  modeled as
\begin{equation}
{\bf c_m}(g_k,b_k) = g_k ({\bf\Delta t_d + \Delta t_v}) + b_k,
\end{equation}
where ${\bf\Delta t_v}$ is the additional dipole moment induced by the motion 
of \map\ relative to the solar system barycenter (the rest frame of the \cobe\ 
dipole).

We fit for the gain and baseline in each calibration period, $k$, by minimizing
\begin{equation}
\chi^2 = \sum_{i \in k} \frac{\left[{\bf c}(t_i) - {\bf c_m}(t_i\vert g_k,b_k)\right]^2}
                        {\sigma_0^2},
\end{equation}
where $i$ is a time-ordered datum index. We omit data that are flagged as
unusable, and data when either the A or B-side beam points within a Galactic
pixel mask.  The mask used for this latter application is the Kp8 mask defined
in \citet{bennett/etal:2003c}, without edge smoothing.  This mask is used
throughout the  map-making pipeline. The fit is performed for each of  the 40
\map\ channels independently.  To minimize the covariance between  the
recovered gain and the baseline, it is necessary to have a scan strategy  such
that the time average of the sky signal, ${\bf \Delta t}$, is  approximately
zero in one calibration period. The combined spin and precession  of \map\ is
designed to produce time-ordered data that satisfies this  requirement. For
example, in K band, which has the largest sky signal, a 1-hr running mean of
the differential sky signal has an rms fluctuation of 14  $\mu$K, compared to
a dipole signal of greater than 3 mK. After each hour of  data is processed for
the calibration solution, the data are accumulated as  per
equation~(\ref{eq:accum_map}) to develop the sky map solution.

The largest source of error in the calibration fit is due to un-modeled
sky signal from the CMB anisotropy and Galactic foreground emission, 
${\bf \Delta t_a} \equiv {\bf \Delta t - \Delta t_d}$.  This projects onto
the dipole signal and, as shown below, causes errors in the gain solution as 
large as 5-10\% in K band, where the Galactic signal is strongest. The 
calibration fit may be iteratively improved by subtracting an estimate of the 
anisotropy from the raw data prior to fitting.  In particular, let $g'_k$
be the gain inferred for calibration period $k$ from the previous iteration 
of the calibration fit.  Then minimize
\begin{equation}
\chi^2 = \sum_{i \in k} \frac{\left[{\bf c'}(t_i) - {\bf c_m}(t_i\vert g_k,b_k)\right]^2}
                             {\sigma_0^2},
\label{eq:fit_gain_base}
\end{equation}
where
\begin{equation}
{\bf c'} = {\bf c} - g'_k {\bf \Delta t'_a}
\label{eq:correct_cal}
\end{equation}
and ${\bf \Delta t'_a}$ is the differential sky signal (less the dipole 
component) computed from the previous sky map iteration. This process is
repeated until the calibration solution is sufficiently converged.

Note that the absolute calibration is tied to the time-dependent portion of the
dipole signal, ${\bf \Delta t_v}$; we use the fixed dipole as a short-term 
transfer standard only.  In particular, when we update the  sky model and apply
the anisotropy correction in equation~(\ref{eq:correct_cal}), any error in the
fixed dipole moment, ${\bf \Delta t_d}$, is assumed to be anisotropy, and is
applied as a correction in the same way.  For a data set  of at least one year
in length (one full cycle of ${\bf \Delta t_v}$),  the error in the absolute
calibration will be essentially orthogonal to any  error in the fixed dipole
${\bf \Delta t_d}$.

\subsubsection{Performance of the Dipole-Based Gain Solution}
\label{sec:dipole_gain_perf}

\begin{figure}[hbt]
\begin{center}
\includegraphics[width=1.0\textwidth]{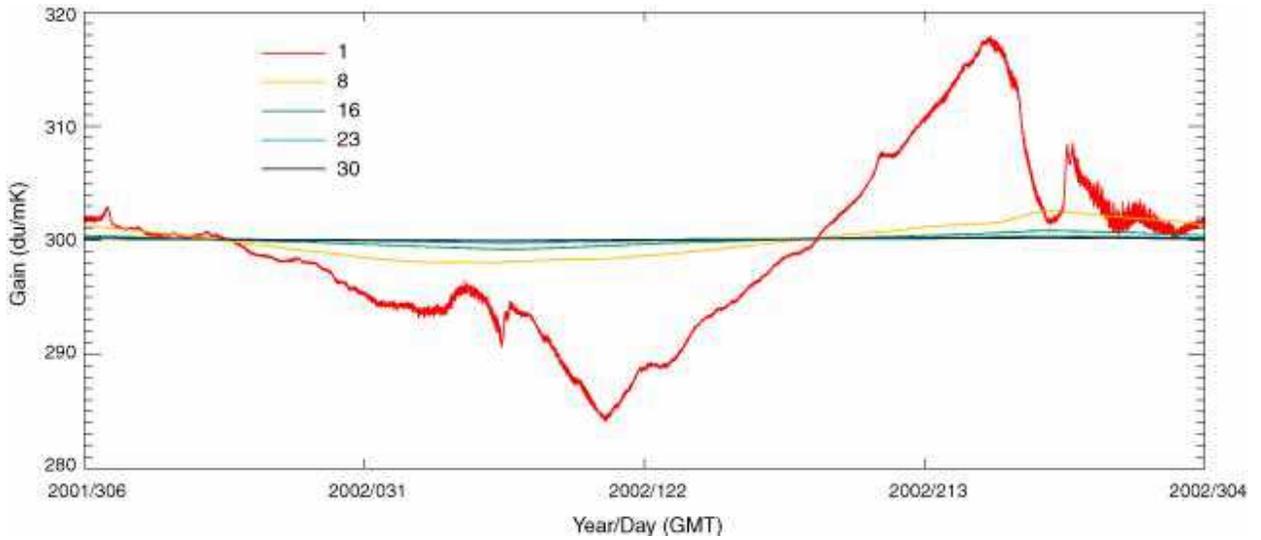}
\end{center}
\caption{Convergence of the dipole-based gain solution for a selected \map\
radiometer channel (K113) based on a one-year simulation.  This simulation was
generated with an input gain of 300.0 du mK\per, and minimal noise. The  first
iteration, which assumes the sky model has only a dipole component,  leaves
residual gain errors of up to 7\%, due to the projection of the relatively
bright Galactic emission onto the dipole model.  After 30  iterations of the
simultaneous fit described in \S\ref{sec:dipole_cal}, the residual errors in
the gain solution are less than 0.1\%.}
\label{fig:cal_conv}
\end{figure}

As an illustration of the systematic gain error induced by higher-order 
($l>1$) anisotropy, Figure~\ref{fig:cal_conv} shows an example of the 
gain solution convergence from a one-year low-noise simulation.  This 
simulation implements the simultaneous calibration and sky map estimation
discussed above and was run for 30 iterations.  The example shown is for one 
channel of K band data (the worst case) which exhibits $\sim$7\% errors after
one iteration, corresponding to a sky model that has only a dipole component.
After 30 iterations, the residual errors are $<$0.1\% over the entire 
year. Similar, or better, performance is achieved for all other \map\
channels.

In processing the final first-year maps, the combined  calibration and
map-making code was run for 10 iterations.  However the initial sky model was
based on an earlier  ``pathfinder'' run of the pipeline that ran for a total of
30 iterations of combined calibration and map-making plus an  additional 20
iterations of sky map convergence.  Thus we conservatively estimate that the
combined absolute and relative calibration  errors due to incomplete
calibration convergence to be $<$0.1\%.  We defer a discussion of the final
calibration uncertainty to \S\ref{sec:cal_map_sims}.

\begin{figure}[hbt]
\begin{center}
\includegraphics[width=1.0\textwidth]{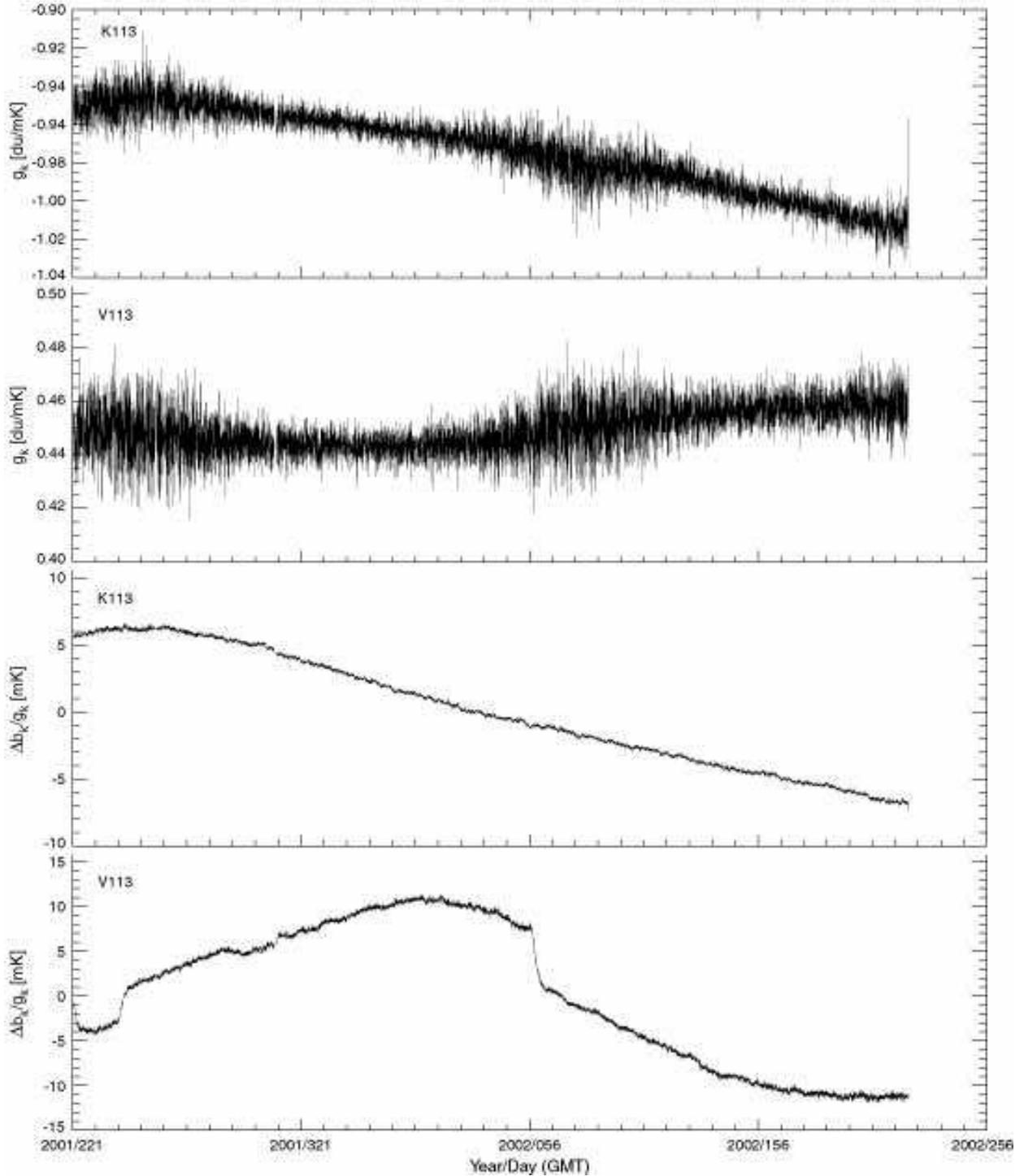}
\end{center}
\caption{The hourly gain and baseline fit described in \S\ref{sec:dipole_cal} 
from the flight data for channels K113 and V113. The top two panels show
the gain solution, the bottom two the baseline.  Note that the gain is stable
to $\sim$5\% over the first year (see also Table~\ref{tab:gain_summary}).
The variable noise is due to the changing projection of the scan pattern
on the CMB dipole over the course of a year.  The instrument baselines have
a typical drift of 5-10 mK over the year.  The channel V113 exhibits one of 
the clearest thermal susceptibilities of the \map\ radiometers, though
we show in \S\ref{sec:thermal} that the induced systematic signal is 
negligible. See also Figure~\ref{fig:baseline_susc}.}
\label{fig:dipole_cal}
\end{figure}

Figure~\ref{fig:dipole_cal} shows a sample of the converged gain solution  from
equation~(\ref{eq:fit_gain_base}) for two \map\ channels, K113 and V113.  Note 
that the  radiometer gains are typically drifting by a few percent over the 
course of  the first year.  As we show below, the dipole-based fits easily
track drifts at this level.  The noise in the gain solution is typically a few
percent per hourly calibration period, though, as is readily seen in the
figure, the noise level changes with time of year as the scan pattern sweeps
around the CMB dipole.  The V113 gain exhibits an additional modulation that is
clearly correlated with the physical temperature of the instrument.  However,
the time scale of the temperature change is slow enough that the corresponding
gain changes are well tracked by the dipole fits.  Quantitative limits on
thermally  induced gain and baseline errors are discussed in
\citet{jarosik/etal:2003b}  and in \S\ref{sec:thermal}.  A summary of the gain
statistics from the flight data is given in Table~\ref{tab:gain_summary}.

\subsubsection{The Initial Baseline Solution}
\label{sec:dipole_base_perf}

The bottom two panels of Figure~\ref{fig:dipole_cal} show the converged  baseline
solution resulting from the fit in equation~(\ref{eq:fit_gain_base}) for  one
year of K113 and V113 flight data. The fits have had a mean subtracted, and
have been divided by the gain to convert to temperature.  These plots, which
are representative of all 40 channels, show that the offsets of the 
radiometers are typically stable to $\pm$5 mK over the course of the first 
year. Simulations demonstrate that the hourly baseline solution is unbiased. 
However, it is also clearly noisier than optimal, consistent with the flight
measurements of the noise power spectral density \citep{jarosik/etal:2003b}. In
\S\ref{sec:filter} we describe an improved baseline model that is based on the
application of a pre-whitening filter tailored to the measured  noise spectrum
of each channel.

\subsection{Improving the Calibration Model}
\label{sec:cal_2}

The sky maps obtained with the hourly calibration are reasonable; however the
noise in the calibration solution, particularly in the baseline, is
significantly higher than optimal, and the use a piecewise continuous 
calibration in the final maps would introduce striping in its own right.  In
the following subsections, we present  the steps undertaken to filter the gain
and baseline solutions that enter into the final sky maps. Prior to generating
the final maps, this refined calibration is applied to the data  and written to
disk as the final first year calibrated time-ordered archive.

\subsubsection{The Gain Model}
\label{sec:gain_model}

\citet{jarosik/etal:2003b} present a physical model for the gain that is based
on the RF bias, or ``total power'' measured in each channel, and on the 
physical temperature of the instrument cold stage, which is monitored with high
resolution platinum resistance thermistors (PRTs).  Each of these quantities 
is recorded once every 23 s in the engineering telemetry with a relative 
noise that is substantially lower than the noise in the dipole-based gain 
solution.  Thus, if the model fits the dipole-based data satisfactorily, 
it offers a means for measuring the gain with more precision, and on time 
scales shorter than the spin period.  The model for the gain, ${\bf g}(t)$, 
has  the form
\begin{equation}
{\bf g}(t) = g_0 \frac{{\bar V}(t) - V_0}{T_{\rm FPA}(t)-T_0},
\label{eq:gain_model}
\end{equation}
where ${\bar V}$ is the measured RF detector bias, $T_{\rm FPA}$ is the measured
temperature of the FPA, and $g_0$, $V_0$, and $T_0$ are fit constants.  See
\citet{jarosik/etal:2003b} for more detail.

Figure~8 in \citet{jarosik/etal:2003b} shows the performance of the gain model
when fit to the dipole-based gain solution (see also
Figure~\ref{fig:dipole_summary} in this paper).  In \S\ref{sec:cal_map_sims} we
evaluate the overall performance of the gain model and the hourly gain 
solution in the context of an end-end simulation designed to place limits on
combined calibration and map-making errors.  We will conclude that the model
provides an excellent description of the radiometer gain, and here we adopt it
as the final gain solution for further processing. The gain model fits into the
data processing sequence as follows.  After we iterate the simultaneous
calibration and sky map solution long enough for the calibration to converge
(10 iterations when starting with a good sky  model), we freeze the
dipole-based calibration and fit the gain model  parameters in
equation~(\ref{eq:gain_model}).  {\em All} subsequent data products are
produced with data calibrated using this gain solution, including the
time-ordered archive, the final sky maps, and the Jupiter beam maps.

\subsubsection{Baseline Filtering}
\label{sec:filter}

The baseline that results from the initial calibration solution is not
optimal.  This is due to the  fact that the initial baseline is sampled once
per hour (0.28 mHz), while \citet{jarosik/etal:2003b} show that the power
spectral density of the noise has a $1/f$ knee frequency of a few mHz,
typically.  If the initial baseline estimate were used in the final sky maps, 
it would generate weak stripes of correlated  noise along the scan paths, as
per equation~(\ref{eq:pixel_noise}).  Even so,  it is important to note that
$1/f$ effects are small relative to the white  noise.  In the {\em worst} \map\
radiometer, W41, the amplitude of the noise covariance ${\bf N}$ at small lag
is $\sim$2\% of the white noise  variance.  Thus we treat $1/f$ noise
iteratively in the data processing by first ignoring it to obtain an estimate 
of the gain, baseline, and sky solution.  Then we subtract the  estimated sky
signal from the time-ordered data, apply a pre-whitening baseline filter to 
the residual noise, add the sky signal back in, and write the data to a final
calibrated, time-ordered data archive. The approach of first subtracting an
estimated sky signal is designed to avoid biasing the gain solution and/or  
removing low-order power from the sky maps.  The noise properties of maps
constructed in this way must  account for the filtering process.  We discuss
the map-making algebra appropriate to our filter implementation in Appendix
\ref{app:map_filter}.

\begin{figure}[hbt]
\begin{center}
\includegraphics[angle=90,width=0.7\textwidth]{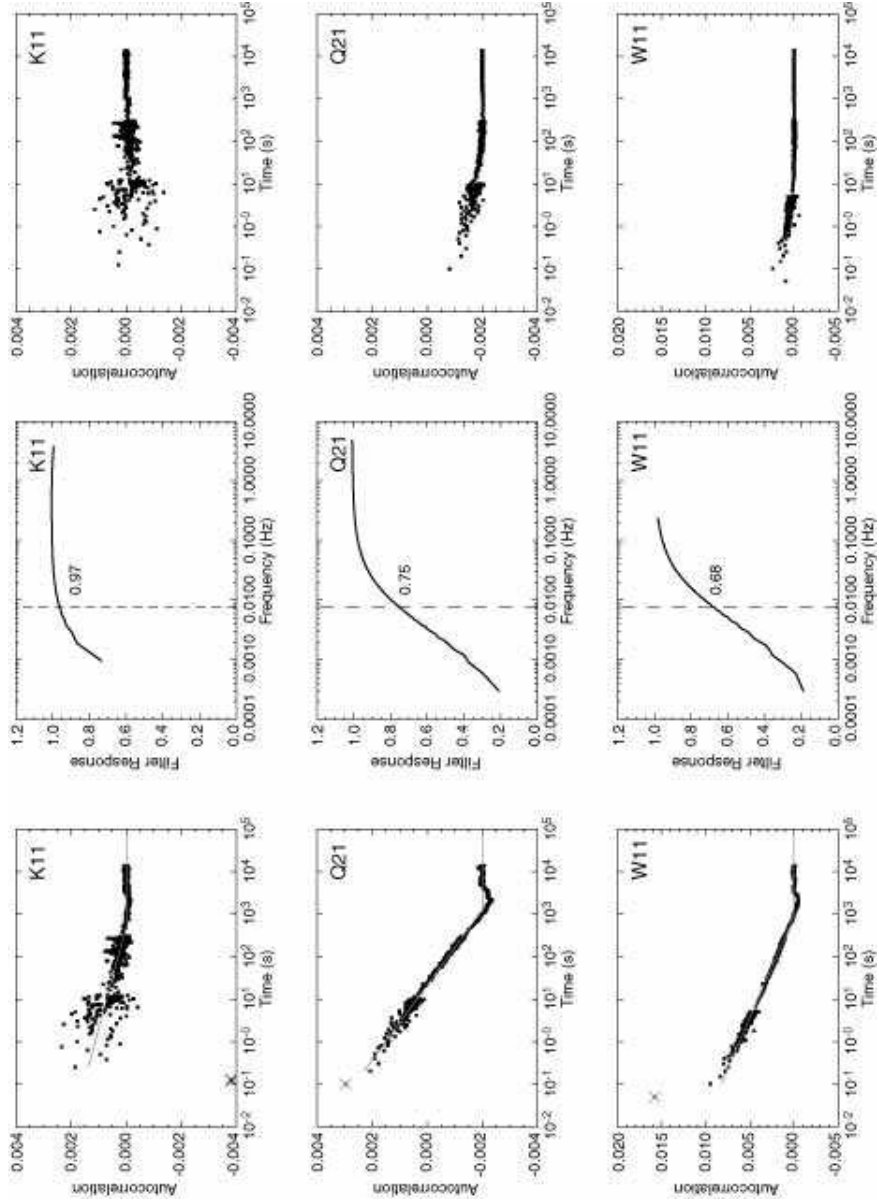}
\end{center}
\caption{The left-hand panels show the measured auto-correlation function,
$C(\Delta t)/C(0)$, for selected radiometers, of \map\ time-ordered data, after
subtracting a model sky signal based on the initial sky maps. The model fits
are indicated by an $\times$ at a lag of 1 observation and by straight lines
for $\Delta t > \tau$  (see \S\ref{sec:filter}).  All of the \map\ DAs except
W4 have a covariance of $<$1\% at non-zero lag (Table~\ref{tab:ac_fit}). The 
anti-correlation at lag $\sim$2000 s is due to the subtraction  of the hourly 
baseline as a pre-filter. The middle  panels show the pre-whitening filter, in
the frequency domain, that is applied  to the time-ordered data  after a model
sky signal has been subtracted.  The vertical dashed line indicates the spin
frequency, 7.7 mHz, and the number indicates the fraction  of power transmitted
by the filter at the spin frequency. The right-hand panels show the  measured
auto-correlation function for selected channels of \map\ time-ordered data,
after pre-whitening, on the same scale as the left panels.  The apparent change 
in noise level at different lags in $C(\Delta t)$ is due to a step-wise change
of bin size in $\Delta t$.}
\label{fig:ac_fit}
\end{figure}

The steps we follow to define and apply the filter are as follows.  We remove 
an estimate of the sky signal, in du, from the raw differential data using
\begin{equation}
{\bf c'}(t) = {\bf c}(t) - {\bf g}(t){\bf \Delta t'}(t) - b_k,
\label{eq:correct_sky}
\end{equation}
where $b_k$ is the hourly baseline point appropriate to the current time, 
${\bf g}$ is the final gain solution from equation~(\ref{eq:gain_model}) and 
${\bf \Delta t'}$ is the differential sky signal computed from the initial sky
map. We then evaluate the auto-correlation function of ${\bf c'}(t)$ to a lag
of $10^4$ sec.  Results for representative radiometers are shown in 
Figure~\ref{fig:ac_fit}.  This range of lags is sufficient to account for both
the long-range correlations due to  $1/f$ noise and the correlation at a lag of
1 observation due to the low-pass post-demodulation filter in the Analog
Electronics Unit \citep{bennett/etal:2003}. The baseline filters are then
defined as follows \citep{wright:1996} \begin{enumerate} \item Fit the
auto-correlation function, $C(\Delta t)$, to the model defined  below.

\item Fourier transform the model correlation function to generate the power 
spectral density $P(f)$.

\item Compute the Fourier space filter $w(f) = 1/\sqrt{P(f)}$ and set
$w(0) = 0$ to produce a zero mean output signal.

\item Fourier transform the filter $w(f)$ to generate the time domain filter 
$w(t)$, normalized to 1 at $t=0$.
\end{enumerate}
By inspection, the auto-correlation functions are well modeled by a log-linear
function
\begin{equation}
C(\Delta t)/C(0) = \left\{ \begin{array}{ll}
   C_1                                 & \mbox{$\Delta t = \tau$} \\
   A - B \log(\Delta t / \mbox{1 s})   & \mbox{$\tau < \Delta t < 10^{A/B}$ s} \\
   0                                   & \mbox{$\Delta t > 10^{A/B}$ s}, \\
                    \end{array}
            \right.
\label{eq:ac_model}
\end{equation}
where $\tau$ is the integration time for a single observation, $C_1$ is the 
correlation at lag $\tau$, measured from the data, and $A$ and $B$ are fit 
parameters.  Note that $A$ gives the typical fractional covariance at small 
lag, while the suppression of correlations at large lag ($\sim$2000 s) is
dictated by the subtraction of the hourly baseline as a pre-filter. The
best-fit parameters are given in Table~\ref{tab:ac_fit}, and  fits for selected
radiometers are shown in Figure~\ref{fig:ac_fit}.

\begin{figure}[hbt]
\begin{center}
\includegraphics[width=1.0\textwidth]{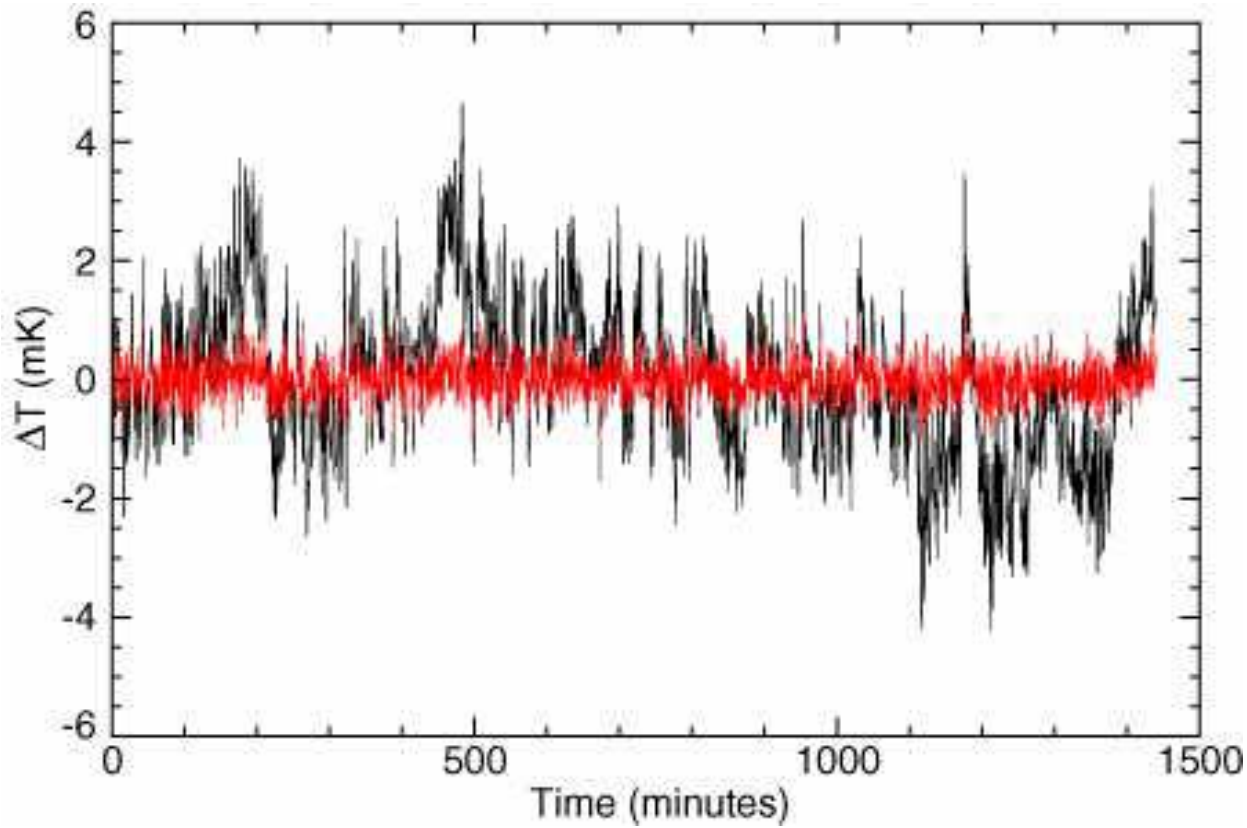}
\end{center}
\caption{The time-ordered data for channel W424 before (black) and after (red)
applying the pre-whitening filter.  The data are boxcar averaged over a 46.08 s
window to show the low frequency noise in the unfiltered data.  Without
averaging, the data before and after filtering are virtually
indistinguishable.  Note that baseline variations in this channel are of order
2 mK on a one-hour time scale, as expected given the measured $1/f$ knee
frequency of this radiometer \citep{jarosik/etal:2003b}.  W4 is the worst 
differencing assembly from the standpoint of $1/f$ noise.}
\label{fig:filtered_dt}
\end{figure}

The derived pre-whitening filters, $w(f)$, are plotted as a function of 
frequency in Figure~\ref{fig:ac_fit}. One point of particular interest  is the
filter response at the spin frequency, 7.7 mHz.  As shown in 
Table~\ref{tab:ac_fit}, the best channels have a transmission of $>$ 95\%,
while the worst case, W41, is just above  35\%.  These values indicate the
amount by which the dipole (calibration)  signal would be suppressed if the
filter were applied prior to calibration and sky signal subtraction. The
convolution of  ${\bf c'}(t)$ with $w(t)$ is performed in Fourier space using
the Numerical  Recipes routine \verb"convlv" \citep{press/etal:NRIC:2e}. The
number of data points convolved at any one time is chosen to be the smallest
power of 2 such that the data span a full day with sufficient padding beyond
the day to guarantee that wrap-around effects are negligible. This is  $2^{20}$
for K--Q bands, and $2^{21}$ for V,W bands, which gives a minimum of  2.9 hours
of padding on each end of a day. Sample auto-correlation functions obtained
from the filtered data are shown  in Figure~\ref{fig:ac_fit}.  The filtering is
clearly effective at removing low frequency noise in the time-ordered data. 
Another example of filtered data is seen in Figure~\ref{fig:filtered_dt}, which
shows 1 day of W42 data, one of the worst radiometers for $1/f$ noise, before
and after filtering. These data are smoothed with a 46 s window to show
structure in the unfiltered data  since plots of  unsmoothed data before and
after filtering are virtually  indistinguishable.  

The above results are encouraging but not definitive, because the process of 
sky signal subtraction and re-addition could introduce correlated  artifacts
that these tests would miss.  The ultimate test of a filter is its ability to
``clean'' the pixel-to-pixel covariance matrix of the final sky maps and the
noise covariance of the angular power spectrum, without altering the underlying
sky signal.  The sky map noise properties are discussed in
\S\ref{sec:diff_maps}, while the noise properties of the power spectrum are
quantified in \citet{hinshaw/etal:2003}.

\subsubsection{Baseline Jumps}
\label{sec:jumps}

\citet{limon/etal:2003} identify 21 instances of sudden baseline jumps,  or
``glitches'', during \map's first year of operation.  These events have been
identified as small shifts in the properties of several  microwave components
resulting from sudden releases of internal mechanical  stress, presumably from
thermal changes.  These events last for less than 1 s, and cause no
discernible  change in the radiometer gain or noise properties.

Care must be taken in the application of the baseline filter in the vicinity 
of these steps to avoid ringing in the filtered data.  Each event is initially 
flagged by the pre-processor for at least 1.2 hours on either side of  the
event. Since the initial hourly baseline is derived entirely in the time 
domain, the $\pm$1.2 hr flagged interval ensures that this baseline estimator
only ``knows'' about data on one side of the jump or the other.  Prior to 
convolving the raw data with the pre-whitening filter, we subtract the initial 
hourly baseline from the data as a pre-conditioner.  Thus all data that is 
input to the convolution routine has approximately zero mean.   On output, the
re-processor expands the flagged interval by 0.5 hr on  either side of the
event to ensure that no edge effects propagate into the  usable data.  In the
first year of operation, a total of 0.13\% of the data was lost to these
steps.  See Table~2 in \citet{bennett/etal:2003b}.

The threshold amplitude for jump detection by visual inspection is  $\sim$0.05
du, which corresponds to a jump of $\sim$150 $\mu$K in the  calibrated output
of radiometer W12, the worst offender. To assess the effect of undetected
baseline steps in the data, we have  generated a test data set in which we take
24 hours of flight W12 data and  insert a step of 0.05 du in each channel. We
then run the data through the pre-whitening filter to see the  magnitude of the
transient response.  The result is a transient baseline  error with a peak
magnitude of 80 $\mu$K, which lasts for less than one  2-minute spin period. 
The total time the baseline error exceeds 10 $\mu$K is  22 minutes, or
approximately 11 spin periods.  We pessimistically assume that there could be
as many as 40 steps at or just below the threshold of detection, and that half
of these are in W12.  If we assume these occur at random times, and note that
\map\ observes $\sim$30\% of the sky in any given hour, then any given sky
pixel is likely to ``see'' approximately $11 \times 20 \times 0.3 \approx 66$
data points with baseline errors greater than 10 $\mu$K.  Since the sign of a
given step is random, and since W11 data is combined with W12 in the sky maps,
we estimate the residual systematic error in a given pixel of the W1 map is
less than $10\mbox{ $\mu$K }/2/\sqrt{20} \sim 1$ $\mu$K.  We emphasize that
{\em no} jumps have been observed in any other W band radiometer, thus DA-DA
consistency is an excellent test of whether any statistic is  sensitive to
baseline errors of this nature.  We have found no evidence that the W1 map is
``out of family'' (\S\ref{sec:diff_maps}).

\subsection{Final Sky Map Processing}
\label{sec:final_maps}

Once the calibrated, time-ordered archive has been written, final sky map
processing commences based on the algorithm presented in
\S\ref{sec:map_making_pre_cal}. At this stage in the sky map processing, we add
a few features to the  algorithm that, for simplicity, are not present in  the
combined calibration and map-making code.  These include: 1) Weighting each
datum by a true weight, $1/\sigma_i^2$ based on an estimate derived from the
physical temperature of the instrument cold stage.  This introduces $\sim$1\%
variations in the data weights over the year, since the instrument noise is a
weak function of temperature, and the temperature varies by $\sim$1\% over the
year.  2) Accounting for loss imbalance, as discussed in  Appendix
\ref{app:map_funct}.  In effect, we model each differential observation as
${\bf \Delta t} = (1+x_{im}){\bf t}(p_A) - (1-x_{im}){\bf t}(p_B)$, where 
$x_{im}$ is the small loss imbalance parameter given by 
\citet{jarosik/etal:2003b}. 3) Computing the planet avoidance flag at run time
to reduce the amount of data lost.  In the  final sky maps, a total of 0.11\%
of the data was lost to planet avoidance.  See Table~2 in
\citet{bennett/etal:2003b}.

The final stage of sky map processing, based on the filtered data, was run for
a total of 20 iterations.  Convergence was determined by measuring the rms
difference between pairs of iterations for a given differencing  assembly.  For
example, the difference between the 10th and 20th iteration of the W2 sky map
is 0.08 $\mu$K rms.  We estimate that artifacts due to lack of solution
convergence are $<$0.1 $\mu$K rms with all of the power being in the low
multipoles, $l < 10$.  We present a final combined estimate of sky map
artifacts due to calibration and map-making  errors in
\S\ref{sec:cal_map_sims}.  This estimate includes the convergence limits given
above.

\begin{figure}
\begin{center}
\includegraphics[width=1.0\textwidth]{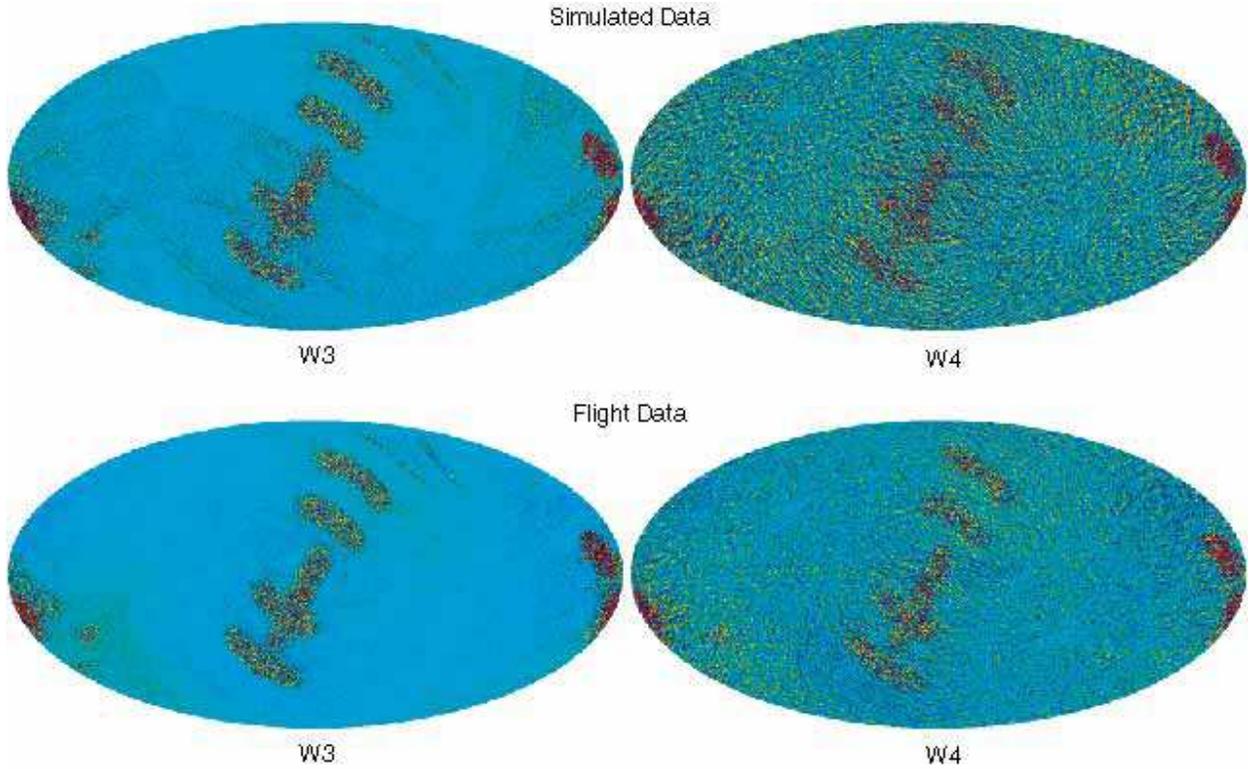}
\end{center}
\caption{Simulated and flight difference maps showing the structure that is
removed from the maps by the pre-whitening filter.  All four maps are 
differences between sky maps generated before and after baseline filtering. 
The maps are projected in Galactic coordinates and the temperature scale on 
each is $\pm$50 $\mu$K.  The ``blobs'' of white noise along the ecliptic plane
can be ignored.  They  arise from differences in the handling of planet flags
in the two forms of the  map-making code.  The top two panels show W3 and W4
data from a one-year  simulation that includes flight-like $1/f$ noise in the
time-ordered data. The bottom two panels are the same for the flight W3 and W4
maps.  Note the very different structure between W3 and W4, due to different
$1/f$ knee  frequencies \citep{jarosik/etal:2003b}.  Note also that the
simulation captures the basic structure of the flight data very well.}
\label{fig:w3w4_sky_filter}
\end{figure}

\begin{figure}[hbt]
\begin{center}
\includegraphics[width=1.0\textwidth]{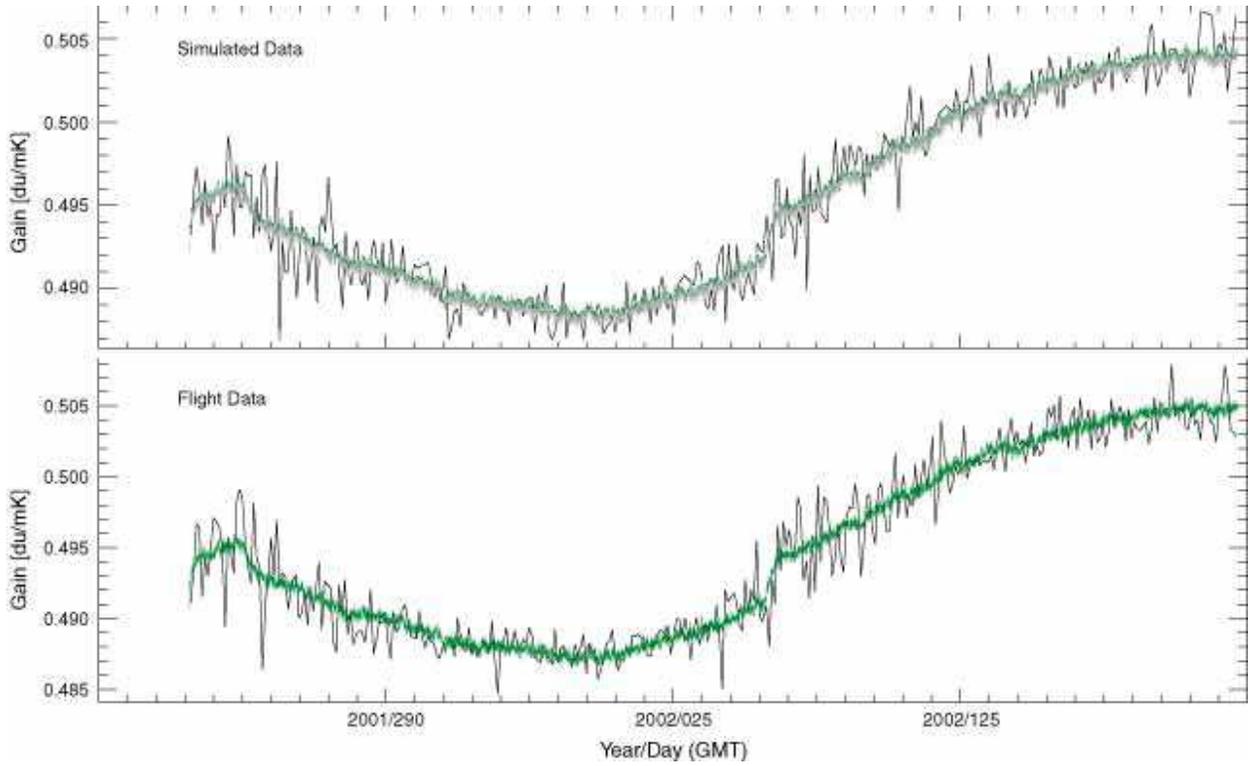}
\end{center}
\caption{The recovered gain solutions for channel V113 in a flight-like 
simulation (top) and in the flight data (bottom).  The ``noisy'' black  traces
show the hourly baseline binned in 24-hr samples (to reduce noise) and the
green traces are the best fit gain model (\S\ref{sec:gain_model}). For the
simulation, the input gain used to generate the simulated data is shown in
grey.  In the simulation, the absolute gain is recovered to better than 0.1\%
in all 40 channels, and the binned hourly gain is everywhere within $\sim$0.2\%
of the gain model, and the input gain.  Gain changes are well tracked by the
pipeline.}
\label{fig:dipole_summary}
\end{figure}

To assess the effect of the improved calibration model on the final sky maps, 
we form differences between the final post-filtered maps and the last iteration
of the intial, pre-filtered maps.  The results for DAs W3 and W4 are shown in 
Figure~\ref{fig:w3w4_sky_filter}.  The top panels in this Figure show the 
difference maps from a one-year simulation (\S\ref{sec:cal_map_sims}) that
included a realistic radiometer noise and gain model. The bottom two panels
show differences from the flight data.  Because these maps are largely based on
the same data, most of the white noise drops out of these differences.  The
remaining ``blobs'' of white noise result from the change in the planet cut and
can be ignored.  The striking  feature is the striping present in the W4
difference, but virtually absent in the W3 difference.  As we show in  Appendix
\ref{app:map_filter}, this is the structure that has been removed from the data
by the pre-whitening  filter, an interpretation that is substantiated by the
analysis of the simulation.  The fact that the W3 difference is very small is
an indication that the level of striping in the unfiltered W3 data was very
small to start with.  We estimate the level of residual striping in the final
maps in \S\ref{sec:diff_maps}. Images of the final maps at each frequency are
presented by \citet{bennett/etal:2003b}.

\section{SYSTEMATIC ERROR ANALYSIS}
\label{sec:syserr}

As discussed in the introduction, systematic errors may be classified by the 
nature of their source.  In this section we place limits on the level of
systematic errors in the final sky maps, using that classification to guide 
the analysis.  In \S\ref{sec:cal_map_sims} we place limits on combined
calibration and map-making artifacts, based largely on a detailed simulation of
the first year of \map\ operation.  In \S\ref{sec:diff_maps} we present null
tests based on difference maps formed from a variety of data combinations, each
of which should yield no sky signal.  We use these maps to measure or place
limits on correlated pixel noise (striping) in the final first-year maps. In 
\S\ref{sec:pointing} we discuss systematic errors relating to pointing and beam
mapping errors.  We conclude by placing stringent limits on residual  errors
due to environmental (thermal and electrical) and other  miscellaneous sources.

\subsection{Calibration and Map-Making Errors}
\label{sec:cal_map_sims}

To assess the combined errors from calibration and map-making artifacts, we
have generated a high fidelity simulation that includes all of the effects we 
believe are important for calibration and map-making.  In particular this 
simulation includes: 1) A sky model that closely mimics the statistical 
properties of the observed sky;  2) A realistic noise model for every channel,
including $1/f$ noise [see \citet{jarosik/etal:2003b} for a tabulation of 
$1/f$ knee frequencies];  3) A model for the thermal drift of the gain,
baseline and offset of each  radiometer, based on measured susceptibility
coefficients, and driven by the actual temperature profile measured in flight.
This simulation generates the sky signal using a circular beam approximation.
The effects of elliptical beams are treated in a separate, noiseless simulation
in \S\ref{sec:elliptical_beam}. We write simulated science data to files that
mimic the raw telemetry, then  process the data using the same pipeline as was
used to process the flight data.

The top panel of Figure~\ref{fig:dipole_summary} shows the converged gain
solution from the simulation for channel V113; the bottom panel shows the
corresponding result from the first year of flight data.  In both panels, the
``noisy'' black traces are the hourly gain data, binned in 24-hr samples to
reduce the noise, and the green traces are the best-fit gain model 
(\S\ref{sec:gain_model}). For the simulation, the input gain used to generate 
the data is shown in grey.  The absolute gain is recovered in the simulation to
better than  0.1\% for all 40 channels.

The dipole signal seen by an observer moving with speed $v$ relative to the
rest frame of the CMB is $T_0\,v/c$, where $T_0$ is the absolute temperature of
the CMB, and $c$ is the speed of light. Thus, additional sources of error that
could affect the absolute calibration of the \map\ data include errors in the
determination of \map's velocity with respect to the solar system barycenter
(the point of reference for the \cobe\ dipole) and errors in the absolute
temperature of the CMB. The velocity of \map\ is routinely measured with
respect to geocentric  inertial coordinates (GCI) with an accuracy of $<$1 cm
s\per.  The velocity of the Earth is determined from the JPL ephemeris with
similar accuracy.  The combined uncertainty from velocity errors is 0.1 nK. The
uncertainty in the absolute temperature of the CMB is 0.1\% 
\citep{mather/etal:1999}.  Combining these uncertainties with the results of
the simulation, we conservatively estimate an absolute calibration error of
0.5\% for the first-year \map\ data.

\begin{figure}
\begin{center}
\includegraphics[width=0.9\textwidth]{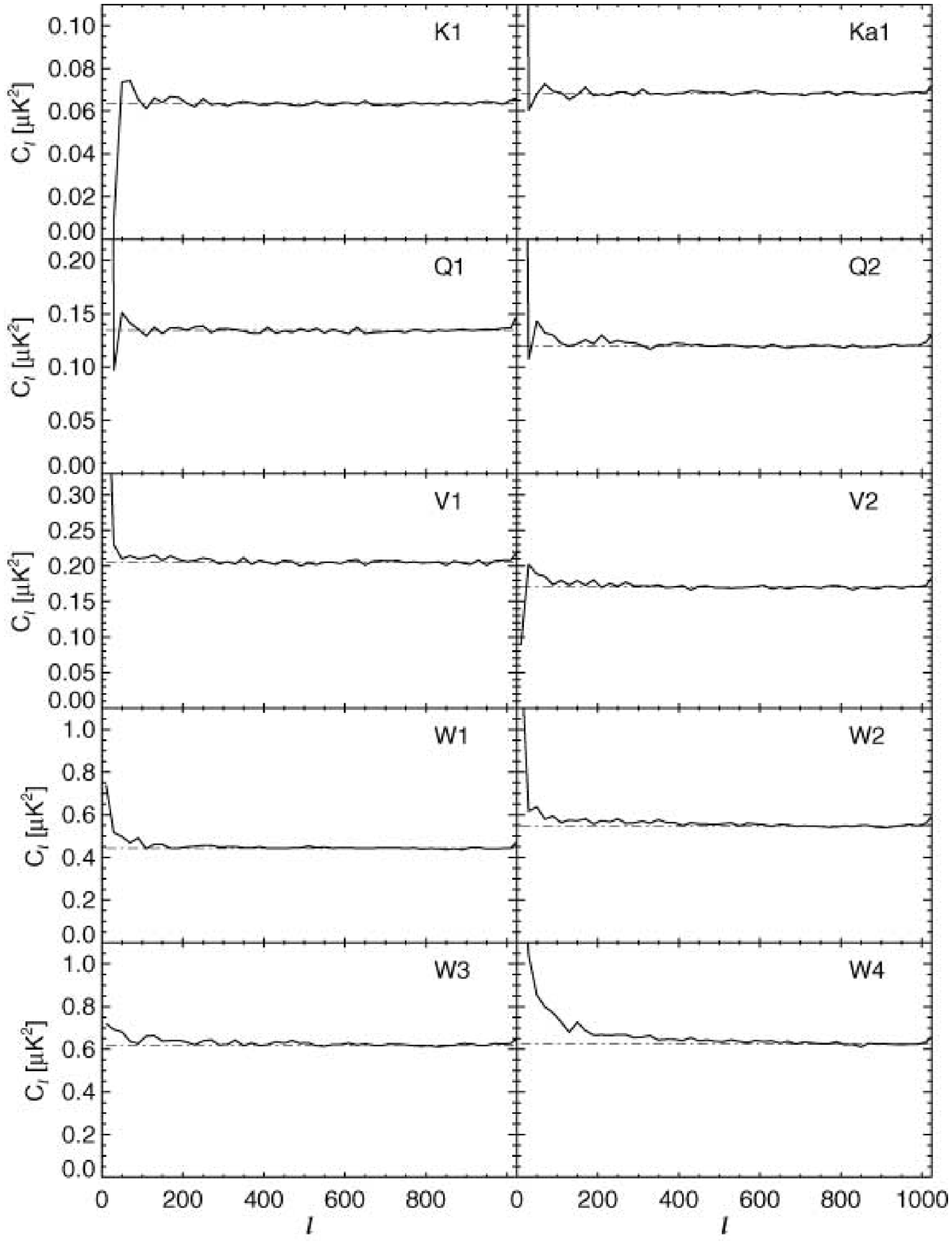}
\end{center}
\caption{Angular power spectra of the 10 resdidual maps ${\bf t}_{\rm out}
- {\bf t}_{\rm in}$ generated from the flight-like one-year simulation.  In
each case the spectra were evaluated in the Kp2 cut sky 
\citep{bennett/etal:2003c}.  Table~\ref{tab:sim_res_map_ps} quantifies 
structure in these maps beyond flat white noise.  Note that most features 
are restricted to $l\lsim25$ but with an amplitude that is still much less than
the sky signal in this range.  The residual effects of $1/f$ noise are seen
in the gradual rise of the noise spectrum at low $l$ in W4.  See
\citet{hinshaw/etal:2003} for further discussion of this.}
\label{fig:sim_res_map_ps}
\end{figure}

Errors in relative calibration can produce structure in the sky maps, beyond an
overall normalization factor.  The largest relative discrepancy between the
dipole gain solution and the gain model in the flight data is $\sim$0.4\%  in K
band, and $\sim$0.2\% in the other bands. Similar deviations are seen in the
simulation, thus we use the simulation as our primary tool for  placing
systematic error limits due to relative calibration and map-making errors. We
have generated residual maps from the simulated data by subtracting the  known
input sky signal from the maps produced by the pipeline.  These residual maps
exhibit no visible structure aside from the pixel noise.  In order to assess
the errors due to map-making artifacts, we compute the angular power spectrum,
$C_l$, of the residual maps and search for features in the spectra beyond a
simple flat, white noise spectrum.  The residual spectra for all 10
differencing assemblies are shown in Figure~\ref{fig:sim_res_map_ps} and
summarized in Table~\ref{tab:sim_res_map_ps}.  In general, the spectra are
consistent with white noise over a wide range of multipole moments, but clearly
show the most variation at low $l$. Because of this, we specifically highlight
these modes in Table~\ref{tab:sim_res_map_ps}, where we give $C_2$,
$\left<C_l\right>_{3-10}$, and $\left<C_l\right>_{11-100}$ for each of the
DA's.  For combined systematic error limits due to calibration and map-making,
we assign twice the excess variance in each $l$  range relative to the white
noise plateau, $\sigma^{sys} \equiv 2\,\left\vert \left<C_l\right>_{\rm band} -
\left<C_l\right>_{700-1000}\right\vert$.  These values are also quoted in
Table~\ref{tab:sim_res_map_ps}. For comparison, the average power in the CMB
in  each band is $C_2 \sim 130$ $\mu$K$^2$, $\left<C_l\right>_{3-10} \sim 
150$ $\mu$K$^2$, and $\left<C_l\right>_{11-100} \sim 6$ $\mu$K$^2$.

Because the simulation includes realistic models of $1/f$ noise and long-term
thermal effects, these limits also implicitly limit artifacts at low $l$ due 
to these effects.  As we demonstrate in subsequent section, we feel this
simulation captures {\em all} of the important radiometric characteristics 
of the instrument.  Potential artifacts due to optical effects, especially
pickup through the far sidelobes, are treated in \citet{barnes/etal:2003},
and are summarized in \S\ref{sec:sidelobe}.

\subsection{Difference Maps and Noise Properties}
\label{sec:diff_maps}

\begin{figure}
\begin{center}
\includegraphics[width=1.0\textwidth]{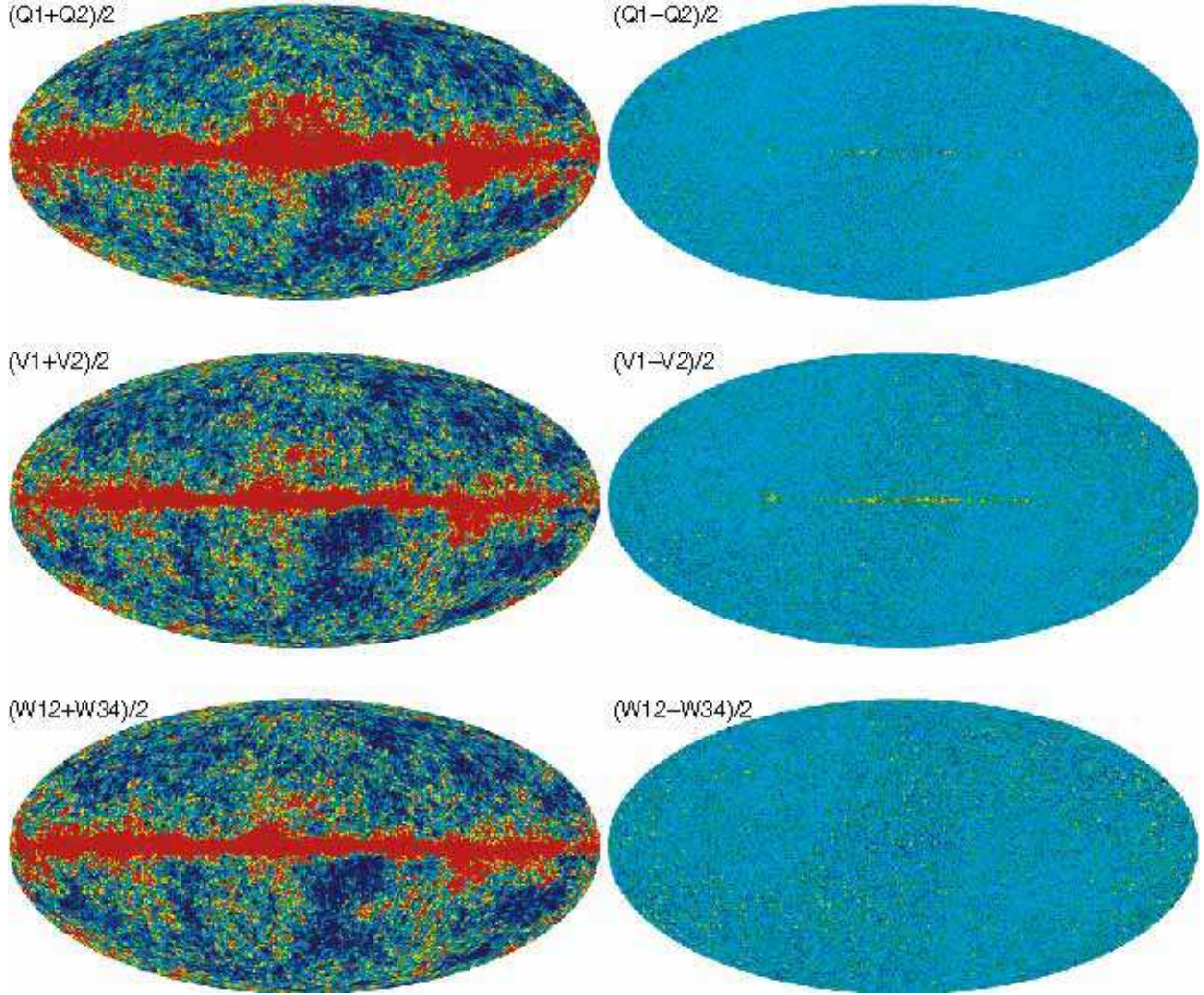}
\end{center}
\caption{Sum and difference maps generated from the flight Q, V, and W band 
data, as indicated.  To reduce the noise, all maps have been binned in larger
pixels (HEALPix $N_{side}=64$) and displayed with a temperature scale of
$\pm$100 $\mu$K.  As discussed in the \S\ref{sec:diff_maps}, the only apparent
structure in the difference maps is due to residual Galactic contamination
owing to the fact that the effective frequencies of the DAs are slightly 
different.  This does {\em not} affect signals with a CMB spectrum, because
the calibration source (the CMB dipole) has the same spectrum. See 
\citet{bennett/etal:2003b} for higher resolution images of the signal maps.}
\label{fig:sum_diff_maps}
\end{figure}

Difference maps are combinations of the data that, ideally, should contain  no
sky signal. They provide insight to potential systematic errors and can be used
to characterize the noise properties of the sky maps. The first set of
difference maps we generate are between DA pairs with the same frequency and
beam response, namely $\case{1}{2}$(Q1$-$Q2), $\case{1}{2}$(V1$-$V2), and
$\case{1}{2}$(W12$-$W34), where W12 = $\case{1}{2}$(W1+W2), and W34 =
$\case{1}{2}$(W3+W4). Images  of these difference maps are shown in
Figure~\ref{fig:sum_diff_maps}, along with low resolution versions of the sum
(signal) maps to give a sense of the signal strength in each map.  Aside from
the pattern of the noise, which follows the sky coverage [see Figure~3 of
\citet{bennett/etal:2003b}], the only visible structure in these difference
maps is in the Galactic plane, especially in V band.  This is understood to be
a result of a small difference in the effective center frequency of the V1 and
V2 differencing assemblies \citep{jarosik/etal:2003}.  In particular, the V1
map has an effective frequency approximately 1 GHz lower than V2. Since the
spectrum of the Galaxy at V band follows $T_A(\nu) \sim \nu^{-2}$ 
\citep{bennett/etal:2003c}, we expect the Galactic signal to be $\sim$3\% 
brighter in V1 than V2, which is consistent with the residual signal seen in
the difference map.  [A complete tabulation of effective center frequencies,
radiometer by radiometer, is given by \citet{jarosik/etal:2003} for diffuse 
sources, and by \citet{page/etal:2003b} for point sources.]  Note that because
the data are calibrated using the CMB dipole, there should be {\em no} residual
CMB signal in such a difference map. A more sensitive comparison of the single
DA maps is  afforded by comparing their angular power spectra.  In that case,
it is easier  to compare across frequencies because differences in beam
response are readily accounted for by deconvolution.  See
\citet{hinshaw/etal:2003b} for such a comparison.

We generate three additional sets of difference maps using different 
combinations of the 4 channels within a DA.  Specifically, we form the 
differences in the time-ordered data then generate maps as follows
\begin{eqnarray}
\frac{1}{2}({\bf d}_{13} + {\bf d}_{14}) - \frac{1}{2}({\bf d}_{23} + {\bf d}_{24})
& \rightarrow & {\bf \Delta}_{12}, \nonumber \\
\frac{1}{2}({\bf d}_{13} - {\bf d}_{14}) + \frac{1}{2}({\bf d}_{23} - {\bf d}_{24})
& \rightarrow & {\bf \Delta}_{34}, \nonumber \\
\frac{1}{2}({\bf d}_{13} - {\bf d}_{14}) - \frac{1}{2}({\bf d}_{23} - {\bf d}_{24})
& \rightarrow & {\bf \Delta}_{1234},
\label{eq:diff_map_def}
\end{eqnarray}
where $\rightarrow$ indicates the map-making process.  The ${\bf \Delta}_{12}$
maps are based on the polarization data, but processed as temperature maps,
i.e., without attempting to demodulate the polarization signal.  Since the 
two radiometers within a DA have completely independent detection chains, 
and since the polarization signal is weak to begin with (and is further 
suppressed by the lack of demodulation) the noise properties of the 
${\bf \Delta}_{12}$ maps should be virtually identical to the nominal signal
maps. The ${\bf \Delta}_{34}$ and ${\bf \Delta}_{1234}$ maps are based on
channel differences, $({\bf d}_{i3}-{\bf d}_{i4})$, and since the two 
channels within a radiometer have partially correlated noise, the noise 
properties of these latter maps will be different than the maps based on
$({\bf d}_{i3}+{\bf d}_{i4})$.  However, these maps do provide a check on the
channel calibration, common-mode thermal effects and other potential artifacts.

\begin{figure}
\begin{center}
\includegraphics[width=1.0\textwidth]{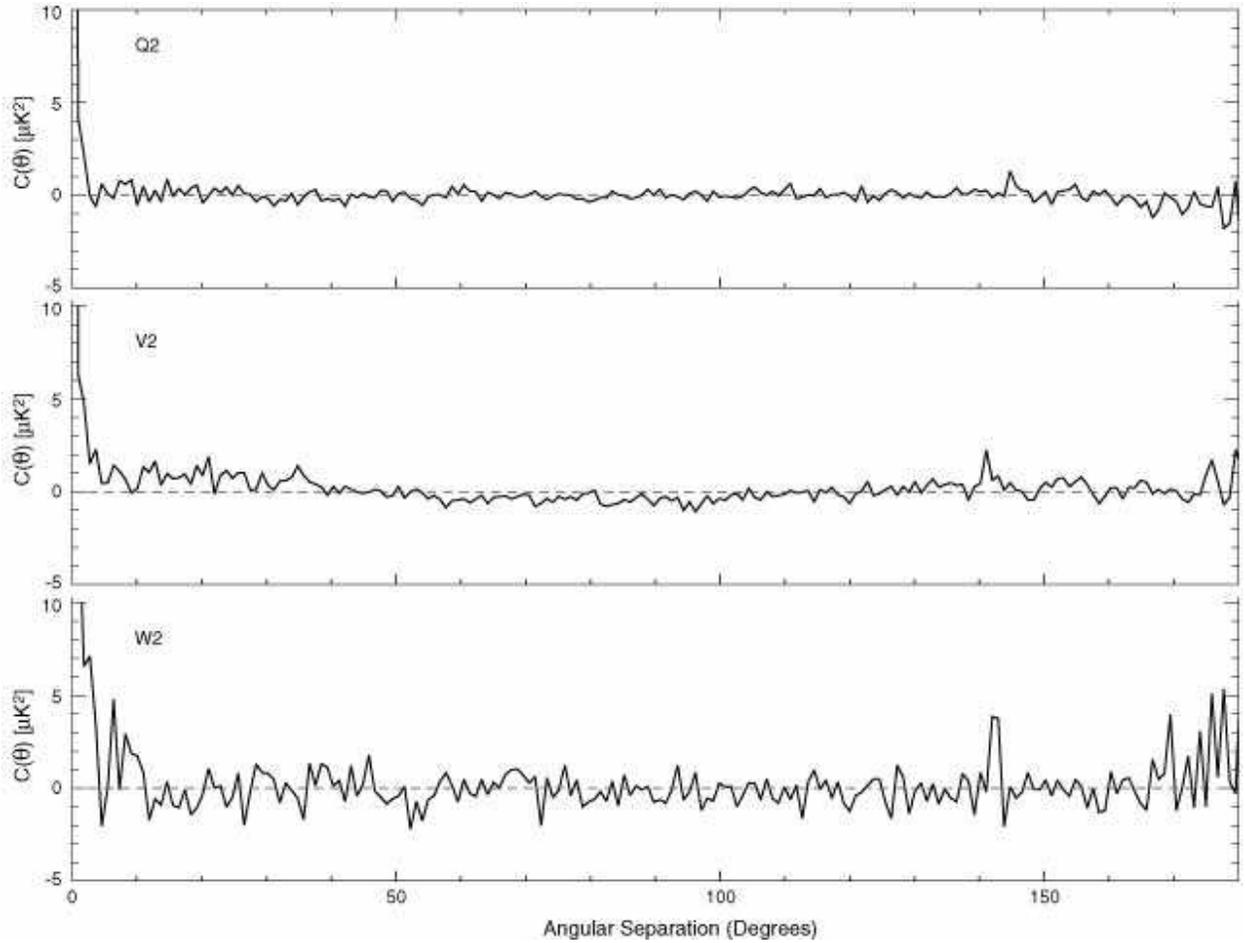}
\end{center}
\caption{Two-point correlation functions of ${\bf \Delta}_{12}$ difference 
maps for three different DAs.  With the exception of a $\sim$0.3\% blip 
at the beam separation angle, $\theta_{\rm beam} \sim 141\dg$, there is 
relatively little structure in the difference maps (see \S\ref{sec:diff_maps}). 
The two-point functions of these maps provide a good representation of the 
angle-averaged pixel-pixel noise covariance in the flight maps.}
\label{fig:q2v2w2_corr}
\end{figure}

\begin{figure}
\begin{center}
\includegraphics[width=0.9\textwidth]{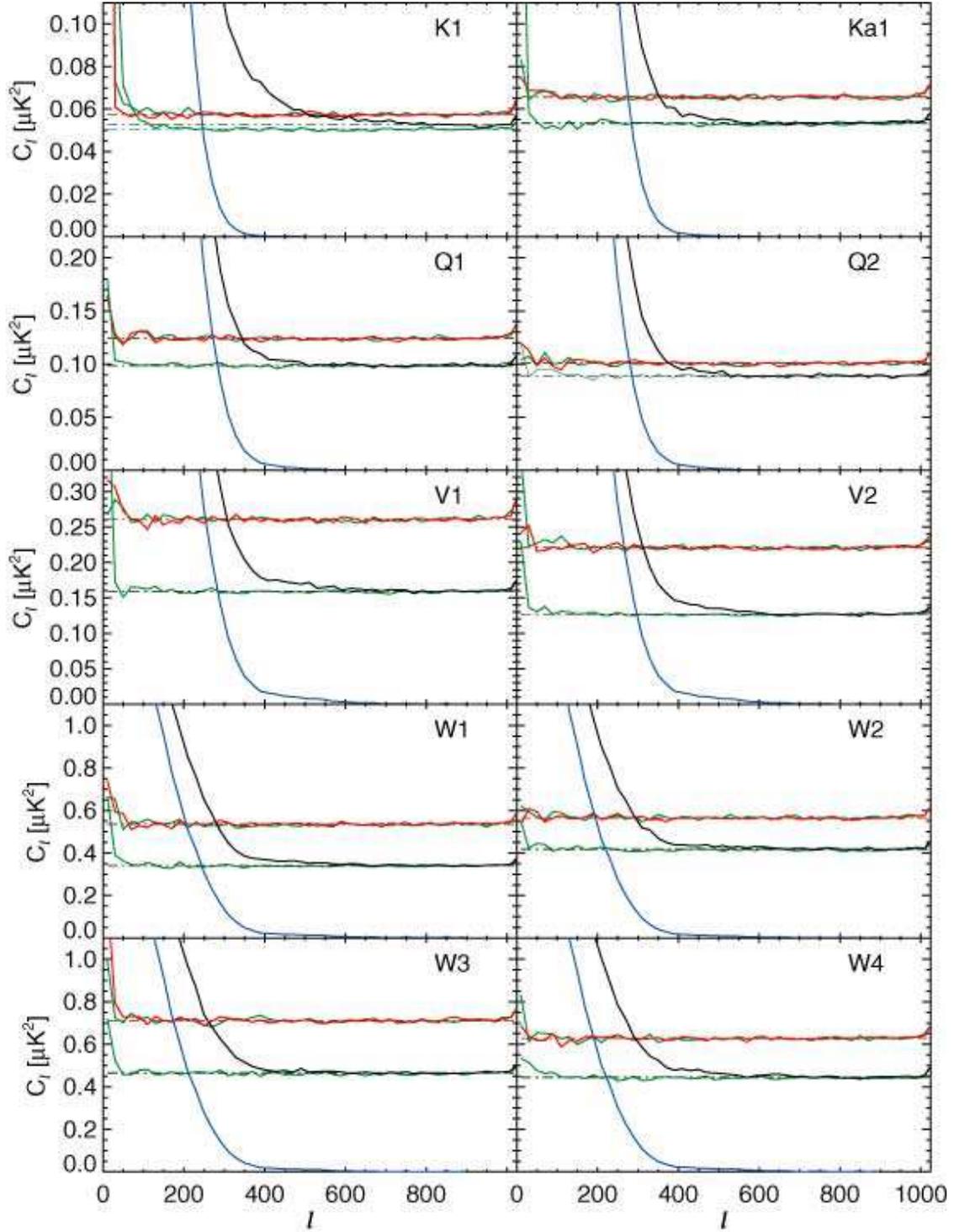}
\end{center}
\caption{Angular power spectra of signal and noise maps for each DA.  In each
panel, the upper red and green traces are the spectra of the null maps, 
${\bf \Delta}_{34}$ and ${\bf \Delta}_{1234}$, respectively.  The lower green
trace is the ${\bf \Delta}_{12}$ map, and the black trace is the signal map.
The blue curve is our best estimate of the underlying CMB signal from 
\citet{hinshaw/etal:2003}.  The pairing of white noise levels is discussed in
\S\ref{sec:diff_maps}, Table~\ref{tab:diff_analysis} presents a measure
of structure in the difference spectra, which are remarkably flat.}
\label{fig:diff_map_ps}
\end{figure}

For each difference map the two-point correlation function and the angular
power spectrum are calculated.  The results are shown in
Figures~\ref{fig:q2v2w2_corr} and \ref{fig:diff_map_ps} and summarized in
Table~\ref{tab:diff_analysis}.  Figure~\ref{fig:q2v2w2_corr} shows the two-point
function computed from the ${\bf \Delta}_{12}$ maps for Q2, V2, and W2.  The
most apparent feature in each of these functions is the slight bump at the beam
separation angle of $\theta_{\rm beam}\sim141\dg$, as expected
(\S\ref{sec:map_making_pre_cal}); the first data column of
Table~\ref{tab:diff_analysis} gives $C(\theta_{\rm beam})/C(0)$, for each DA. 
Note that, with the exception of K band, the ratio is typically 0.3\%. The
larger K band values arise because Galactic leakage in these difference maps is
most severe in this band. This is also the source of the weak large-scale 
feature in the V2 two-point function in Figure~\ref{fig:q2v2w2_corr}. While this
residual signal is small compared to the temperature signal, it is a systematic
error that must be  accounted for in the analysis of polarization data
\citep{kogut/etal:2003}. 

Figure~\ref{fig:diff_map_ps} shows the angular power spectra of the difference
maps for each of the 10 DAs, as well as for the final signal maps.  
Table~\ref{tab:diff_analysis} summarizes their statistics.  Note that the white
noise plateau in the 4 channel combinations per DA divide into two  families,
as noted above, due to the correlations between channels 3 and 4. As a
result, the null combinations, ${\bf \Delta}_{34}$ and ${\bf \Delta}_{1234}$,
{\em cannot} be used to estimate the white noise parameter $\sigma_0$ for the
signal maps.  However, the polarization channel is seen to be in excellent
agreement with the temperature channel in the white noise tail, thus, to the
extent that real polarization signals, or other  systematics, such as bandpass
mismatch are not important, these maps should provide an excellent noise model
for the temperature data.  Table~10 summarizes the shape of the angular power
spectrum at low $l$ in the same way  Table~\ref{tab:sim_res_map_ps} did for the
simulation.  We find the spectra of these difference maps to be remarkably
flat, with residual quadrupole moments of $<$4 $\mu$K$^2$ for all bands except
K (in which the difference is dominated by bandpass mismatch) and a single
combination of W3.  This value is much smaller than the small quadrupole
measured in our sky \citep{bennett/etal:2003b}. The power in the other
multipole ranges is very close to the white noise floor,  as seen in the final
columns of Table~\ref{tab:diff_analysis}.  Since the residual signals seen in
the flight difference maps are somewhat lower than those seen in the simulation
(Table~\ref{tab:sim_res_map_ps}), we adopt the more conservative limits from the
simulation as systematic error limits for  structure at low $l$.  This allows
for the possibility than some of the error seen in the simulation comes from,
e.g., common-mode calibration errors that cancel in the difference maps.

\begin{figure}
\begin{center}
\includegraphics[angle=90,width=0.9\textwidth]{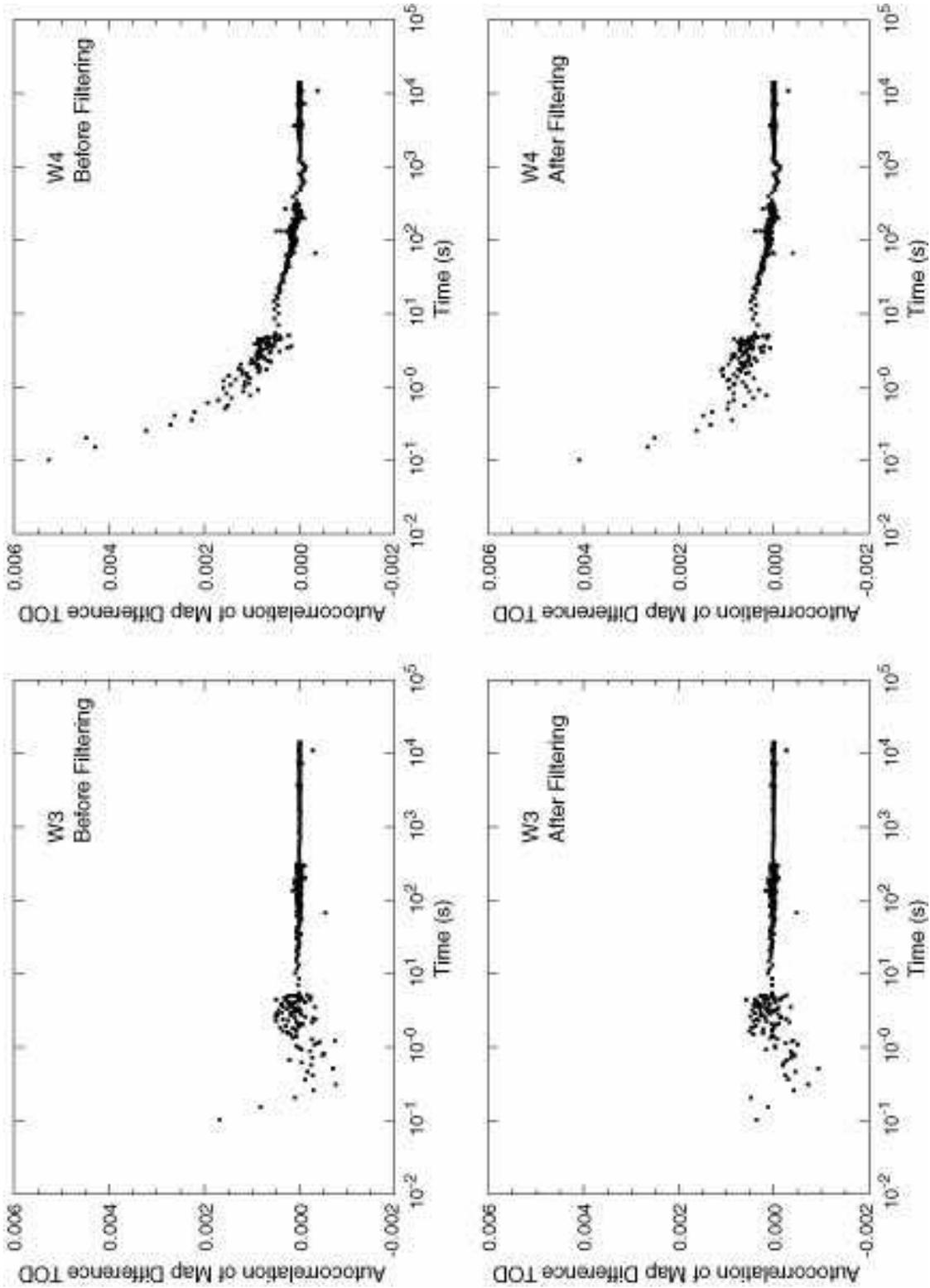}
\end{center}
\caption{An estimate of pixel-pixel noise covariance in a W3 and W4 noise 
map along the scan direction before and after filtering (top and bottom,
respectively). See \S\ref{sec:diff_maps} for a description of the processing 
steps used to produce these data.  The stripe covariance is negligible in W3,
and $<$0.2\% in W4 for lags $>$0.1 s (pixel separation $>$0\ddeg25).  All other
DAs will have at least 2-3 times lower covariance than W4.}
\label{fig:ac_mapdiff}
\end{figure}

The two-point correlation function of the ${\bf \Delta}_{12}$ maps demonstrates
that the angle-averaged off-diagonal terms of the pixel-pixel covariance 
matrix are less than $\sim$0.3\%. However, the maps in 
Figure~\ref{fig:w3w4_sky_filter} indicate the potential for stripes along the 
scan directions for which the covariance can be locally higher than the angle-
averaged value.  In order to determine the magnitude of the covariance along
the scan directions we perform the following computation.  We form W band 
difference maps: W$i-$W$i'$, where $i$ = 1-4, and W$i'$ is the average of the 3
other W band maps, e.g., W1$'= \case{1}{3}$(W2+W3+W4).  We then form
time-ordered data from this map using the pointing appropriate to DA W$i$ and
compute the auto-correlation function, $C(\Delta t)$, from 30 days of data.
This provides a measure of the pixel-pixel covariance along a stripe.  The
results for W3 and W4 are shown in  Figure~\ref{fig:ac_mapdiff}; the top panels
show the covariance, normalized to one at lag zero, computed from the
unfiltered maps, while the bottom panels  show the results for the filtered
maps.  In order to convert the time axis to angular displacement along a scan,
recall that the 2\ddeg784 s\per spin rate translates to a 2\ddeg6\ s\per rate
for the beams in either focal plane (the second decimal place depends on 
position in the focal plane, and time in the precession cycle). The W3 result
shows a hint of covariance ($\sim$0.1\%) at lag 0.1 s, or 0\ddeg26, prior to
filtering, but {\em none} after ($<$0.05\%).  Prior to filtering, the W4 result
shows clear covariance of up to 0.5\% at small lag, decaying to $<$0.1\% at
lags of $\sim10^2$ s, roughly one full spin.  After filtering, the covariance
is reduced by nearly a factor  of two, but is still clearly detectable.  This
is understood to be residual covariance that survives the filtering process
because of the fact that we subtract an estimated sky signal, based on the
data, prior to filtering the noise, then add it back in to restore the signal. 
The algebra of this process is presented in Appendix \ref{app:map_filter}.  We
emphasize  that W4 is the worst DA for $1/f$ stripes by at least a factor of 3
\citep{jarosik/etal:2003b}, and we limit covariance along scan directions to be
$<$0.1\% for all other \map\ first-year sky maps.

\subsection{Pointing and Beam Determination}
\label{sec:pointing}

\subsubsection{Spacecraft Attitude Control and Determination}
\label{sec:acs_quat}

The spacecraft attitude is determined from a combination of two autonomous 
star trackers (ASTs) with boresights perpendicular to the spin axis (along 
the spacecraft $\pm y$ axes), two rate gyroscopes, and two digital sun
sensors.  The sensor outputs are combined using a Kalman filter to determine
the aspect  solution. The sensor noise parameters and offsets were initially
calibrated in flight during the in-orbit checkout (IOC) period in July 2001.  
By the end of IOC, the final tables were uploaded to the spacecraft. 

Spacecraft quaternions output by the Kalman filter provide the definitive 
transformation from the spacecraft reference frame to the J2000 geocentric
inertial (GCI) system. Errors in the attitude solution are estimated using the
residuals of the  individual sensor signals and propagated to the quaternions.
After the final  Kalman filter parameters were loaded, quaternion differences
show a noise-like  error with a $10\asec$ rms. In addition to the sensor
noise, there is an apparent spin-synchronous error of $\sim10\asec$ that is
believed to be due to propagation errors in the Kalman filter. As discussed
below, this error  is apparently seen in the instrument boresight determination
using Jupiter  observations.  Since the pointing performance exceeds the
requirement of  $0\damin9$ (root-sum-square for three axes), no correction of
the spacecraft  quaternions is attempted for the first-year processing. 
Sufficient information exists in the raw telemetry to attempt a correction in
the future, if it is warranted.  Note that random quaternion errors are
automatically  accounted for in the flight beam response maps generated from
the Jupiter  observations \citep{page/etal:2003b}.  

\subsubsection{Instrument Boresight Determination}
\label{sec:boresight}

As mentioned above, the spacecraft quaternions provide the definitive reference
frame for the spacecraft.  The instrument boresights, 10 each on the A and B 
sides, are determined from the Jupiter beam maps \citep{page/etal:2003b},  
which are generated with respect to the spacecraft frame provided by the 
quaternions. The boresight is defined as the location of the peak of a circular
Gaussian fit to the main beam.  The results of this fitting are given in 
Table~\ref{tab:boresight} as 20 unit vectors in spacecraft coordinates.  These
are  the values used to determine instrument pointing in the first-year data 
processing. The uncertainty in the boresight position is $\sim2\asec$ per beam
in both spacecraft azimuth and elevation. In addition to statistical
uncertainty in the boresight fits, there are two other potential sources of
error in the boresight determination: changes (drifts) with time, and errors in
the relative time-tagging of quaternion  data and science data.

To test stability, we note that \map\ is in a position to see Jupiter twice per
year for about 45 days each time. We refer to each $\sim$45 day period as a
Jupiter ``season''.  During the first season of each year, the boresights scan
across Jupiter from roughly ecliptic north to ecliptic south, and vice-versa in
the second season.  As a test of  boresight stability, we have generated beam
maps from each of the first two  seasons of data separately, and have fit
boresight directions to each.  We  find the azimuth positions agree to better
than $3\asec$ on both the A and B  sides, but the elevation positions differ by
$\sim10\asec$ on the A side, and a smaller amount on the B side.  This 
difference between seasons is consistent with the $\sim10\asec$
spin-synchronous error in the spacecraft quaternions discussed above.  We
ignore this small effect in the first year processing, and subsume the 
small systematic error that results into our error estimate for the beam
transfer functions, as discussed in \citet{page/etal:2003b}.

The relative time-tag accuracy of telemetry packets was tested on the ground. 
A timing computer was set up to simultaneously receive test pulses from both 
the Attitude Control Electronics (ACE) box and the Digital Electronics Unit 
(DEU), the two computers that tag the attitude and science data packets, 
respectively.  Each of these boxes in turn derives its time from the main 
``Mongoose'' computer on \map\ \citep{bennett/etal:2003}.  This test
demonstrates a relative time-tag  accuracy of 30 $\mu$s between the quaternion
packets and the science packets.  In observing mode, the boresights sweep the
sky at a rate of $2\ddeg6$ s$^{-1}$,  so a  30 $\mu$s time error produces a
negligible pointing error of $< 0.3\asec$.

Uncertainty in the spacecraft position is another potential source of boresight
determination error.  For the first year processing we use a predicted
ephemeris that is uploaded to the spacecraft approximately weekly for on-board
use by the Attitude Control System.  This solution is returned in telemetry and
is the basis for the ephemeris data supplied with the first-year release of
time-ordered data.  The uncertainty in these predictions are $<7$ km in
position and $<$1 cm s$^{-1}$ in velocity (3 $\sigma$), relative to the Earth. 
An error of 7 km in \map's position would result in error of $\sim$2 mas in the
apparent position of  Jupiter as seen from \map\ and is thus completely
negligible.

\subsubsection{Beam and Window Function Determination}
\label{sec:beam_window}

Along with gain calibration and noise properties, knowledge of the beam shapes
and window functions are among the most important aspects of the instrument to
characterize for accurate measurements of the CMB.  \citet{page/etal:2003b}
describe in detail the process by which beam maps are formed from in-flight 
observations of Jupiter, and how those maps are transformed to determine the
beam window functions.  The primary result they derive are a set of 10 beam
transfer functions, $b_l$, one per DA, based on azimuthally-averaged beam 
radial profiles.  These transfer functions are included in the first-year data
release \citep{limon/etal:2003}.  In addition, they derive a full covariance
matrix for each  transfer function, $\Sigma^b_{ll'}$, which characterizes the
uncertainty in $b_l$.  Typically, the uncertainty for a single  DA is about 
1-2\%, with moderate covariance in $l$. See Figure~5 of
\citet{page/etal:2003b}.  These estimates already include a systematic error
allowance to bound the small range of results obtained from different analysis
methods.  As described in \citet{hinshaw/etal:2003b}, the window function
covariance matrices are propagated into the Fisher matrix (inverse covariance
matrix) for the final combined angular power spectrum.  Thus the final  power
spectrum, and the parameter fits based on it, already include statistical and
systematic window function uncertainties
\citep{spergel/etal:2003,verde/etal:2003,peiris/etal:2003}.

\subsubsection{Effects from Elliptical Beams}
\label{sec:elliptical_beam}

The \map\ beams are moderately elliptical, so the use of azimuthally-averaged
radial profiles to describe the beam response is an approximation.  This
approximation is justified in \citet{page/etal:2003b} by noting that the \map\
scan pattern produces excellent azimuthal averaging of the beam response in a
large fraction of the sky.  They have placed limits on the variation of the 
window function across the sky by comparing the effective window function in
the ecliptic plane, based on a full two-dimensional transform of the beam 
response averaged over the flight range of scan angles, to the fully averaged 
transform, $b_l$.  For the three highest frequency cosmology bands, these 
variations range from 2-3\% at Q band to $\sim$1\% in V and W bands.  These
variations are consistent with estimates of the angular power spectrum using 
data at high and low ecliptic latitudes separately \citep{hinshaw/etal:2003b}.
Since most of the statistical weight at high-$l$ resides in the V and W band 
data at high ecliptic latitudes, the use of fully averaged beam transforms is 
appropriate, and the systematic error estimate incorporated into 
$\Sigma^b_{ll'}$ should encompass any error in this approximation.

Elliptical beams can also produce errors in the sky maps that are difficult to
characterize in a simple way.  We can define the sky map error due to 
non-circular beam response as
\begin{equation}
{\bf \Delta t}_{\rm asym} \equiv {\bf t}_{\rm obs} - {\bf t}_{\rm circ},
\end{equation}
where ${\bf t}_{\rm obs}$ is the hypothetical noise-free sky map obtained with the 
actual experimental beam and scan pattern, and ${\bf t}_{\rm circ}$ is the ideal 
sky map obtained by convolving the true sky with the averaged beam transform, 
$b_l$.  For a differential experiment like \map, there are two effects that
contribute to ${\bf \Delta t}_{\rm asym}$. The first, as noted, is incomplete 
azimuthal coverage in a given pixel, which gives rise to slightly elliptical
peak structure at low ecliptic latitudes (see below), the second is a
localized  effect due to echoes from bright Galactic sources propagating to
other pixels in the map. Specifically, as a bright source is observed in
different orientations, the differential signal changes with orientation.  Since
the map-making algorithm must assign one  average value to the pixel with the
bright source, the ring of pair pixels at  the beam separation will see an echo
with a quadrupolar temperature  distribution around the ring.  We mitigate this
effect by incorporating a  bright source mask in the map-making algorithm, 
which is invoked as follows.  If side A observes a pixel in the bright source 
mask, we only update the sky map accumulator for  pixel A, but {\em not} for 
pixel B
\begin{eqnarray}
{\bf n}_{\rm obs}\cdot{\bf t}_{n+1}(p_A) \rightarrow {\bf n}_{\rm obs}\cdot{\bf t}_{n+1}(p_A) 
+ w_i\,[ {\bf d}(t_i) + {\bf t}_n(p_B) ] 
& \mbox{     } & {\bf n}_{\rm obs}(p_A) \rightarrow {\bf n}_{\rm obs}(p_A) + w_i 
\nonumber \\
{\bf n}_{\rm obs}\cdot{\bf t}_{n+1}(p_B) \not\rightarrow {\bf n}_{\rm obs}\cdot{\bf t}_{n+1}(p_B) 
- w_i\,[ {\bf d}(t_i) - {\bf t}_n(p_A) ]
& \mbox{     } & {\bf n}_{\rm obs}(p_B) \not\rightarrow {\bf n}_{\rm obs}(p_B) + w_i,
\label{eq:accum_map_ab}
\end{eqnarray}
where the terms are as defined in after equation~(\ref{eq:accum_map}). In this
way we obtain an estimate of ${\bf t}(p_A)$, but we do not propagate bright
echoes to the ring of neighbor pixels, of which  $p_B$ is one.  The mask we use
for assigning this cut is the same Kp8  processing mask we used for the
calibration fits (\S\ref{sec:dipole_cal}).

\begin{figure}[hbt]
\begin{center}
\includegraphics[width=1.0\textwidth]{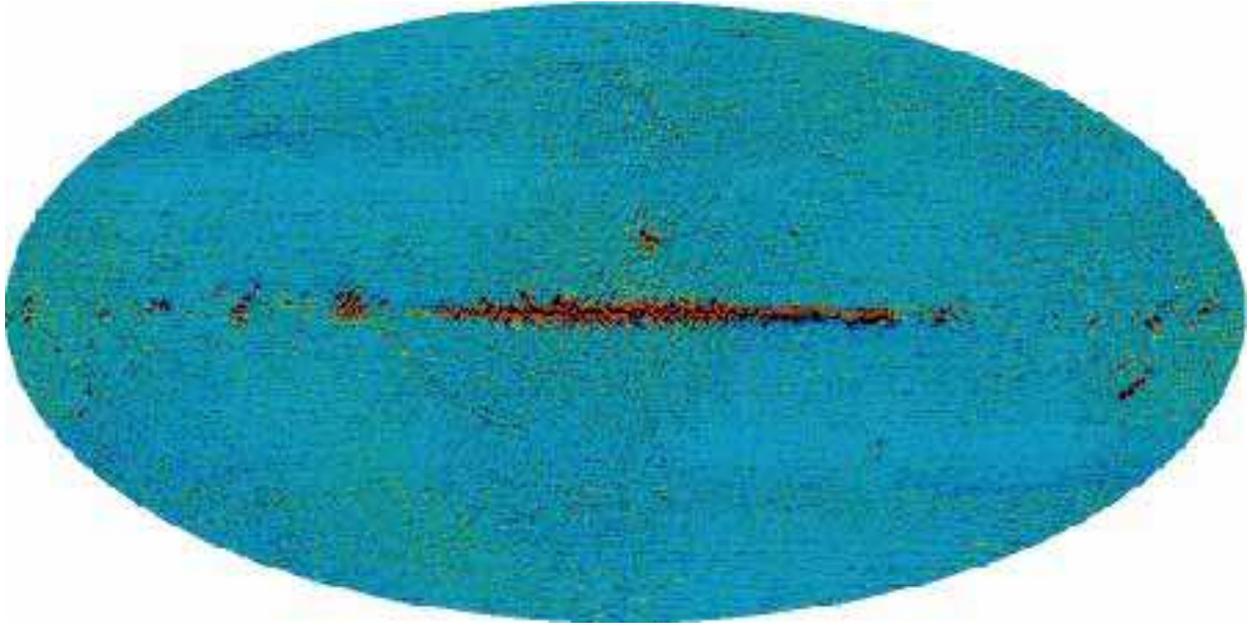}
\end{center}
\caption{Residual map from a K band elliptical beam simulation. The output
map was generated from a one-year simulation of data with an elliptical beam
response. The residual map shown was generated by subtracting the underlying 
sky signal convolved with the nearest effective circular beam response. 
This remaining structure contributes to the four-point fluctuation spectrum. 
The scale of the color range is $\pm$10 $\mu$K.  The rms structure in the 
Kp2 cut sky is 2 $\mu$K.  See \S\ref{sec:elliptical_beam}.}
\label{fig:beam_asym}
\end{figure}

We have generated a simulation to quantify the errors from both of these 
effects.  Specifically, the simulation generates one year of noise-free 
differential sky signal which includes a model for the flight beam
ellipticity.  We run this data through the flight map-making pipeline to
generate sky maps, ${\bf t}_{\rm obs}$.  We also generate convolved maps ${\bf
t}_{\rm circ}$ using the azimuthally-averaged beam transforms appropriate to
the beam model.  The  residual map, ${\bf \Delta t}_{\rm asym}$, for DA K1 is
shown in  Figure~\ref{fig:beam_asym}.  The K band radiometers have the largest
beam  ellipticity of all the DAs, so this represents a worst case result.  The
general ``mottling'' near the ecliptic plane results from the relatively
limited azimuthal coverage in this region producing elliptical peaks and
anti-peaks which, in ${\bf \Delta t}_{\rm asym}$, are differenced with circular
counterparts.  This is especially noticeable near bright Galactic sources. The
rms amplitude of these fluctuations in the Kp2 cut sky is 2 $\mu$K in K band,
and at least a factor of 2 lower in Q-W bands. The effect of this structure in
the power spectrum is primarily represented as a variation in the window
function across the sky, as discussed above, and in \citet{page/etal:2003b} and
\citet{hinshaw/etal:2003}.  However, this structure also contributes to the
4-point function of the data in the sense that it couples power from different
$l$  ranges.  This effect is potentially important for the interpretation of
any gravitational lensing analysis of the \map\ data.

The figure also exhibits faint echoes of the brightest sources that evade the
map-making cut discussed above. We limit localized features in the Kp2 cut
sky to less than 10 $\mu$K in K band and less than 2 $\mu$K in Q-W bands due
to a combination of dimmer sources and more circular beams.  We estimate that
such features occupy $<$0.1\% of the Kp2 cut sky.

\subsubsection{Far Sidelobe Pick-up}
\label{sec:sidelobe}

The \map\ optical system was designed to produce minimal pick-up from signals
entering the far sidelobes.  \citet{barnes/etal:2003} present a complete 
determination of the \map\ sidelobe response by combining measurements from a 
variety of ground-based sources with in-flight measurements of the Moon.   They
produce response maps covering $4\pi$ sr that are included as part of the
first-year data release. They then use these response maps, with the 
first-year sky maps, to estimate the systematic artifacts remaining in the 
first-year maps based on the well-justified assumption that sidelobe artifacts 
are small relative to the sky signal.  The K band data have the largest
sidelobe signal due both to the largest sidelobe spill, and to the brightest
Galactic signal.  The signal was deemed to be large enough, and well enough 
characterized, to warrant a small post-processing correction to the first-year 
K band map.  Limits on remaining sidelobe induced artifacts in all the bands 
are presented in Table~1 of \citet{barnes/etal:2003}.

\subsection{Environmental Effects}
\label{sec:environmental}

\subsubsection{Thermal Effects}
\label{sec:thermal}

The radiometer gain and offset are dependent on temperature.  There are several
aspects of the \map\ design that are critical to mitigating this source of
systematic error \citep{bennett/etal:2003}. The instrument is differential, so
thermally induced gain changes act on a relatively small offset signal.  The
observatory environment was designed to be as stable as possible, consistent
with other goals. For example, all nominal thermal control is passive to avoid
heaters cycling on and off.  The observatory is placed at the second Earth-Sun
Lagrange point far from the Earth, and the solar panels maintain a fixed
$22\ddeg5$ angle with respect to the Sun during normal observing mode.  The
instrument temperature is monitored with precision platinum resistance
thermistors (PRTs) to verify the degree to which  thermal stability is in fact
achieved.

Temperature variations at the spin period are the most critical since they can
induce signals that couple relatively efficiently to the sky. However, owing to
the relatively fast (129.3 s) spin period and the thermal mass of the
instrument, any induced signals will have a very red spectrum and thus will
couple only to the lowest few harmonic modes on the sky ($l \lsim 10$).  In the
analysis below we use flight data to estimate the susceptibility of the gain
and baseline to temperature variations of the instrument.  In turn we use
limits on the instrument's physical temperature variation at the spin period
from \citet{jarosik/etal:2003b} to put limits on thermally induced artifacts in
the time-ordered data, and hence the sky maps.

Thermally induced signals can enter either through changes in the gain acting
on the instrument offset or through changes in the offset itself.  We show
below that the latter are more significant for \map.

The radiometer gain model presented by \citet{jarosik/etal:2003b}  describes
the gain in terms of the RF bias (``total power'') of the detector, and the
temperature of the FPA. This model tracks thermal variations in the gain on the
time scale of the RF bias readout (23.04 s), and the map-making algorithm
updates the gain on this time scale.  However, since this sample rate is only a
few times per spin, it is possible that a systematic temperature variation at
the spin period could induce gain changes that are not well sampled by this
model.  As a separate check of gain-induced artifacts, we infer the temperature
susceptibility of the gain from data taken over a long time period where gain
changes are measurable.  Results are given in Table~\ref{tab:suscep_raw}.  We
combine these measurements with the limits on temperature modulation at the
spin period derived by \citet{jarosik/etal:2003b} to place limits on gain
induced artifacts, as shown in Table~\ref{tab:suscep_summary}.  We conclude
that gain-induced signals at the spin period, which might be poorly tracked by
the gain model, are $<$20 nK.

\begin{figure}[hbt]
\begin{center}
\includegraphics[width=0.9\textwidth]{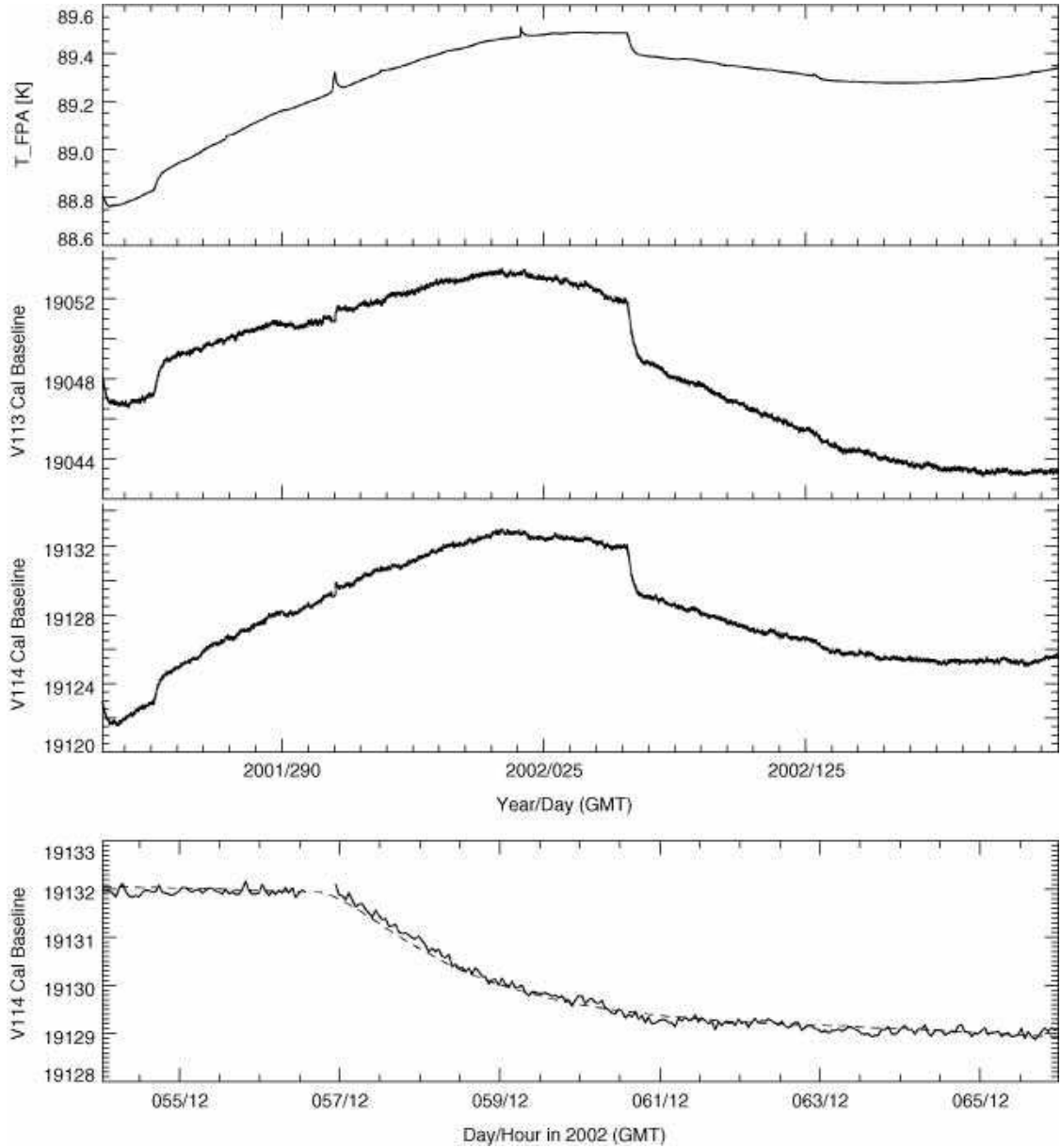}
\end{center}
\caption{An illustration of the mild thermal susceptibility of the instrument
baseline.  The top panel shows the temperature of the instrument Focal Plane
Assembly (FPA) over the course of the first year.  The second and third panels
show the hourly baseline solution for channels V113 and V114,  which are among
the most thermally susceptible.  Note that the thermal baseline response is
mostly common-mode. The channel combination that contains sky signal is the
{\em difference} between channels 3 and 4, thus most of this response cancels. 
On day 2002:054 (GMT) a partial battery cell failure led to a commanded
decrease in spacecraft bus voltage with a corresponding  decrease in overall
power dissipation and spacecraft temperature.  This event provides a clean
measurement of the instrument baseline thermal susceptibility -- the bottom
panel shows a close-up of the V114 baseline near this event.  The dashed line
is a fit to a model including a term proportional to $\partial {\bf b} /
\partial T_{\rm FPA}$. The best-fit susceptibility  values for all channels are
given in Table~\ref{tab:suscep_raw}.  See \citet{limon/etal:2003} for a 
complete discussion of \map's first-year thermal profile.}
\label{fig:baseline_susc}
\end{figure}

The instrument baseline is the product of the gain times the offset. As
described in \S\ref{sec:map_making_cal}, we get an initial estimate of the
baseline from the dipole calibration algorithm.  This gives us an estimate of
the instrument baseline once per hourly precession period.  Sample hourly 
baselines for channels V113 and V114 are shown as a function of time over the
first year of operation in Figure~\ref{fig:baseline_susc}.  Also shown is the
temperature of the instrument FPA over the same time period; there is a clear
temperature dependence in the baseline.  We measure the baseline temperature
susceptibility by fitting the hourly baseline estimates to a model of the form
${\bf b}(t)  = c_0 + c_1 t + c_2 \Delta T_{\rm FPA}(t)$ where the $c_i$ are
model coefficients and $\Delta T_{\rm FPA} = T_{\rm FPA} - \left< T_{\rm FPA}
\right>$ is the deviation of the FPA temperature from its mean.  The most
robust susceptibility results come from fitting a portion of the data near the
time of a partial battery cell failure which occurred on day 2002:054 (GMT)
\citep{limon/etal:2003}.  In response to this event, the  spacecraft bus
voltage was autonomously commanded lower on day 2002:058 (GMT) causing the
spacecraft to dissipate less power and thus cool slightly. The coefficients are
given in Table~\ref{tab:suscep_raw}. We have combined the results for the two
channels in each radiometer because this is the combination that enters into
the final sky maps. This has the effect of canceling some of the common-mode
susceptibility measured in individual channels.  As noted above, we combine
these susceptibility measurements, taken over long time periods, with limits on
the temperature variations at the spin period \citep{jarosik/etal:2003b} to
place limits on induced signals at this time scale.  The results are given in
Table~\ref{tab:suscep_summary}.  We conclude that offset-induced signals at the
spin period are $<$180 nK.

Slow drifts in the instrument temperature will be largely filtered out by the
baseline pre-whitening discussed in \S\ref{sec:filter}.  The steepest 
temperature gradient observed during the first year of observation occurred
just after the above-mentioned battery cell failure.  To assess the efficiency
of the filtering process, we have analyzed the data during this  period as
follows.  We applied the baseline pre-whitening filter to the temperature
signal, $T_{\rm FPA}(t)$, to measure how much the cooling gradient  was
suppressed by the filter.  The input temperature gradient on day 2002:058 (GMT)
was $-1.7$ mK hr\per.  Applying the K11 filter to $T_{\rm FPA}(t)$  yielded an
output gradient of $-0.1$ $\mu$K hr\per, while the W41 filter yielded an upper
limit 10 times smaller.  We conservatively estimate upper limits on residual
drift in the filtered baseline of $<$10 nK hr\per for the most
susceptible channels.

\subsubsection{Electrical Effects}

A variable electrical signal on board the observatory could induce an apparent
signal in the radiometers.  Sources of variable electrical signals include the
reaction wheels, transponder, bus voltage fluctuations, and RF noise coupled to
the instrument through the power bus.  During the final observatory thermal
vacuum test, in which the observatory was operating at temperatures close to
those achieved in flight, searches for such electrically induced radiometric
artifacts were conducted \citep{jarosik/etal:2003}.  Upper limits on radiometer
bus voltage susceptibility, based on ground tests, are given in
Table~\ref{tab:suscep_raw}. We combine these results with an upper limit on bus
voltage variations of 3.0 mV rms, measured  on-orbit, to conclude that
electrically-induced signals at the spin period are $<$40 nK.  See
Table~\ref{tab:suscep_summary}.

\subsection{Miscellaneous Effects}

\subsubsection{Radiometer Cross-talk}

A large signal in one radiometer could induce an erroneous signal at the output
of another radiometer due to electrical cross talk. Such cross talk is not
expected but could arise from, e.g., non-ideal amplifier behavior, or other
parasitic effects, such as pickup in the wiring harnesses.

A careful search was made for this effect during the instrument ground tests.
Noise diodes were used to inject a large signal into one radiometer at a time
while the input feeds of all other radiometers were covered by absorptive
loads. The outputs of the 9 non-driven differencing assemblies were searched
for any evidence of the injected signal. The tests were run with the amplifiers
in the passive channels both on and off in order to distinguish pickup
mechanisms. No pickup was found in any test. Table~\ref{tab:xtalk_summary}
gives  $2\sigma$ upper limits on the pickup by any DA due to any of the 9
other DAs.  The column labeled Electrical gives the results obtained from the
test with the amplifiers turned off, and the column labeled Radiometric gives
weaker limits from the test with the HEMTs turned on.  The latter limits are
weaker because the output noise levels are higher.

This level of pickup is quite small and could only be of potential concern when
\map\ scans across Jupiter, the brightest source in the sky for \map\ at L2.
The values in the Table~\ref{tab:xtalk_summary} are thus referred to peak
Jupiter signals in each band.  For example, the first entry indicates that when
Jupiter induces a signal of 185 mK in the W1 radiometer, the pickup in the K1
differencing assembly is  $<$30 $\mu$K (95\% confidence) which is $-$26.8 dB below
the peak Jupiter signal of 14 mK in K band. This signal occurs when the beam of
the pickup channel is within a few degrees of Jupiter, depending on channel
separation in the focal plane, and in every case is less than direct
radiometric detection of Jupiter in the near side lobes.

This limit on cross talk implies pickup is completely inconsequential in normal
observing mode.  Using the same Table example, a 100 $\mu$K signal in W1 could
cause at most a 200 nK signal in the most susceptible the four K band
differencing assemblies.

\subsubsection{Source Variability}
\label{sec:var_source}

Time variable objects are a potential source of contamination for observations
of the CMB; see for example \citet{sokasian/gawiser/smoot:2001} and references
therein. One concern is that an object may grow in brightness over the course
of \map's observations, avoid detection during an initial source survey, and
remain unmasked during subsequent data analysis. For example, blazars produce
relatively rapid and large amplitude variability in all wavebands. Long term
observations of such objects show that increases in flux by a factor of up to a
few over a time scale of years can be anticipated \citep{flett/henderson:1983,
ennis/neugebauer/werner:1982,stevens/etal:1994,bower/etal:1997}. Observations
of Zw 2 by \citet{Falcke/etal:1999}, provide an extreme example: a greater than
20-fold increase in brightness, from $\sim$0.1 Jy to $\sim$2 Jy, over a period
of less than two years.  While  this object could produce a temperature
response of a few hundred $\mu$K in the \map\ data, such objects are rare and,
if left undetected, would have a minimal effect on cosmological inferences.
Tests for point source contamination in the \map\ data are given by
\citet{bennett/etal:2003c} and \citet{hinshaw/etal:2003}. These tests will need
to be revisited on an annual basis.

Another source of concern is that a time variable source in the \map\ data  has
the effect of broadcasting noise to the ring of $\sim$1000 pixels which  are
separated from the variable source by the beam separation angle ($\theta_{\rm
beam}\sim141\dg$). The point source list derived from the \map\ first-year data
is 98\%  reliable with $\sim$5 spurious detections at the $\sim$0.5 Jy flux
limit of the  survey \citep{bennett/etal:2003c}. The nominal point source
sensitivity of the \map\ telescope is $\Gamma \sim 200$ $\mu$K/Jy, thus a noise
level of $\sim$0.1 $\mu$K  is expected from variable sources that evade
detection.

\section{CONCLUSIONS}
\label{sec:conclude}

The processing steps used to produce the first-year \map\ sky maps include an
initial simultaneous estimate of the sky map and the instrument  calibration. 
The instrument gain is then refined using a model based on engineering
telemetry, and the instrument baseline is refined by the  application of a
pre-whitening filter.  A final archive of calibrated data is produced and used
to generate final sky maps using a slightly refined iterative algorithm.

We limit systematic artifacts due to calibration, map-making and environmental 
disturbances to less than $\sim$15 $\mu$K$^2$ in the quadrupole $C_2$, with 
tighter limits at higher multipole moments (Table~\ref{tab:sim_res_map_ps}).
Beam transfer functions are measured for each beam with 1-3\% over the entire
range of  multipole moments that \map\ is sensitive to \citep{page/etal:2003b}.
The covariance matrix of the beam transfer function is propagated through to
the final power spectrum error matrix. We characterize pixel-pixel covariance
matrix and place limits on residual stripes in the final maps.  

All major data products from the first year of \map\ observations are being
released through NASA's new Legacy Archive for Microwave Background Data 
Analysis (LAMBDA) at \verb"http://lambda.gsfc.nasa.gov/".

\acknowledgements

The \map\ mission is made possible by the support of the Office of Space 
Sciences at NASA Headquarters and by the hard and capable work of scores of 
scientists, engineers, technicians, machinists, data analysts, budget analysts, 
managers, administrative staff, and reviewers.  We acknowledge use of the
HEALPix package.

\appendix

\section{THE MAPPING FUNCTION}
\label{app:map_funct}

Equation~(\ref{eq:map_funct_def}) defines the continuous form of the mapping
function, which encodes both the scan strategy of an experiment, and
convolution due to the beam response.  We can relate this to the matrix form,
in equation~(\ref{eq:map_funct_mat}) as follows.  The mapping function evaluated
at time $t_i$ for a finite integration time $\tau$ may be written in terms of
the beam response function as
\begin{equation}
{\bf M}({\bf n},t_i) = \frac{1}{\tau} \int_{t_i}^{t_i+\tau} dt \,\,
\left[ \alpha\,B_A({\bf R}(t)\cdot{\bf n}) - \beta\,B_B({\bf R}(t)\cdot{\bf n})\right],
\label{eq:map_funct_1}
\end{equation}
where $B_A({\bf n})$ is the beam response of the A-side beam, in spacecraft
coordinates, normalized to
unit integral
\begin{equation}
\int d\Omega_{{\bf n}} \, B_A({\bf n}) \equiv 1,
\end{equation} 
similarly for the B side, and ${\bf R}(t)$ is the rotation matrix from sky-fixed
(Galactic) coordinates to spacecraft coordinates at time $t$.  The terms
$\alpha$ and $\beta$ in equation~(\ref{eq:map_funct_1}) account for possible ohmic
losses in the A and B-side optics that are not necessarily equal 
\citep{jarosik/etal:2003b}.  Since the data are calibrated using the modulation
of the CMB dipole, we only need to parameterize the loss {\em imbalance} which,
following \citet{jarosik/etal:2003b}, we parameterize as
\begin{eqnarray}
\alpha & \equiv & 1 + x_{\rm im}, \nonumber \\
\beta  & \equiv & 1 - x_{\rm im}.
\end{eqnarray}
Note that loss imbalance is separate from lossless differences in the beam
response function, e.g. differences in the solid angle of the A and B-side
beams. Once the calibration is applied, the differential sky signal is a
measurement of the form
\begin{equation}
{\bf \Delta t}(t) = \int d\Omega_{{\bf n}} \, {\bf t}({\bf n}) \,
\left[ (1 + x_{\rm im})\,B_A({\bf R}(t)\cdot{\bf n}) - (1 - x_{\rm im})\,B_B({\bf R}(t)\cdot{\bf n})\right],
\end{equation}
which still includes the effects of any loss imbalance.  We now separately
consider how this calibrated differential data propagates into the sky maps 
and the Jupiter beam maps. 

When making sky maps from the calibrated data, each datum is modeled simply as
\begin{equation}
{\bf \Delta t}(t_i) = (1 + x_{\rm im}) \, {\bf t}(p_A) - (1 - x_{\rm im}) \, {\bf t}(p_B),
\end{equation}
where $p_A$ is the pixel observed by the A-side beam at time $t_i$, and 
similarly for $p_B$. That is, each row of the mapping matrix in 
equation~(\ref{eq:map_funct_mat}) has the form
\begin{equation}
{\bf M}(p,t_i) = [\ldots,(1 + x_{\rm im}), \ldots , - (1 - x_{\rm im}), \ldots ],
\end{equation}
with non-zero entries in pixel columns $p_A$ and $p_B$ only. Upon solving
for the sky map, this ideally leads to an effective beam response of the form
\begin{equation}
B({\bf n}) = \frac{1}{2} \left[
    \frac{B_A^{(s)}({\bf n})}{(1 + x_{\rm im})}
  + \frac{B_B^{(s)}({\bf n})}{(1 - x_{\rm im})} \right],
\label{map_def}
\end{equation}
where $B_A^{(s)}$ is the symmetrized beam response for the A-side beam and
similarly for the B-side. For this ideal case to obtain, the following must
hold: 1) each pixel must be observed equally by the A and B-side beams, which
is true to a very good approximation for \map, and 2) each pixel must be
observed with uniform azimuthal coverage. Deviations from these assumptions 
are considered in the text.

The beam mapping data is compiled from calibrated observations of the bright
source Jupiter.  The calibrated data taken when side A is observing Jupiter 
has the form
\begin{equation}
{\bf \Delta t} = \int d\Omega_{{\bf n}} \, {\bf t}_J({\bf n}) \,
 (1 + x_{\rm im})\,B_A({\bf R}\cdot{\bf n}) + {\bf \Delta t}_{\rm sky},
\end{equation}
where ${\bf t}_J({\bf n})$ is the brightness temperature of Jupiter in the
direction  ${\bf n}$ and ${\bf \Delta t}_{\rm sky}$ is the background sky
temperature difference, which is subtracted during processing.  An analogous
equation holds for the B-side data.  Assuming the beam response is constant
over the extent of  Jupiter, the integral reduces to
\begin{equation}
{\bf \Delta t} =  T_J \Omega_J \, (1 + x_{\rm im})\,B_A({\bf R}\cdot{\bf n}_J),
 + {\bf \Delta t}_{\rm sky},
\end{equation}
where $T_J$ is the disk brightness temperature of Jupiter, and $\Omega_J$ is 
its solid angle.  Beam maps are compiled by binning the corrected data 
${\bf \Delta t} - {\bf \Delta t}_{\rm sky}$ as a function of ${\bf n}$, 
in spacecraft coordinates.  This produces maps proportional to the beam 
response
\begin{equation}
T_A({\bf n}) = T_J \Omega_J (1 + x_{\rm im})\,B_A({\bf n}),
\end{equation}
and similarly for the B-side.  

Ultimately, we wish to compute the transfer function of the symmetrized beam 
response. This may be obtained from the symmetrized beam maps as
\begin{equation}
B({\bf n}) = \frac{1}{2 T_J \Omega_J} \left[
    \frac{T_A^{(s)}({\bf n})}{(1 + x_{\rm im})}
  + \frac{T_B^{(s)}({\bf n})}{(1 - x_{\rm im})} \right].
\end{equation}
We don't know the brightness temperature of Jupiter {\em a priori}, but since
this is an overall normalization factor, we are free to normalize the final
transfer function to 1 at $l = 0$.

\section{MAP-MAKING WITH FILTERED DATA}
\label{app:map_filter}

\S\ref{sec:filter} presents the filtering algorithm used to determine the final
instrument baseline.  This process included an estimated sky signal 
subtraction based on the initial sky maps produced with the hourly calibration.
In this Appendix we derive the noise properties of sky maps produced with this
filtered data. In the following, we assume the time-ordered data has a noise
covariance ${\bf N} = \left<{\bf n n}^T\right>$ that includes a  $1/f$
component, and that we have a pre-whitening filter ${\bf F}$ such that 
\begin{equation}
\left<({\bf F\,n})({\bf F\,n})^T\right> \propto {\bf I}.
\end{equation}

\subsection{Map-making with filtered signal + noise}

We could filter the full data prior to any calibration or sky map estimation, 
then deconvolve the effects of the filter in the subsequent data processing.
The input data would have the form
\begin{equation}
{\bf d'} = {\bf F\,d} = {\bf F\,M t} + {\bf F\,n}.
\end{equation}
Then, in order to obtain an unbiased sky map estimate, we would need to 
evaluate the sky map estimator
\begin{equation}
{\bf t'} = ({\bf M}^T {\bf F}^T {\bf F}{\bf M})^{-1}
\cdot ({\bf M}^T {\bf F}^{T} {\bf d'}),
\end{equation}
which deconvolves the action of the filter on the sky signal.  Since the $1/f$ 
noise in the \map\ data is relatively small, implementing this estimator for the 
first-year sky maps was deemed unnecessary and would likely have delayed the
release of the maps.  The alternative is to filter only the noise by 
subtracting an estimate of the sky signal prior to filtering, then adding it
back in to the time-ordered data prior to making new maps.

\subsection{Map-making with filtered noise}

Let ${\bf t}_0$ be the sky map estimated from unfiltered data, using the hourly
calibration.  This is related to the true sky signal by
\begin{eqnarray}
{\bf t}_0 & = & {\bf W d} \\
          & = & {\bf W} \cdot ({\bf M t} + {\bf n}) \\
          & = & {\bf t} + {\bf W n}
\end{eqnarray}
where ${\bf W} = ({\bf M}^T {\bf M})^{-1} \cdot {\bf M}^T$ is the
map-making operator defined in \S\ref{sec:map_making_pre_cal}, and we have 
used the fact that ${\bf W} \cdot {\bf M} = 1$. We use this sky map to 
subtract a sky signal from the time-ordered data prior to filtering, then we 
add it back in after filtering. This produces a filtered data set
\begin{eqnarray}
{\bf d}_1 & = & {\bf F} \cdot ({\bf d} \, - \, {\bf M t}_0) 
                + {\bf M t}_0 \\
          & = & {\bf F} \cdot ({\bf M t} + {\bf n} 
                - {\bf M t} - {\bf M W n}) 
                + {\bf M t} + {\bf M W n} \\
          & = & {\bf M t} + {\bf F n} + ({\bf 1-F}) \cdot {\bf M W n}.
\end{eqnarray}
This time series data set consists of an unbiased sky signal ${\bf M t}$, a
white noise term  ${\bf F n}$, and a residual noise term $({\bf 1-F}) 
\cdot {\bf M W n}$ which is due to the off-diagonal ``wings'' of the filter 
$({\bf 1-F})$ acting on the noise from the initial sky map estimate,
 ${\bf M W n}$.

We can make a map ${\bf t}_1$ from the data ${\bf d}_1$ using the algorithm
of \S\ref{sec:map_making_pre_cal}
\begin{eqnarray}
{\bf t}_1 & = & {\bf W d}_1 \\
          & = & {\bf t} + {\bf W F n} 
              + {\bf W} \cdot ({\bf 1-F}) \cdot {\bf M W n} \\
          & = & {\bf t} + {\bf W F n} 
              + ({\bf 1- W F M}) \cdot {\bf W n} \\
          & = & {\bf t} + {\bf W F n} + {\bf R} \cdot {\bf W n},
\label{eq:filter_pix_noise}
\end{eqnarray}
where we have again used ${\bf W} \cdot {\bf M} = 1$ and we have defined a
``residual'' operator ${\bf R} \equiv ({\bf 1- W F M})$ which is small in the
sense that only off-diagonal terms in ${\bf F}$ contribute to it. This is most
easily seen if we recall that the filter operator is 1 on the diagonal and has
small off-diagonal terms.  We can then write ${\bf F} \equiv {\bf 1 - E}$
from which it follows  ${\bf R} = {\bf W E M}$.  It follows that ${\bf t}_1$ 
is an unbiased estimate of ${\bf t}$ that includes a white noise term 
${\bf W F n}$ (this noise still contains the small beam separation covariance),
and a residual noise term, ${\bf R} \cdot {\bf W n}$, due to the noisy sky 
signal estimator used in the filtering process. This latter term is the 
``excess'' noise seen in the W band single DA maps after filtering 
\citep{hinshaw/etal:2003b}.

The residual noise term can be reduced somewhat by iterating the filter 
algorithm a second time, using ${\bf t}_1$ as a sky signal estimator.
It is straightforward to show that, after some algebra, the resulting map is
\begin{equation}
{\bf t}_2 = {\bf t} + {\bf W F n} + {\bf R} \cdot {\bf W F n} 
          + {\cal O}({\bf R}^2),
\end{equation}
where we have neglected a term of order ${\bf R}^2 \cdot {\bf W n}$. The
residual noise is reduced slightly since ${\bf R}$ is now acting on the white
noise ${\bf W F n}$ instead of the full noise ${\bf W n}$.  But since the white
noise dominates, this is a relatively insignificant improvement.  It is also
clear that subsequent iterations of the filtering only contribute higher order
corrections which are negligible.  This convergence has been verified  with the
flight data. Another feature we have verified with the flight data is the fact
that the excess noise term decreases with time.  The reason for this is simply
that  ${\bf R} = ({\bf 1 - W F M}) = {\bf W E M}$ gets smaller with additional
data  because the map-making operator ${\bf W}$ gets smaller as more
observations  accumulate in its ``denominator'', $({\bf M}^T {\bf M})^{-1}$.
The first-year maps were only filtered once, as per
equation~(\ref{eq:filter_pix_noise}), because the improvement in noise
properties was not deemed sufficient to warrant the additional processing time.

\section{MAP-MAKING WITH POLARIZATION}
\label{app:map_pol}

\map\ observes the sky with two orthogonal linear polarization modes per feed,  
thus it is sensitive to the 3 Stokes parameters $I$, $Q$, and  $U$.  This
Appendix outlines the algorithm with which these  parameters can be mapped
using the differential data from \map.  The approach is an extension of the
iterative method in \S\ref{sec:iterate_map} introduced by 
\citet{wright/hinshaw/bennett:1996a}.

\subsection{Polarization Mapping with Total Power Data}

Suppose we observed the sky with a single beam, total power radiometer that is
sensitive to a single linear polarization, denoted mode \#1.  In terms of the
Stokes  parameters, the temperature observed by the instrument at time $t$ in 
pixel $p$ would be
\begin{equation}
{\bf d}_1(t) = {\bf i}(p) + {\bf q}(p) \cos 2\gamma + {\bf u}(p) \sin 2\gamma,
\label{eq:t1_pol}
\end{equation}
where ${\bf i}$, ${\bf q}$, and ${\bf u}$ are Stokes parameter maps in units 
of temperature, and $\gamma$ is the angle between the polarization  axis of the
beam and the chosen reference direction for pixel $p$ (the choice of reference
direction is discussed below).  Note that we adopt the common convention that the
instrument response  reduces to the total temperature in the limit of an
unpolarized  source.

The signal in the orthogonal polarization channel, which is fed by the other 
port of the ortho-mode transducer (OMT) and denoted mode \#2, is given by

\begin{eqnarray}
{\bf d}_2(t) & = & {\bf i}(p) + {\bf q}(p) \cos 2(\gamma + \frac{\pi}{2})
                 + {\bf u}(p) \sin 2(\gamma + \frac{\pi}{2}) \\
             & = & {\bf i}(p) - {\bf q}(p) \cos 2\gamma - {\bf u}(p) \sin 2\gamma.
\label{eq:t2_pol}
\end{eqnarray}
We can isolate the intensity and polarization signals by taking sums and 
differences 
\begin{eqnarray}
{\bf d}(t) & \equiv & \frac{1}{2}\left({\bf d}_1 + {\bf d}_2 \right) = {\bf i}(p) \\
{\bf p}(t) & \equiv & \frac{1}{2}\left({\bf d}_1 - {\bf d}_2 \right) = 
 {\bf q}(p) \cos 2\gamma + {\bf u}(p) \sin 2\gamma.
\end{eqnarray}
Given noisy data, we can estimate the intensity ${\bf i}(p)$ by averaging all 
the data ${\bf d}(t_i)$.  For the polarization we can only  estimate a linear 
combination of ${\bf q}$ and ${\bf u}$ from a single observation.   However, 
if we have several observations of pixel $p$ with a variety of  polarization 
angles $\gamma$, we can estimate ${\bf q}$ and ${\bf u}$ in a given pixel by  
minimizing $\chi^2$, defined as
\begin{equation}
\chi^2 \equiv \sum_{i \in p} \frac{\left[ {\bf p}(t_i)
 - {\bf q}(p) \cos 2\gamma_i - {\bf u}(p) \sin 2\gamma_i \right]^2}{\sigma_i^2},
\label{eq:pol_chi2}
\end{equation}
where $i$ is a time-ordered data index and the sum is over observations within
pixel $p$, and $\gamma_i$ is the polarization angle for the $i\uth$ observation. 
The best-fit values for ${\bf q}$ and ${\bf u}$ are given by
\begin{equation}
\left( \begin{array}{c}
{\bf q}(p) \\ {\bf u}(p) \end{array} \right) =
\frac{1}{\Delta}
\left( \begin{array}{cc}
\sum_i s_i^2/\sigma_i^2 & -\sum_i c_i s_i/\sigma_i^2 \\
-\sum_i c_i s_i/\sigma_i^2 & \sum_i c_i^2/\sigma_i^2 \end{array} \right)
\left( \begin{array}{c}
\frac{1}{2}\sum_i c_i \, {\bf p}(t_i)/\sigma_i^2 \\ 
\frac{1}{2}\sum_i s_i \, {\bf p}(t_i)/\sigma_i^2 \end{array} \right).
\end{equation}
where $c_i \equiv \cos 2\gamma_i$, $s_i \equiv \sin 2\gamma_i$, and $\Delta 
\equiv \sum_i c_i^2/\sigma_i^2 \sum_i s_i^2/\sigma_i^2 -  \left(\sum_i c_i
s_i/\sigma_i^2 \right)^2$ is the determinant of the normal equations matrix. 
The standard errors for ${\bf q}$ and ${\bf u}$ are given by the inverse of the
normal equations matrix
\begin{eqnarray}
\sigma^2_{\bf q}  & = & \frac{1}{\Delta}\,\sum_i s_i^2/\sigma_i^2, \\
\sigma^2_{\bf u}  & = & \frac{1}{\Delta}\,\sum_i c_i^2/\sigma_i^2, \\
\sigma^2_{{\bf qu}} & = & -\frac{1}{\Delta}\,\sum_i c_i s_i/\sigma_i^2.
\end{eqnarray}
In the limit of uniform azimuthal coverage and constant noise per observation
($\sigma_i = \sigma_0$), the matrix elements  in the linear system reduce to
\begin{equation}
\left( \begin{array}{cc}
\sum_i c_i^2 & \sum_i c_i s_i \\
\sum_i c_i s_i & \sum_i s_i^2 \end{array} \right)
\longrightarrow
\left( \begin{array}{cc}
N/2 & 0 \\
0 & N/2 \end{array} \right),
\end{equation}
where $N$ is the number of observations of pixel $p$.  In this limit, the 
noise in ${\bf q}$ and ${\bf u}$ is equal and uncorrelated and reduces to
\begin{equation}
\sigma_{\bf q} = \sigma_{\bf u} \longrightarrow \sqrt{\frac{2}{N}} \sigma_0.
\end{equation}
Thus the noise in each polarization component is $\sqrt{2}$ times noisier 
than in the intensity ${\bf i}$.

\subsection{Polarization Mapping with Differential Data}

We now generalize to the case of polarization mapping with differential input 
data. For clarity, we first consider the case where the loss in the two sides A
and B are equal.  We generalize to the case with unbalanced loss in the next 
subsection. In the case of \map, one radiometer in a differencing assembly
(radiometer \#1) is fed from the axial port of the OMT and the other (\#2) is 
fed by the lateral port \citep{jarosik/etal:2003}.  Following 
equations~(\ref{eq:t1_pol},\ref{eq:t2_pol}) the differential signals from
radiometers  1 and 2 are
\begin{eqnarray}
{\bf d}_1(t) = \frac{1}{2}({\bf d}_{13} + {\bf d}_{14})
        & = & {\bf i}(p_A) + {\bf q}(p_A) \cos 2\gamma_A + {\bf u}(p_A) \sin 2\gamma_A \\
        & - & {\bf i}(p_B) - {\bf q}(p_B) \cos 2\gamma_B - {\bf u}(p_B) \sin 2\gamma_B,
\end{eqnarray}
and
\begin{eqnarray}
{\bf d}_2(t) = \frac{1}{2}({\bf d}_{23} + {\bf d}_{24})
        & = & {\bf i}(p_A) - {\bf q}(p_A) \cos 2\gamma_A - {\bf u}(p_A) \sin 2\gamma_A \\
        & - & {\bf i}(p_B) + {\bf q}(p_B) \cos 2\gamma_B + {\bf u}(p_B) \sin 2\gamma_B,
\end{eqnarray}
where $\gamma_A$ is the angle between the axial polarization plane and the 
reference direction in the pixel seen by the A beam, and similarly for 
$\gamma_B$. We take sums and differences of the two signals to isolate the 
unpolarized and polarized portions of the signal
\begin{equation}
{\bf d}(t) \equiv \frac{1}{2}({\bf d}_1 + {\bf d}_2) = {\bf i}(p_A) - {\bf i}(p_B)
\end{equation}
\begin{equation}
{\bf p}(t) \equiv \frac{1}{2}({\bf d}_1 - {\bf d}_2)
  = {\bf q}(p_A) \cos 2\gamma_A + {\bf u}(p_A) \sin 2\gamma_A 
  - {\bf q}(p_B) \cos 2\gamma_B - {\bf u}(p_B) \sin 2\gamma_B.
\end{equation}

An iterative scheme for making maps of ${\bf q}$ and  ${\bf u}$ follows the
form used for intensity maps (\S\ref{sec:iterate_map}).  Let ${\bf q}_n$ and
${\bf u}_n$ be the $n\uth$ estimates of ${\bf q}$ and  ${\bf u}$ respectively. 
Estimates for ${\bf q}_{n+1}$ and ${\bf u}_{n+1}$ are obtained by combining the
per-pixel fitting algorithm in equation~(\ref{eq:pol_chi2})  with the iterative
map-making algorithm, as follows
\begin{equation}
\chi^2 \equiv \sum_{i \in p} \frac{\left[ {\bf p'}(t_i) 
 - {\bf q}_{n+1}(p) \cos 2\gamma_i - {\bf u}_{n+1}(p) \sin 2\gamma_i \right]^2}
 {\sigma_i^2},
\end{equation}
where the sum is over all observations of pixel $p$ by either the A- or B-side 
beam, and ${\bf p'}(t)$ is the polarization data corrected with an estimate
of the signal in the opposite beam
\begin{equation}
{\bf p'}(t_i) \equiv \left\{ \begin{array}{ll}
{\bf p}(t_i) + {\bf q}_n(p_B) \cos 2\gamma_B + {\bf u}_n(p_B) \sin 2\gamma_B 
& \mbox{beam } A \in p \\
-{\bf p}(t_i) + {\bf q}_n(p_A) \cos 2\gamma_A + {\bf u}_n(p_A) \sin 2\gamma_A 
& \mbox{beam } B \in p.
\end{array} \right.
\end{equation}
The best-fit solution for ${\bf q}_{n+1}$ and ${\bf u}_{n+1}$ is then
\begin{equation}
\left( \begin{array}{c}
{\bf q}_{n+1}(p) \\ {\bf u}_{n+1}(p) \end{array} \right) =
\frac{1}{\Delta}
\left( \begin{array}{cc}
\sum_i s_i^2/\sigma_i^2 & -\sum_i c_i s_i/\sigma_i^2 \\
-\sum_i c_i s_i/\sigma_i^2 & \sum_i c_i^2/\sigma_i^2 \end{array} \right)
\left( \begin{array}{c}
\sum_i c_i \,{\bf p'}(t_i)/\sigma_i^2 \\ \sum_i s_i \,{\bf p'}(t_i)/\sigma_i^2 \end{array} 
\right),
\end{equation}
where the sum on $i$ is as defined above.  The uncertainties are as given for
the total power case, where $\sigma_i$ is now the uncertainty per differential
observation, ${\bf p}(t_i)$.

\subsection{Map Making with Unbalanced Differential Data}

We now generalize to the case of map-making with unbalanced differential input
data. In this case, losses in the A and B-side front ends are different and the
differential signal is of the form
\begin{eqnarray}
{\bf d}_1(t) & = & (1+x_{\rm im,1}) \left[{\bf i}(p_A) + {\bf s}(p_A,\gamma_A) \right] 
                 - (1-x_{\rm im,1}) \left[{\bf i}(p_B) + {\bf s}(p_B,\gamma_B) \right] \\
{\bf d}_2(t) & = & (1+x_{\rm im,2}) \left[{\bf i}(p_A) - {\bf s}(p_A,\gamma_A) \right] 
                 - (1-x_{\rm im,2}) \left[{\bf i}(p_B) - {\bf s}(p_B,\gamma_B) \right],
\end{eqnarray}
where $x_{\rm im,1}, x_{\rm im,2}$ are the loss imbalance parameters in
radiometers 1 and 2, respectively, as defined in \citet{jarosik/etal:2003b}, 
and ${\bf s}(p,\gamma)$ is short-hand for the linear combination of Stokes 
parameters
\begin{equation}
{\bf s}(p,\gamma) \equiv {\bf q}(p) \cos 2\gamma + {\bf u}(p) \sin 2\gamma.
\end{equation}
As before, we take sums and differences of the two signals to isolate the 
unpolarized and polarized portions of the signal. First, define the mean 
imbalance and the ``imbalance in the imbalance'' as
\begin{eqnarray}
\bar{x}_{\rm im}  & \equiv & \frac{1}{2}(x_{\rm im,1}+x_{\rm im,2}) \\
\delta x_{\rm im} & \equiv & \frac{1}{2}(x_{\rm im,1}-x_{\rm im,2}),
\end{eqnarray}
then
\begin{eqnarray}
{\bf d}(t) = \frac{1}{2}({\bf d}_1+{\bf d}_2) & = & 
{\bf i}(p_A)-{\bf i}(p_B)
 + \bar{x}_{\rm im}\left[{\bf i}(p_A)+{\bf i}(p_B)\right]
 + \delta x_{\rm im}\left[{\bf s}(p_A,\gamma_A) + {\bf s}(p_B,\gamma_B)\right] 
\nonumber \\
 & = & (1+\bar{x}_{\rm im})\,{\bf i}(p_A) - (1-\bar{x}_{\rm im})\,{\bf i}(p_B) 
 \,+\, {\cal O}(\delta x_{\rm im}) \label{eq:d_pol_loss} \\
{\bf p}(t) = \frac{1}{2}({\bf d}_1-{\bf d}_2) & = & 
{\bf s}(p_A,\gamma_A)-{\bf s}(p_B,\gamma_B)
 + \bar{x}_{\rm im} \left[{\bf s}(p_A,\gamma_B)+{\bf s}(p_B,\gamma_B)\right]
 + \delta x_{\rm im}\left[{\bf i}(p_A)+{\bf i}(p_B)\right]
 \nonumber \\
 & = & (1+\bar{x}_{\rm im})\,{\bf s}(p_A,\gamma_A) 
     - (1-\bar{x}_{\rm im})\,{\bf s}(p_B,\gamma_B) 
 \,+\, {\cal O}(\delta x_{\rm im}). 
\label{eq:p_pol_loss}
\end{eqnarray}
Note that the term of ${\cal O}(\delta x_{\rm im})$ in
equation~(\ref{eq:d_pol_loss}) is negligible, because ${\bf i} \gg {\bf s}$,
but the term in  equation~(\ref{eq:p_pol_loss}) must be considered more
carefully.  First note that  $\delta x_{\rm im}$ is small -- from Table~3 in
\citet{jarosik/etal:2003b}, the largest value is 0.35\% in W2, with most values
being $\sim$0.1\%.  Second, the multiplier, $({\bf i}(p_A)+{\bf i}(p_B))$ does
not modulate with polarization angle, $\gamma$, thus it is effectively an
offset term that is further  suppressed by the map-making demodulation. 
Finally, the term is out of phase with the differential signal $({\bf
i}(p_A)-{\bf i}(p_B))$ so it does not  effectively couple to the sky.  The
effect of this term in the first-year data  is further considered by
\citet{kogut/etal:2003}.

We generalize the differential map-making algorithm to account for loss 
imbalance as follows.  For intensity, let ${\bf i}_n$ be the $n\uth$ estimate 
of ${\bf i}$, then
\begin{equation}
{\bf d'}(t_i) \equiv \left\{ \begin{array}{ll}
\left[+{\bf d}(t_i) + (1-\bar{x}_{\rm im})\,{\bf i}_n(p_B)\right]/(1+\bar{x}_{\rm im})
& \mbox{beam A } \in p \\
\left[-{\bf d}(t_i) + (1+\bar{x}_{\rm im})\,{\bf i}_n(p_A)\right]/(1-\bar{x}_{\rm im})
& \mbox{beam B } \in p. 
\end{array} \right.
\end{equation}
The updated intensity map is then estimated by binning the corrected data
\begin{equation}
{\bf i}_{n+1}(p) = \frac{\sum_i w_i \, {\bf d'}(t_i)}{\sum_i w_i},
\end{equation}
where $w_i$ is the normalized statistical weight of each observation
\begin{equation}
w_i = \left\{ \begin{array}{ll}
(1+\bar{x}_{\rm im})^2\sigma_0^2/\sigma_i^2 & \mbox{beam A }\in p \\
(1-\bar{x}_{\rm im})^2\sigma_0^2/\sigma_i^2 & \mbox{beam B }\in p. 
\end{array} \right.
\label{eq:pol_wt}
\end{equation}

For polarization, let ${\bf q}_n$ and ${\bf u}_n$ be the $n\uth$ estimates of ${\bf q}$
and  ${\bf u}$ respectively, and define 
${\bf s}_n(p,\gamma) \equiv {\bf q}_n(p)\cos 2\gamma + {\bf u}_n(p)\sin 2\gamma$.
Estimates for ${\bf q}_{n+1}$ and ${\bf u}_{n+1}$ are obtained by combining the 
per-pixel ${\bf q}$ and ${\bf u}$ demodulation with the iterative map-making
algorithm.  We define $\chi^2$ as follows
\begin{equation}
\chi^2 \equiv \sum_{i \in p} \frac{\left[ {\bf p'}(t_i) 
 - {\bf q}_{n+1}(p)\cos 2\gamma_i - {\bf u}_{n+1}(p)\sin 2\gamma_i \right]^2}
 {\sigma_i^2},
\end{equation}
where the sum is over all observations of pixel $p$ by either the A- or B-side
beam, and
\begin{equation}
{\bf p'}(t_i) \equiv \left\{ \begin{array}{ll}
\left[+{\bf p}(t_i) + (1-\bar{x}_{\rm im}){\bf s}_n(p_B,\gamma_B) \right]/(1+\bar{x}_{\rm im})
& \mbox{beam A } \in p \\
\left[-{\bf p}(t_i) + (1+\bar{x}_{\rm im}){\bf s}_n(p_A,\gamma_A) \right]/(1-\bar{x}_{\rm im})
& \mbox{beam B } \in p. 
\end{array} \right.
\end{equation}
The best-fit solution for ${\bf q}_{n+1}$ and ${\bf u}_{n+1}$ is then
\begin{equation}
\left( \begin{array}{c}
{\bf q}_{n+1}(p) \\ {\bf u}_{n+1}(p) \end{array} \right) =
\frac{1}{\Delta}
\left( \begin{array}{cc}
\sum_i w_i s_i^2 & -\sum_i w_i c_i s_i \\
-\sum_i w_i c_i s_i & \sum_i w_i c_i^2 \end{array} \right)
\left( \begin{array}{c}
\sum_i w_i c_i \,{\bf p'}(t_i) \\ \sum_i w_i s_i \,{\bf p'}(t_i) \end{array} 
\right),
\end{equation}
where the weight $w_i$ is the same as equation~(\ref{eq:pol_wt}).

\subsection{The Choice of Reference Direction}

We adopt the sign convention of \citet{zaldarriaga/seljak:1997} in which the
polarization components are defined in a right-handed coordinate system with 
the z-axis pointed outward towards the sky.  The Stokes parameters are defined
with respect to a fiducial direction in each pixel on the sky. We adopt 
the convention in which the reference direction is aligned with the local 
Galactic meridian, i.e., the great circle connecting a given point to the 
North Galactic Pole.  The unit vector, $\hat{n}$, tangent to this great circle, pointing in the direction 
of the North Pole, is given by
$$
\hat{n} = \hat{l} \times \hat{e} 
        = \hat{l} \times \frac{(\hat{z} \times \hat{l})}{\sin\theta}
$$
where $\hat{l}$ is the unit vector along the line of sight of the current
observation, $\hat{e}$ is a unit vector pointing east from $\hat{l}$, $\hat{z}$ 
is the unit vector to the North Galactic Pole, and $\theta$ is the polar angle 
(co-latitude) of $\hat{l}$.

For reference, we give formulae for computing the factors $\cos 2\gamma$ and
$\sin 2\gamma$ here.  Let $\hat{l}$ be the unit vector along the line of sight,
$\hat{w}$ be the unit vector pointing west from $\hat{l}$, $\hat{n}$ be the unit
vector pointing north from $\hat{l}$ (the polarization reference direction) and 
$\hat{p}$ be the unit vector along the polarization plane defined by the axial 
port of the OMT.  Then, for both the $A$ and $B$ sides, we have
\begin{eqnarray*}
\cos\gamma & = & \hat{p} \cdot \hat{n} \\
\sin\gamma & = & \hat{p} \cdot \hat{w} \\
\cos 2\gamma & = & 2 \cos^2\gamma - 1 \\
\sin 2\gamma & = & 2 \sin\gamma \cos\gamma.
\end{eqnarray*}

Note that this defines a right-handed coordinate system with $(x,y,z)$ axes
$(\hat{n},\hat{w},\hat{l})$ whose $z$ axis is oriented outward following the
sign conventions of \citet{zaldarriaga/seljak:1997}.

\clearpage

\clearpage

\clearpage

\clearpage
\begin{deluxetable}{ll}
\tablecaption{\map\ data processing notation}
\tablewidth{0pt}
\tablehead{
\colhead{Symbol(s)} & \colhead{Description}}
\startdata
$t$, $t_i$             &  Time, time of $i$th observation, in s \\
$\tau$                 &  Integration time per observation, in s \\
$N_p$                  &  Number of pixels in a map, $0-(N_p-1)$ \\
$N_t$                  &  Number of time-ordered data points \\
$p$                    &  HEALPix pixel number \\
$p_A$, $p_B$           &  A \& B-side pixels at time $t$ \\
$\theta_{\rm beam}$    &  Separation of A \& B-side beams, in degrees \\
$\gamma$               &  Polarization angle \\
$\gamma_A$, $\gamma_B$ &  A \& B-side polarization angles, at time $t$ \\
$c_i$, $s_i$, $c_is_i$ &  $\cos 2\gamma$, $\sin 2\gamma$,
                          $\cos 2\gamma \cdot \sin 2\gamma$, at time $t_i$ \\
${\bf t}(p)$           &  Sky map, in mK \\
${\bf i}(p)$           &  Sky map, Stokes parameter $I$, in mK \\
${\bf q}(p)$, ${\bf u}(p)$ &  Sky map, Stokes parameters $Q$, $U$, in mK \\
${\bf \tilde t}(p)$, ${\bf t}_n(p)$ 
                       &  Estimated sky maps, in mK \\
${\bf n}_{\rm obs}(p)$ &  Number of observation of pixel $p$ \\
${\bf t_c}(p)$         &  CMB anisotropy map, in mK \\
${\bf t_g}(p)$         &  Galactic foreground map, in mK \\
${\bf t_s}(p,t)$       &  Time-dependent source map (Sun, Earth, Moon), in mK \\
${\bf \Delta t}(t)$    &  Time-ordered differential sky signal, in mK \\
${\bf \Delta t_d}(t)$  &  Time-ordered \cobe\ dipole signal, in mK \\
${\bf \Delta t_v}(t)$  &  Time-ordered local velocity dipole signal, in mK \\
${\bf \Delta t_a}(t)$  &  Time-ordered anisotropy signal, 
                          ${\bf \Delta t}-{\bf \Delta t_d}$, in mK \\
${\bf c}(t)$           &  Time-ordered raw data, single channel, in du \\
${\bf d}_{ij}(t)$      &  Time-ordered data, radiometer $i$, channel $j$, in mK \\
${\bf d}(t)$           &  Time-ordered intensity data, co-added channels, in mK \\
${\bf p}(t)$           &  Time-ordered polarization data, co-added channels, in mK \\
${\bf g}(t)$           &  Rad. responsivity ($\propto$ gain), single channel, in du mK\per \\
${\bf n}(t)$           &  Rad. noise, single channel in calibration, \\
                       &  4 co-added channels in map-making, in mK \\
${\bf b}(t)$           &  Rad. baseline, single channel, in du \\
$g_k$, $b_k$           &  Hourly gain, baseline, single channel, k\uth precession \\
$\sigma_i$             &  rms noise, $i$\uth observation, single channel in calibration, \\
                       &  4 co-added channels in map-making, in mK \\
$\sigma_0$             &  Mean rms noise, single or co-added channels, in mK \\
$x_{\rm im}$, $x_{{\rm im},j}$
                       &  Loss imbalance parameter (radiometer $j$) \\
$C(\Delta t)$          &  Auto-correlation function of noise, in mK$^2$ \\
$C_1$, $A$, $B$        &  Auto-correlation function model parameters \\
$w(f)$                 &  Pre-whitening filter, Fourier space \\
${\bf F}$              &  Pre-whitening filter, time domain, $N_t \times N_t$ matrix \\
${\bf N}$              &  Time-ordered noise covariance, $N_t \times N_t$ matrix, in mK$^2$ \\
${\bf M}$              &  Mapping function, $N_t \times N_p$ matrix \\
${\bf W}$              &  Map-making operator, $({\bf M}^T {\bf M})^{-1} \cdot {\bf M}^T$,
                          $N_p \times N_t$ matrix \\
${\bf \Sigma}$         &  Pixel-pixel noise covariance, $N_p \times N_p$ matrix, in mK$^2$ \\
${\bf D}$              &  Reduced inverse noise, 
                         $({\bf M}^T {\bf M}) = \sigma_0^2\,{\bf \Sigma}^{-1}$, 
                          $N_p \times N_p$ matrix \\
${\bf \Delta}_{ij}$    &  Difference map from channel combination $ij$
\enddata
\label{tab:notation}
\end{deluxetable}

\clearpage
\begin{deluxetable}{lcc}
\tablecaption{\map\ Attitude Control System Requirements}
\tablewidth{0pt}
\tablehead{\colhead{Parameter} & \colhead{Requirement} & \colhead{Performance}}
\startdata
Precession rate ($d\phi/dt$) & $-0\ddeg1{\rm s}^{-1}\pm6.3\%$   & $-0\ddeg1{\rm s}^{-1}\pm 3.6\%$   \\
Spin rate ($d\psi/dt$)       & $ 2\ddeg784{\rm s}^{-1}\pm 5\%$  & $ 2\ddeg784{\rm s}^{-1}\pm 0.13\%$ \\
Sun--spin angle ($\theta$)   & $22\ddeg5 \pm 0\ddeg25$          & $22\ddeg5 \pm 0\ddeg023$           \\
\enddata
\label{tab:acs_req}
\end{deluxetable}

\begin{deluxetable}{lcccccc}
\tablecaption{\map\ Dipole-Based Gain Summary}
\tablewidth{0pt}
\tablehead{
\colhead{} & \multicolumn{3}{c}{Channel 3} & \multicolumn{3}{c}{Channel 4} \\
\colhead{Radiometer} & \colhead{$\left<g_k\right>$} & \colhead{$\Delta g_k$\tablenotemark{a}} & \colhead{$\sigma_{g_k}$\tablenotemark{b}}
                     & \colhead{$\left<g_k\right>$} & \colhead{$\Delta g_k$\tablenotemark{a}} & \colhead{$\sigma_{g_k}$\tablenotemark{b}} \\
\colhead{} & \colhead{(du mK$^{-1}$)} & \colhead{(\%)} & \colhead{(\%)}
           & \colhead{(du mK$^{-1}$)} & \colhead{(\%)} & \colhead{(\%)}}
\startdata
 K11 & $-0.974$ & $7.4$ & $0.66$ & $+0.997$ & $6.8$ & $0.66$ \\
 K12 & $+1.177$ & $6.2$ & $0.75$ & $-1.122$ & $6.4$ & $0.75$ \\
Ka11 & $+0.849$ & $4.7$ & $0.75$ & $-0.858$ & $5.0$ & $0.75$ \\
Ka12 & $-1.071$ & $5.1$ & $0.75$ & $+0.985$ & $5.3$ & $0.75$ \\
 Q11 & $+1.015$ & $4.5$ & $0.94$ & $-0.948$ & $4.4$ & $0.94$ \\
 Q12 & $+0.475$ & $5.3$ & $1.03$ & $-0.518$ & $5.5$ & $1.03$ \\
 Q21 & $-0.958$ & $5.8$ & $0.94$ & $+0.986$ & $6.0$ & $0.94$ \\
 Q22 & $-0.783$ & $2.9$ & $1.22$ & $+0.760$ & $2.8$ & $1.22$ \\
 V11 & $+0.449$ & $4.5$ & $1.50$ & $-0.494$ & $4.5$ & $1.50$ \\
 V12 & $-0.532$ & $4.0$ & $1.40$ & $+0.532$ & $4.7$ & $1.40$ \\
 V21 & $-0.450$ & $4.7$ & $1.22$ & $+0.443$ & $5.1$ & $1.22$ \\
 V22 & $+0.373$ & $3.2$ & $1.59$ & $-0.346$ & $3.0$ & $1.59$ \\
 W11 & $+0.311$ & $5.1$ & $2.25$ & $-0.332$ & $4.1$ & $2.43$ \\
 W12 & $+0.262$ & $3.5$ & $2.62$ & $-0.239$ & $6.0$ & $2.71$ \\
 W21 & $-0.288$ & $4.6$ & $3.09$ & $+0.297$ & $3.8$ & $2.53$ \\
 W22 & $+0.293$ & $6.1$ & $2.43$ & $-0.293$ & $6.3$ & $2.62$ \\
 W31 & $-0.260$ & $3.3$ & $2.25$ & $+0.281$ & $3.8$ & $2.34$ \\
 W32 & $-0.263$ & $3.6$ & $2.62$ & $+0.258$ & $3.4$ & $2.43$ \\
 W41 & $+0.226$ & $6.0$ & $4.40$ & $-0.232$ & $5.7$ & $4.21$ \\
 W42 & $+0.302$ & $6.3$ & $3.28$ & $-0.286$ & $5.9$ & $3.37$ \\
\enddata
\tablenotetext{a}{Peak-peak variation in the daily mean gain, indicates
range of gain drift during first year.}
\tablenotetext{b}{Mean statistical uncertainty per hour.}
\label{tab:gain_summary}
\end{deluxetable}

\begin{deluxetable}{lrrrr}
\tablecaption{Auto-correlation Model Parameters\tablenotemark{a}}
\tablewidth{0pt}
\tablehead{
\colhead{Radiometer} & \colhead{$C_1$} & \colhead{$A$} & \colhead{$B$}
 & \colhead{$w(f_{\rm spin})$\tablenotemark{b}}}
\startdata
 K11 & $-0.0038$ & $0.0011$ & $0.00042$ & $0.966$  \\
 K12 & $ 0.0008$ & $0.0011$ & $0.00040$ & $0.963$  \\
Ka11 & $-0.0075$ & $0.0015$ & $0.00048$ & $0.960$  \\
Ka12 & $-0.0031$ & $0.0006$ & $0.00019$ & $0.984$  \\
 Q11 & $ 0.0044$ & $0.0018$ & $0.00053$ & $0.934$  \\
 Q12 & $-0.0088$ & $0.0007$ & $0.00024$ & $0.978$  \\
 Q21 & $ 0.0124$ & $0.0088$ & $0.00282$ & $0.754$  \\
 Q22 & $ 0.0178$ & $0.0128$ & $0.00415$ & $0.686$  \\
 V11 & $ 0.0010$ & $0.0001$ & $0.00005$ & $0.989$  \\
 V12 & $ 0.0034$ & $0.0014$ & $0.00048$ & $0.925$  \\
 V21 & $-0.0038$ & $0.0010$ & $0.00033$ & $0.951$  \\
 V22 & $ 0.0087$ & $0.0093$ & $0.00320$ & $0.689$  \\
 W11 & $ 0.0158$ & $0.0062$ & $0.00211$ & $0.680$  \\
 W12 & $ 0.0048$ & $0.0005$ & $0.00019$ & $0.950$  \\
 W21 & $ 0.0207$ & $0.0071$ & $0.00262$ & $0.644$  \\
 W22 & $ 0.0167$ & $0.0053$ & $0.00187$ & $0.701$  \\
 W31 & $ 0.0062$ & $0.0006$ & $0.00021$ & $0.943$  \\
 W32 & $ 0.0077$ & $0.0002$ & $0.00007$ & $0.975$  \\
 W41 & $ 0.0562$ & $0.0323$ & $0.01152$ & $0.374$  \\
 W42 & $ 0.0393$ & $0.0194$ & $0.00692$ & $0.461$  \\
\enddata
\tablenotetext{a}{See equation~\ref{eq:ac_model} for model definition.  
All parameters are dimensionless.}
\tablenotetext{b}{Derived filter response at the spin frequency, 7.7 mHz.}
\label{tab:ac_fit}
\end{deluxetable}

\begin{deluxetable}{lrccrcc}
\tablecaption{Calibration and Map-Making Error Limits\tablenotemark{a}}
\tablewidth{0pt}
\tablehead{
\colhead{DA} & \colhead{$C_2$} & \colhead{$\left<C_l\right>_{3-10}$}
 & \colhead{$\left<C_l\right>_{11-100}$} & \colhead{$\sigma^{sys}\vert_2$} 
 & \colhead{$\sigma^{sys}\vert_{3-10}$} & \colhead{$\sigma^{sys}\vert_{11-100}$} \\
\colhead{} & \colhead{($\mu$K$^2$)} & \colhead{($\mu$K$^2$)} & \colhead{($\mu$K$^2$)}
 & \colhead{($\mu$K$^2$)} & \colhead{($\mu$K$^2$)} & \colhead{($\mu$K$^2$)}}
\startdata
K1  & -21.4\phm{a} &  0.6 &  0.08 &  42.9\phm{a} &  1.1 &  0.03  \\
Ka1 &  18.5\phm{a} &  1.3 &  0.06 &  37.0\phm{a} &  2.5 &  0.01  \\
Q1  &  59.6\phm{a} &  1.2 &  0.14 & 118.9\phm{a} &  2.2 &  0.01  \\
Q2  &   7.3\phm{a} &  0.9 &  0.13 &  14.4\phm{a} &  1.6 &  0.02  \\
V1  &   3.9\phm{a} &  0.6 &  0.21 &   7.4\phm{a} &  0.7 &  0.01  \\
V2  &  -6.1\phm{a} &  0.8 &  0.19 &  12.6\phm{a} &  1.2 &  0.03  \\
W1  &  -2.6\phm{a} &  1.4 &  0.49 &   6.0\phm{a} &  2.0 &  0.10  \\
W2  &  12.0\phm{a} &  0.7 &  0.62 &  22.9\phm{a} &  0.4 &  0.15  \\
W3  &   4.3\phm{a} &  0.4 &  0.65 &   7.3\phm{a} &  0.4 &  0.07  \\
W4  &  -6.6\phm{a} &  3.3 &  0.90 &  14.5\phm{a} &  5.4 &  0.55  \\
\enddata
\tablenotetext{a}{All values derived from a one-year simulation of \map\ data.
The first 3 data columns give the mean power in the residual map ${\bf t}_{\rm out}
-{\bf t}_{\rm in}$ from the simulation.  The last 3 columns give an estimate 
of the systematic error due to calibration and map-making, as defined in 
\S\ref{sec:cal_map_sims}. For comparison, the average power in the CMB in 
each band is $C_2 \sim 130$ $\mu$K$^2$, $\left<C_l\right>_{3-10} \sim 
150$ $\mu$K$^2$, and $\left<C_l\right>_{11-100} \sim 6$ $\mu$K$^2$.}
\label{tab:sim_res_map_ps}
\end{deluxetable}

\begin{deluxetable}{llccccccc}
\tablecaption{Difference Map Statistics}
\tablewidth{0pt}
\tablehead{
\colhead{DA} & \colhead{Diff. map\tablenotemark{a}} & \colhead{$C(\theta_{\rm beam})/C(0)$}
 & \colhead{$C_2$} & \colhead{$\left<C_l\right>_{3-10}$}
 & \colhead{$\left<C_l\right>_{11-100}$} & \colhead{$\left\vert\Delta C_l\right\vert_2$\tablenotemark{b}} 
 & \colhead{$\left\vert\Delta C_l\right\vert_{3-10}$\tablenotemark{b}} 
 & \colhead{$\left\vert\Delta C_l\right\vert_{11-100}$\tablenotemark{b}} \\
\colhead{} & \colhead{} & \colhead{} & \colhead{($\mu$K$^2$)} & \colhead{($\mu$K$^2$)} & \colhead{($\mu$K$^2$)}
 & \colhead{($\mu$K$^2$)} & \colhead{($\mu$K$^2$)} & \colhead{($\mu$K$^2$)}}
\startdata
K1  & $\phm{aa}{\bf\Delta}_{12}$   & $0.160$  & 107.30 & 1.77  &  0.12 & 107.25 &   1.72 &    0.071  \\
K1  & $\phm{aa}{\bf\Delta}_{34}$   & $0.014$  &   6.43 & 0.11  &  0.06 &   6.37 &   0.05 &    0.003  \\
K1  & $\phm{aa}{\bf\Delta}_{1234}$ & $0.030$  &  13.56 & 0.30  &  0.07 &  13.50 &   0.24 &    0.011  \\
Ka1 & $\phm{aa}{\bf\Delta}_{12}$   & $0.0057$ &   2.11 & 0.14  &  0.06 &   2.06 &   0.08 &    0.002  \\
Ka1 & $\phm{aa}{\bf\Delta}_{34}$   & $0.0022$ &   0.11 & 0.08  &  0.07 &   0.05 &   0.01 &    0.002  \\
Ka1 & $\phm{aa}{\bf\Delta}_{1234}$ & $0.0028$ &   0.01 & 0.09  &  0.07 &   0.05 &   0.03 &    0.003  \\
Q1  & $\phm{aa}{\bf\Delta}_{12}$   & $0.0035$ &   1.08 & 0.14  &  0.10 &   0.98 &   0.04 &    0.004  \\
Q1  & $\phm{aa}{\bf\Delta}_{34}$   & $0.0032$ &   0.49 & 0.16  &  0.13 &   0.37 &   0.03 &    0.003  \\
Q1  & $\phm{aa}{\bf\Delta}_{1234}$ & $0.0044$ &   0.31 & 0.17  &  0.13 &   0.19 &   0.05 &    0.003  \\
Q2  & $\phm{aa}{\bf\Delta}_{12}$   & $0.0031$ &   0.09 & 0.14  &  0.09 &   0.00 &   0.05 &    0.005  \\
Q2  & $\phm{aa}{\bf\Delta}_{34}$   & $0.0030$ &   0.28 & 0.13  &  0.10 &   0.18 &   0.03 &    0.001  \\
Q2  & $\phm{aa}{\bf\Delta}_{1234}$ & $0.0025$ &   0.03 & 0.11  &  0.10 &   0.07 &   0.01 &    0.004  \\
V1  & $\phm{aa}{\bf\Delta}_{12}$   & $0.0038$ &   4.40 & 0.35  &  0.16 &   4.24 &   0.19 &    0.006  \\
V1  & $\phm{aa}{\bf\Delta}_{34}$   & $0.0032$ &   0.22 & 0.35  &  0.28 &   0.05 &   0.09 &    0.016  \\
V1  & $\phm{aa}{\bf\Delta}_{1234}$ & $0.0024$ &   0.26 & 0.30  &  0.27 &   0.01 &   0.04 &    0.006  \\
V2  & $\phm{aa}{\bf\Delta}_{12}$   & $0.0043$ &   1.72 & 0.15  &  0.13 &   1.59 &   0.02 &    0.006  \\
V2  & $\phm{aa}{\bf\Delta}_{34}$   & $0.0026$ &   0.57 & 0.25  &  0.22 &   0.35 &   0.03 &    0.003  \\
V2  & $\phm{aa}{\bf\Delta}_{1234}$ & $0.0033$ &   1.05 & 0.38  &  0.23 &   0.83 &   0.16 &    0.008  \\
W1  & $\phm{aa}{\bf\Delta}_{12}$   & $0.0036$ &   5.45 & 0.36  &  0.36 &   5.11 &   0.02 &    0.025  \\
W1  & $\phm{aa}{\bf\Delta}_{34}$   & $0.0033$ &   2.10 & 0.74  &  0.56 &   1.56 &   0.21 &    0.019  \\
W1  & $\phm{aa}{\bf\Delta}_{1234}$ & $0.0034$ &   0.14 & 0.75  &  0.57 &   0.39 &   0.22 &    0.037  \\
W2  & $\phm{aa}{\bf\Delta}_{12}$   & $0.0029$ &   1.20 & 0.51  &  0.44 &   0.79 &   0.10 &    0.021  \\
W2  & $\phm{aa}{\bf\Delta}_{34}$   & $0.0026$ &   0.12 & 0.57  &  0.57 &   0.44 &   0.00 &    0.009  \\
W2  & $\phm{aa}{\bf\Delta}_{1234}$ & $0.0028$ &   0.26 & 0.56  &  0.59 &   0.31 &   0.01 &    0.025  \\
W3  & $\phm{aa}{\bf\Delta}_{12}$   & $0.0035$ &   4.95 & 0.49  &  0.47 &   4.49 &   0.03 &    0.013  \\
W3  & $\phm{aa}{\bf\Delta}_{34}$   & $0.0031$ &   8.84 & 0.90  &  0.75 &   8.12 &   0.19 &    0.039  \\
W3  & $\phm{aa}{\bf\Delta}_{1234}$ & $0.0039$ &   2.91 & 1.11  &  0.72 &   2.20 &   0.40 &    0.007  \\
W4  & $\phm{aa}{\bf\Delta}_{12}$   & $0.0030$ &   1.50 & 0.44  &  0.48 &   1.05 &   0.00 &    0.040  \\
W4  & $\phm{aa}{\bf\Delta}_{34}$   & $0.0025$ &   0.36 & 0.73  &  0.64 &   0.27 &   0.10 &    0.006  \\
W4  & $\phm{aa}{\bf\Delta}_{1234}$ & $0.0027$ &   0.73 & 0.98  &  0.65 &   0.10 &   0.35 &    0.021  \\
\enddata
\tablenotetext{a}{Difference maps from linear combinations of channels within 
a single DA, defined in equation~(\ref{eq:diff_map_def}).}
\tablenotetext{b}{Power in difference map in excess of white noise,
$\left\vert\left<C_l\right>_{\rm band} - \left<C_l\right>_{700-1000}\right\vert$.}
\label{tab:diff_analysis}
\end{deluxetable}

\begin{deluxetable}{lrrr}
\tablecaption{\map\ Boresight Pointing Vectors\tablenotemark{a}}
\tablewidth{0pt}
\tablehead{
\colhead{DA/Side} & \colhead{$n_x$} & \colhead{$n_y$} & \colhead{$n_z$}}
\startdata
K1A   &  $0.0399374$ & $ 0.9244827$ & $-0.3791264$ \\
Ka1A  & $-0.0383635$ & $ 0.9254372$ & $-0.3769539$ \\
Q1A   & $-0.0315719$ & $ 0.9521927$ & $-0.3038624$ \\
Q2A   &  $0.0319339$ & $ 0.9522016$ & $-0.3037965$ \\
V1A   & $-0.0331733$ & $ 0.9415643$ & $-0.3351958$ \\
V2A   &  $0.0333767$ & $ 0.9414947$ & $-0.3353711$ \\
W1A   & $-0.0091894$ & $ 0.9394385$ & $-0.3425944$ \\
W2A   & $-0.0095070$ & $ 0.9458644$ & $-0.3244228$ \\
W3A   &  $0.0098004$ & $ 0.9457678$ & $-0.3246956$ \\
W4A   &  $0.0098081$ & $ 0.9393480$ & $-0.3428252$ \\
K1B   &  $0.0379408$ & $-0.9239176$ & $-0.3807057$ \\
Ka1B  & $-0.0400217$ & $-0.9246344$ & $-0.3787473$ \\
Q1B   & $-0.0334030$ & $-0.9517688$ & $-0.3049925$ \\
Q2B   &  $0.0301434$ & $-0.9519277$ & $-0.3048361$ \\
V1B   & $-0.0350363$ & $-0.9409454$ & $-0.3367405$ \\
V2B   &  $0.0314445$ & $-0.9411385$ & $-0.3365553$ \\
W1B   & $-0.0114732$ & $-0.9388325$ & $-0.3441830$ \\
W2B   & $-0.0115900$ & $-0.9453501$ & $-0.3258511$ \\
W3B   &  $0.0076818$ & $-0.9454070$ & $-0.3258014$ \\
W4B   &  $0.0075141$ & $-0.9388923$ & $-0.3441291$ \\
\enddata
\tablenotetext{a}{Beam line-of-sight unit vectors in spacecraft coordinates.
Available in full precision in the released time-ordered data.}
\label{tab:boresight}
\end{deluxetable}

\begin{deluxetable}{lrrr}
\tablecaption{Measured Gain and Baseline Susceptibilities\tablenotemark{a}}
\tablewidth{0pt}
\tablehead{
\colhead{} & \colhead{$\partial {\bf g}/\partial T_{\rm FPA}$} 
           & \colhead{$\partial {\bf b}/\partial T_{\rm FPA}$} 
           & \colhead{$\partial {\bf b}/\partial V_{\rm bus}$} \\
\colhead{Radiometer} & \colhead{(du mK\per) K\per} & \colhead{mK K\per} & \colhead{$\mu$K V\per}}
\startdata
K11  & $-0.0021\phm{aaaaa}$ & $   3.52\phm{aaaaa}$ & $  0.1\phm{aaa}$ \\
K12  & $-0.0185\phm{aaaaa}$ & $   5.05\phm{aaaaa}$ & $  3.1\phm{aaa}$ \\
Ka11 & $-0.0024\phm{aaaaa}$ & $  -1.47\phm{aaaaa}$ & $  0.2\phm{aaa}$ \\
Ka12 & $ 0.0077\phm{aaaaa}$ & $   2.00\phm{aaaaa}$ & $ -3.2\phm{aaa}$ \\
Q11  & $-0.0037\phm{aaaaa}$ & $   3.79\phm{aaaaa}$ & $ -1.1\phm{aaa}$ \\
Q12  & $-0.0016\phm{aaaaa}$ & $  -3.52\phm{aaaaa}$ & $ -1.6\phm{aaa}$ \\
Q21  & $ 0.0086\phm{aaaaa}$ & $  -1.00\phm{aaaaa}$ & $ -2.1\phm{aaa}$ \\
Q22  & $ 0.0058\phm{aaaaa}$ & $  -0.57\phm{aaaaa}$ & $ -4.6\phm{aaa}$ \\
V11  & $ 0.0018\phm{aaaaa}$ & $  57.4\phm{aaaaaa}$ & $ 32.9\phm{aaa}$ \\
V12  & $-0.0045\phm{aaaaa}$ & $  -6.23\phm{aaaaa}$ & $ 17.2\phm{aaa}$ \\
V21  & $ 0.0029\phm{aaaaa}$ & $   6.10\phm{aaaaa}$ & $  3.4\phm{aaa}$ \\
V22  & $-0.0002\phm{aaaaa}$ & $  -9.43\phm{aaaaa}$ & $ -3.8\phm{aaa}$ \\
W11  & $ 0.0004\phm{aaaaa}$ & $ -14.7\phm{aaaaaa}$ & $ -4.3\phm{aaa}$ \\
W12  & $ 0.00002\phm{aaaaa}$ & $ -61.9\phm{aaaaaa}$ & $ -11.3\phm{aaa}$ \\
W21  & $ 0.0007\phm{aaaaa}$ & $-127.\phm{aaaaaaa}$ & $ -7.4\phm{aaa}$ \\
W22  & $-0.0004\phm{aaaaa}$ & $ -58.1\phm{aaaaaa}$ & $  0.4\phm{aaa}$ \\
W31  & $ 0.0003\phm{aaaaa}$ & $   4.49\phm{aaaaa}$ & $  5.3\phm{aaa}$ \\
W32  & $ 0.0021\phm{aaaaa}$ & $ -20.2\phm{aaaaaa}$ & $ 19.0\phm{aaa}$ \\
W41  & $ 0.0006\phm{aaaaa}$ & $  41.4\phm{aaaaaa}$ & $ 16.7\phm{aaa}$ \\
W42  & $-0.0011\phm{aaaaa}$ & $  19.9\phm{aaaaaa}$ & $  8.0\phm{aaa}$ \\
\enddata
\tablenotetext{a}{The thermal values are based on fits to a 10-day cooling 
period following a partial battery cell failure.  See 
Figure~\ref{fig:baseline_susc} and \S\ref{sec:thermal}.}
\label{tab:suscep_raw}
\end{deluxetable}

\begin{deluxetable}{lrrr}
\tablecaption{Limits on Spin-Synchronous Environmental Effects\tablenotemark{a}}
\tablewidth{0pt}
\tablehead{
\colhead{Radiometer/} & \colhead{Gain} & \colhead{Thermal} & \colhead{Voltage} \\
\colhead{Band} & \colhead{nK} & \colhead{nK} & \colhead{nK}}
\startdata
K11   &  $-1.2\phm{aa}$  &   $22\phm{aa}$ &   $0.3\phm{aa}$ \\
K12   & $-11.1\phm{aa}$  &   $32\phm{aa}$ &   $9.3\phm{aa}$ \\
Ka11  &  $-4.2\phm{aa}$  &   $-9\phm{aa}$ &   $0.5\phm{aa}$ \\
Ka12  &  $ 5.8\phm{aa}$  &   $13\phm{aa}$ &  $-9.7\phm{aa}$ \\
Q11   &  $-1.3\phm{aa}$  &   $24\phm{aa}$ &  $-3.3\phm{aa}$ \\
Q12   &  $-1.2\phm{aa}$  &  $-23\phm{aa}$ &  $-4.7\phm{aa}$ \\
Q21   &  $31.0\phm{aa}$  &   $-6\phm{aa}$ &  $-6.3\phm{aa}$ \\
Q22   &  $35.2\phm{aa}$  &   $-4\phm{aa}$ & $-13.8\phm{aa}$ \\
V11   &  $ 0.7\phm{aa}$  &  $367\phm{aa}$ &  $98.8\phm{aa}$ \\
V12   &  $-6.0\phm{aa}$  &  $-40\phm{aa}$ &  $51.7\phm{aa}$ \\
V21   &  $ 5.9\phm{aa}$  &   $39\phm{aa}$ &  $10.3\phm{aa}$ \\
V22   &  $-1.7\phm{aa}$  &  $-60\phm{aa}$ & $-11.3\phm{aa}$ \\
W11   &  $ 4.0\phm{aa}$  &  $-94\phm{aa}$ & $-13.0\phm{aa}$ \\
W12   &  $-0.1\phm{aa}$  & $-396\phm{aa}$ & $-33.9\phm{aa}$ \\
W21   &  $ 9.1\phm{aa}$  & $-812\phm{aa}$ & $-22.3\phm{aa}$ \\
W22   &  $-5.8\phm{aa}$  & $-372\phm{aa}$ &   $1.1\phm{aa}$ \\
W31   &  $ 1.1\phm{aa}$  &   $29\phm{aa}$ &  $16.0\phm{aa}$ \\
W32   &  $ 1.1\phm{aa}$  & $-129\phm{aa}$ &  $57.1\phm{aa}$ \\
W41   &  $21.1\phm{aa}$  &  $265\phm{aa}$ &  $50.1\phm{aa}$ \\
W42   & $-30.1\phm{aa}$  &  $128\phm{aa}$ &  $24.0\phm{aa}$ \\
\\
K     &  $ 6.2\phm{aa}$  &   $27\phm{aa}$ &   $4.8\phm{aa}$ \\
Ka    &   $0.8\phm{aa}$  &    $2\phm{aa}$ &   $4.6\phm{aa}$ \\
Q     &  $15.9\phm{aa}$  &    $2\phm{aa}$ &   $7.0\phm{aa}$ \\
V     &   $0.3\phm{aa}$  &   $77\phm{aa}$ &  $37.4\phm{aa}$ \\
W     &   $0.1\phm{aa}$  & $ 173\phm{aa}$ &   $9.9\phm{aa}$ \\
\enddata
\tablenotetext{a}{1$\sigma$ upper limits derived from measured gain and baseline 
susceptibilities in Table~\ref{tab:suscep_raw}, combined with upper limits
on temperature and voltage fluctuations at the spin period.  Sign is preserved
for each radiometer for roll-up by band.}
\label{tab:suscep_summary}
\end{deluxetable}

\begin{deluxetable}{lcc}
\tablecaption{Upper Limits on Radiometer Cross Talk}
\tablewidth{0pt}
\tablehead{
\colhead{DA} & \colhead{Electrical} & \colhead{Radiometric} \\
\colhead{} & \colhead{dB} & \colhead{dB}}
\startdata
K1  &  $-37.7$ &  $-26.8$ \\
Ka1 &  $-39.5$ &  $-30.4$ \\
Q1  &  $-41.6$ &  $-32.3$ \\
Q2  &  $-41.5$ &  $-32.2$ \\
V1  &  $-43.1$ &  $-35.2$ \\
V2  &  $-42.8$ &  $-35.4$ \\
W1  &  $-48.8$ &  $-48.3$ \\
W2  &  $-47.1$ &  $-43.5$ \\
W3  &  $-38.6$ &  $-42.6$ \\
W4  &  $-46.1$ &  $-47.5$ \\
\enddata
\label{tab:xtalk_summary}
\end{deluxetable}

\end{document}